\documentclass[twocolumn,times, trackchanges]{aastex62}

\usepackage{amsmath,amssymb}
\usepackage{graphicx}
\usepackage{CJKutf8}
\usepackage{gensymb}
\usepackage{enumitem}
\usepackage{longtable}
\usepackage{makecell}

\PassOptionsToPackage{hyphens}{url}\usepackage{hyperref}
\newcommand{\starbreak}{%
\begin{center}
    {*~~~~~~~~~~~*~~~~~~~~~~~*}
\end{center}
}

\usepackage[all]{hypcap} %Links go to figures;breaks on deluxetables

\newcommand{\mum}{\ifmmode{\rm \mu m}\else{$\mu$m}\fi}

\newcommand{\cosmology}{$\Omega _m = 0.27$, $\Omega_{\Lambda}=0.73$ and $H_0=71$km~s$^{-1}$Mpc$^{-1}$}

\submitjournal{ApJ}

\received{April 1, 2022}
\accepted{September 12, 2022}
\begin{document}

\title{\bf\Large  
    AGN Selection and Demographics in GOODS-S/HUDF from X-ray to Radio
}

%\correspondingauthor{Jianwei Lyu}
%\email{jianwei@email.arizona.edu}

\author[0000-0002-6221-1829]{Jianwei Lyu (\begin{CJK}{UTF8}{gbsn}吕建伟\end{CJK})}
\affiliation{Steward Observatory, University of Arizona, 
933 North Cherry Avenue, Tucson, AZ 85721, USA}
\author[0000-0002-8909-8782]{Stacey Alberts}
\affiliation{Steward Observatory, University of Arizona, 
933 North Cherry Avenue, Tucson, AZ 85721, USA}

\author[0000-0003-2303-6519]{George H. Rieke}
\affiliation{Steward Observatory, University of Arizona, 
933 North Cherry Avenue, Tucson, AZ 85721, USA}

\author[0000-0002-0303-499X]{Wiphu Rujopakarn}
\affiliation{
Department of Physics, Faculty of Science, Chulalongkorn University
, 254 Phayathai Road, Pathumwan, Bangkok 10330, Thailand}
\affiliation{
National Astronomical Research Institute of Thailand, 260 Moo 4, T. Donkaew, A. Maerim, Chiangmai 50180, Thailand
}

\begin{abstract}
  We present a comprehensive census of the AGNs in the GOODS-S/HUDF region from
  the X-ray to the radio, covering both the obscured and unobscured
  populations.  This work includes a robust analysis of the source
  optical-to-mid-IR SEDs featuring (semi-)empirical AGN and galaxy dust
  emission models and Baysian fitting techniques, ultra-deep VLA 3 and 6 GHz
  observations, and an integrated analysis of various AGN selection techniques,
  including X-ray properties, UV-to-MIR SED analysis, optical spectral
  features, mid-IR colors, radio loudness and spectral slope, and AGN
  variability. In total, we report $\sim$900 AGNs over the $\sim$170 arcmin$^2$
  3D-HST GOODS-S footprint, which has doubled the AGN number identified in the
  previous X-ray sample with $\sim$26\% of our sample undetected in the deepest
  Chandra image.  With a summary of AGN demographics from different selection
  methods, we find that no one single band or technique comes close to
  selecting a complete AGN sample despite the great depth of the data in
  GOODS-S/HUDF. We estimate the yields of various approaches and explore the
  reasons for incompleteness. We characterize the statistical properties, such
  as source number density, obscuration fraction and luminosity function of the
  AGN sample in this field and discuss their immediate implications. We also
  provide some qualitative predictions of the AGN sample that might be
  discovered by the upcoming JWST surveys.
\end{abstract}

% no longer useful

\section{Introduction}

%comment examples: \added{something to be added}, \deleted{something to be deleted}, \replaced{old text}{new text}, \explain{explanatory text}.

A comprehensive census of the active galactic nucleus (AGN) population is
essential to understand the growth of supermassive black holes (SMBHs) and its
relation with galaxy evolution. Due to the diversity of source properties,
manifold approaches have been proposed across the electromagnetic spectrum to
search for AGNs from the local Universe to the reionization age. However,
selection biases are inherent in each method and there is no convincing case
for a complete sample with single band selections \citep[e.g.,][]{Padovani2017,
Hickox2018}. How to fill the various holes in our current knowledge of AGN
demographics is a key question waiting to be addressed before reaching robust
conclusions on the BH-galaxy relation as well as the physical structure of
AGNs. { In this work, we will address the challenge of completing the AGN 
census by applying a comprehensive set of AGN identification 
tools on the deepest available relevant data, i.e., in the Great Observatories 
Origins Deep Survey - South (GOODS-S)/Hubble Ultra Deep Field (HUDF).}

{ Taking advantage of the available multi-wavelength datasets of deep sky surveys such as GOODS \citep{Dickinson2003}, COSMOS \citep{Scoville2007} and AEGIS \citep{Davis2007}, substantial 
efforts have been made previously to constrain the AGN census across wide ranges of source redshift and luminosity \citep[e.g.,][]{Alexander2003,Elvis2009, Laird2009, smolcic2009, Comastri2011,stern2012,smolcic2017,delvecchio2017,  Luo2017,algera2020}. Among them, the most extensive searches have 
focused on GOODS-South as it contains the deepest data from X-ray to radio, providing the best
possibility to characterize most, if not all, of the diversity in the AGN
population through the full wavelength range. This comprehensive set of data has supported searches for AGNs with a broad variety of 
approaches,} e.g.,  AGN identified from mid-IR color selection \citep{donley2008},
optical spectra \citep{Santini2009, Silverman2010}, variability
\citep{Villforth2010, Sarajedini2011, Pouliasis2019}, and multi-wavelength AGN
identifications among X-ray detected sources \citep{Luo2017} and radio-detected
galaxies \citep{Alberts2020}.  However, all these previous works have either
focused on one single selection technique or required the sample to be flux
limited at one wavelength, failing to reach a complete AGN sample in terms of
source bolometric luminosity.

\begin{figure}[htp!]
    \begin{center}
  \includegraphics[width=1.0\hsize]{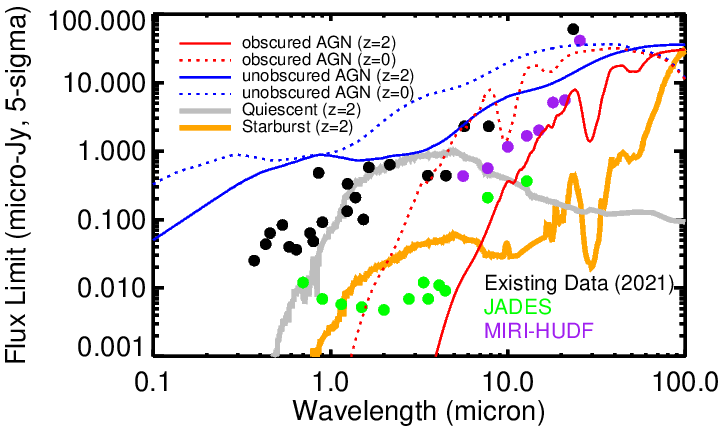}
    \caption{
    Five-sigma flux sensitivities as a function of band wavelength for the data
    used in this work and sensitivities predicted for the planned JWST GTO
    surveys.  A few AGN templates are also plotted to compare against the
    survey limits to demonstrate the importance of the mid-IR wavelength
    coverage in the identification of obscured AGNs. Two galaxy templates
    shifted to a redshift of 2, one for star-bursting galaxies (Arp 220) and
    another for quiescent galaxies (a 2-Gyr old elliptical galaxy), are shown
    for comparison.  At a typical redshift of $z \sim 2$, the 4--6 $\mu$m range
    provides good potential for AGN detection for both existing data and JWST
    MIRI-HUDF. 
    }
  \label{fig:mission_sen}
    \end{center}
\end{figure}

Another key question yet to be addressed is how to select obscured AGNs. The
widely applied approach to find AGNs through UV/optical color is insensitive to
obscured objects due to their reddened SEDs and strong galaxy contamination.
Identification with emission line diagnostics such as BPT diagrams
\citep{Baldwin1981} requires observationally expensive spectral observations
and performs unsatisfactorily for systems where host-galaxy emission lines
dominate either due to e.g., large-scale dust extinction or the AGN intrinsic
properties \citep[e.g.,][]{castello-mor2012,agostino2019}.  { Deep X-ray
surveys are widely employed for more efficient searches.} However, even in the
X-ray band, where the obscuration is believed to be minimal, a large number of
Compton-thick AGN can be missed \citep[e.g.,][]{buchner2021} and some AGNs
might be intrinsically X-ray weak \citep[e.g.,][]{leighly2004,simmonds2016}.
{ To help complete the sample, the X-ray searches have been supplemented in
    the mid-IR, where circumnuclear dust in almost all types of
    AGNs\footnote{ One possible exception is jet-mode AGNs where the dusty
    torus is may be insignificant \citep[e.g.,][]{heckman2014} in some
examples.} produces a distinct emission SED  shape compared with typical
galaxies, as illustrated in Figure~\ref{fig:mission_sen}. } Mid-IR missions
like {\it Spitzer} and {\it WISE} offer opportunities to find AGNs through
simple color cuts, which have proven to be quite successful
\citep[e.g.,][]{lacy2004, stern2005, Donley2012, stern2012, assef2018}.
Nonetheless, such techniques are limited largely to lightly obscured objects
and are biased toward high-luminosity AGNs. 
 Therefore, all these methods are likely to miss the most heavily obscured cases, which are of high interest as they may represent a unique (and early) stage in AGN evolution, for example as
predicted by the popular model in which AGNs initially form at the cores of
ultra-luminous infrared galaxies (ULIRGs) and are revealed as the surrounding
interstellar clouds dissipate \citep[e.g.,][]{hopkins2006}.

There are currently multiple
estimates of the fraction of heavily obscured AGNs, ranging from $\sim$ 10\%
\citep{mendez2013} to 30\% \citep{delmoro2016} to $>$ 50\% \citep{ananna2019}.
Some of this large variation results from different definitions and/or ambiguities and
degeneracies in identifications, but they alone cannot conceal the large set of
possibilities that can be clarified by a more thorough search.
Indeed, 
in recent years,
efforts have been made to identify AGNs more thoroughly using IR SED fitting
\citep[e.g.,][]{assef2011, Pozzi2012, Chung2014, Alberts2020, algera2020}.  However, most of these
works did not include the full range of behavior of obscured objects,
making even the SED-identified samples biased and incomplete.

{ To advance our characterization of the AGN census, this paper improves on some of the previous methods 
and introduces new ones to search for AGNs in GOODS-S.} We use this experience to discuss how to conduct more thorough searches
with the upcoming deep multi-band SED data from the James Webb Space Telescope
(JWST) that provides continuous coverage of the IR over 1--26 $\mu$m in multiple bands.
Our major goals are to build up a
comprehensive census of the AGN in GOODS-S/HUDF with (almost) all the available
data, characterize the performance of different selection methods, and discuss
the AGN demographics and what can be learned from these results. Compared with our previous work 
\citep{Alberts2020}: (1) we have included the full area of
the newly reduced, deeper VLA 3 GHz map, increasing the number of AGN detected
in the radio-band; (2) we have added a model of obscured AGN SEDs to an
improved set of unobscured ones as the basis for our SED fits; (3) we use a
state-of-the-art Bayesian fitting approach to constrain the SED model, to
identify AGN candidates, and to measure the relevant source properties; 
and (4) our AGN sample is not limited to the radio-detected population but
contains sources identified by any of radio, X-ray, UV, or mid-IR properties,
with the goal to build a bolometric luminosity limited sample with a large
range of source properties. All these efforts will be integrated together to make
the best estimates of the AGN statistical properties (e.g., number density,
obscured AGN fraction, luminosity { function}) covering a wide range of 
redshift and to establish a foundation for future AGN study.

This paper is organized as follows. Section~\ref{sec:data} describes the data
and parent sample and Section~\ref{sec:agn-select} introduces the SED analysis
as well as other AGN selection methods.  Section~\ref{sec:agn-sample}
summarizes the final AGN sample and presents a value-added GOODS-S AGN catalog.
We put in-depth analysis and discussions of our results in
Section~\ref{sec:discuss}, which includes a summary of AGN demographics
selected by different methods, the various reasons behind the failures of AGN
identification by some methods, the statistical properties of the AGN sample
(e.g., sky number density, obscuration fraction, luminosity function) and
some qualitative predictions for the AGN sample that can be discovered by the
upcoming JWST GTO surveys of this field.  Section~\ref{sec:summary} is a final
summary.

Throughout this paper, we adopt the cosmology \cosmology.

\section{Data and Parent Sample}\label{sec:data}

A major part of our analysis focuses on radio sources detected in ultra-deep
JVLA imaging at 3 and 6 GHz. In this section, we first describe the new radio
data and source extraction.  We then discuss the parent sample and the
multi-wavelength counterparts used to identify AGNs both detected and
undetected in the radio images.

\subsection{New JVLA Observations and Radio Source Catalog}

\begin{figure*}[htp]
    \begin{center}
        \includegraphics[width=1.0\hsize]{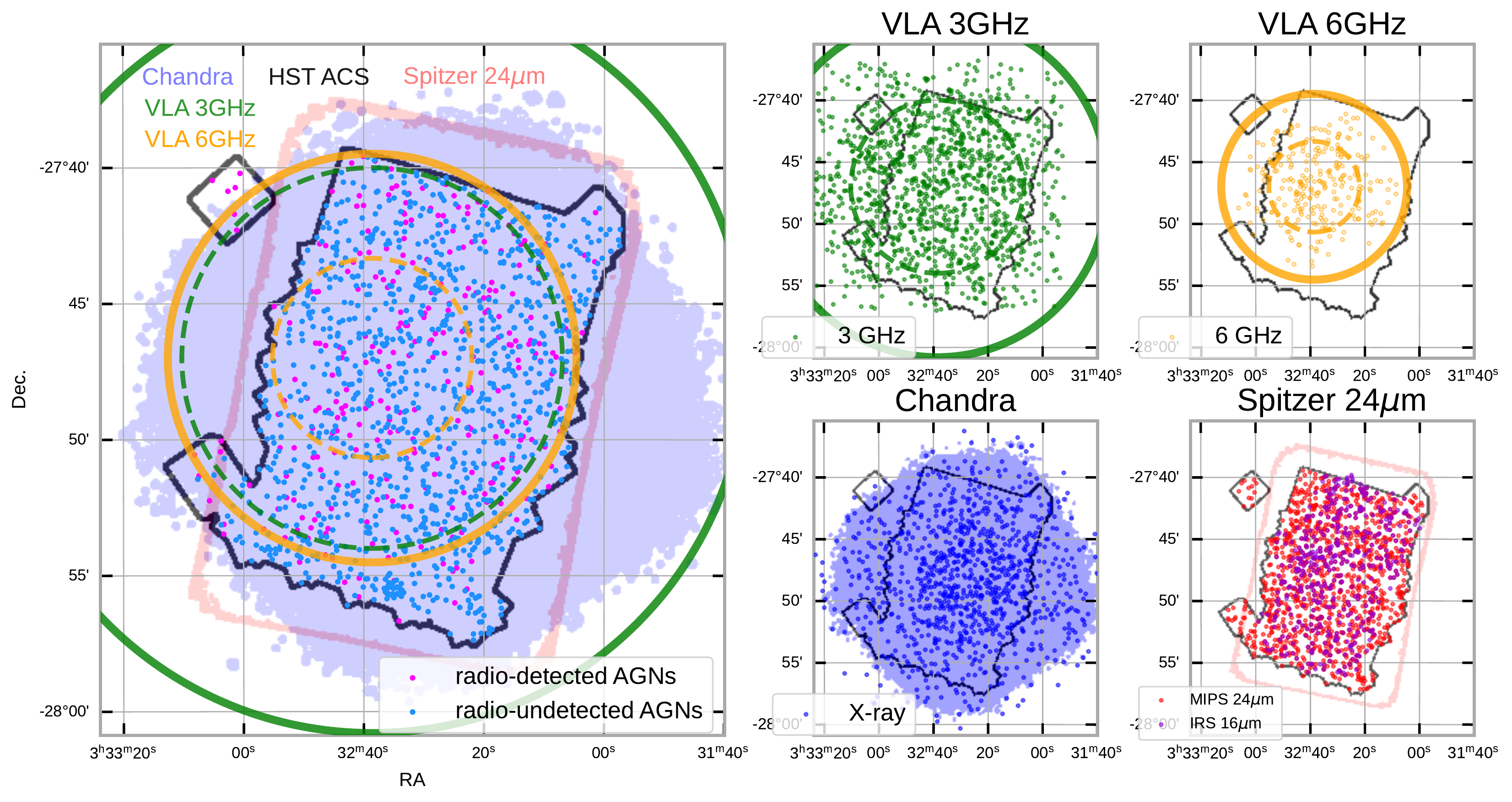}
        \caption{
            Left: Survey footprints of the GOODS-S: {\it Chandra} 2--7 KeV
            (blue), 3D-HST 0.3--8.0$\mum$ (black), {\it Spitzer}/MIPS 24 $\mum$
            (red) and VLA 3 GHz (green) and 6 GHz (orange) observations. For
            the VLA surveys, we have denoted the sizes of the 3 and 6 GHz
            half-power primary beam (HPBW) in dashed circles and the primary
            beam cutoffs in solid circles.  We have also highlighted the
            location of radio-detected and -undetected AGNs in this area.
            Right: Four small panels show the footprints of different surveys
            and the distributions of corresponding sources.
        }
  \label{fig:map}
    \end{center}
\end{figure*}

Radio imaging in S-Band (2--4 GHz) and C-band (4--8 GHz)  centered on
GOODS-S/HUDF (3:32:38.6, $-$27:46:59.89) was obtained with the Karl G. Jansky
Very Large Array (JVLA) with half-power primary beam (HPBW) sizes of 420\arcsec
and 220\arcsec, respectively, as shown in Figure~\ref{fig:map}.  The total
177-hour 6 GHz data set has been described in \citet{Alberts2020}. Partial
depth 3 GHz (S-band) data were also used there; however, our current work uses
the full 3 GHz depth obtained from programs VLA/18A-199 and VLA/19A-242 (PI:
Wiphu Rujopakarn), totaling 190 hours. 

The reader is referred to \cite{Alberts2020} for a summary of the data
reduction using CASA \citep{McMullin2007}.  For source extraction, two 3 GHz
images were made, one at the native resolution (1.16\arcsec$\times$0.55\arcsec
synthesized beam) and the other with a 100k$\lambda$ taper resulting in a
synthesized beam of 1.49\arcsec$\times$1.10\arcsec,  to capture extended
sources. The rms noise at phase center at native resolution is 0.51$\,\mu$Jy
beam$^{-1}$, corresponding to a point source 5-$\sigma$ sensitivity of
2.55$\,\mu$Jy. Full details of the 3 GHz imaging and catalogs will be presented
in W. Rujopakarn, in prep.

The 6 GHz (C-band) imaging used in this work is as described in
\citet{Alberts2020} with the following differences: a re-reduction was
performed resulting in a marginal increase in the rms sensitivity
(0.29$\,\mu$Jy beam$^{-1}$ at phase center).  To recover extended sources,
images were created at native resolution (0.62\arcsec$\times$0.31\arcsec
synthesized beam) and three tapers: 300k$\lambda$, 200k$\lambda$, 100k$\lambda$
resulting in synthesized beam sizes of 0.74\arcsec$\times$0.45\arcsec,
0.83\arcsec$\times$0.59\arcsec, 1.20\arcsec$\times$1.05\arcsec, respectively. 

Source extraction was done down to S/N$_{\rm peak}$= 5 independently on all
native resolution and tapered images using the {\tt Python Blob Detection and
Source Finder} (PyBDSF; \citealt{PyBDSF}) software package. The extraction was
performed on images uncorrected for the primary beam. PyBDSF extracts the integrated 
fluxes after source detection and the flux reported is always integrated flux.  The fluxes are then
corrected using the primary beam profile. The S/N will peak in the image where 
the tapering scale matches the intrinsic source extent. Source fluxes were adopted for the
highest S/N detection among the different resolutions. The 3 and 6 GHz catalogs
were matched using a search radius of 1.5\arcsec\footnote{This large search
radius helps match extended sources that are broken into multiple islands in
the 6 GHz map. The map is sufficiently sparse that this does not lead to
mismatch, as confirmed by visually inspecting the cataloged sources.}. 1276 3
GHz sources were detected over the full area, with 278 detected at both 3 and
GHz and 998 at 3 GHz only.  The primary beam is four times smaller in area at 6
GHz, so of the total of 1276 detected sources, one would predict 319 to be
detected at the higher frequency. That only 87\% of this number are  detected
indicates that a significant fraction, about 1/8, have radio SED slopes steeper
than $-$0.8, which is the canonical value for star-forming galaxies
\citep{condon1992}. The final radio sources are distributed over the primary
beams with cut-offs at radii of $\sim$830\arcsec at 3~GHz and $\sim$450\arcsec
at 6~GHz, corresponding to areas of $\sim$600~arcmin$^2$ and
$\sim$180~arcmin$^2$.

\subsection{Parent Sample and Multi-wavelength Counterparts}

We have generated our prime sample from this very deep radio image, which
brings a number of advantages to our science. { At our detection limits,} most AGNs themselves are not
evident in the radio band as the emission is dominated by the host galaxy star
forming activity \citep[e.g.,][]{smolcic2017,algera2020, Alberts2020}.  Since for most objects, we are not
detecting signals from the AGNs but from their host galaxies, this selection
mitigates biases resulting from radio AGN properties\footnote{Although we will show that the radio-detected AGNs are similar to those not detected, we expect a small bias in this regard due to the proportionalities between star formation rate, stellar mass, and supermassive black hole mass.}. In addition, the
optical depths in the GHz range are extremely low, so the deep JVLA surveys
sample AGNs at different obscuration levels in an unbiased way. 

We first build a 3GHz-flux-limited sample of 759 radio-detected { objects}
within the GOODS-S 3D-HST footprint { and search their multi-wavelength data for
AGN identification, as discussed in Section~\ref{sec:sample-radio}. To complete the AGN census, we will
also search for AGN in another sample of 1129
radio-undetected { objects}, defined in Section~\ref{sec:sample-norad}.} This secondary sample  can include bolometrically luminous AGNs with un-detected radio emission from either their nuclei or host
galaxies. We will show that, { besides some Malmquist bias}, the intrinsic properties (e.g.,
bolometric luminosity, redshift and obscuration level) of the AGNs in the two samples are similar { (see Sections~\ref{sec:agn-measure} and \ref{sec:bias_detection})}.

\subsubsection{Radio-Detected Sample}\label{sec:sample-radio}

The GOOD-S 3D-HST (v4.1) photometric catalog \citep{Brammer2012, Skelton2014}
provides 20 photometric bands at observed-frame 0.3 -- 8 $\mu$m in this field
and builds the foundation of our multi-wavelength SED analysis. After
correcting the coordinate errors of this catalog \citep[see Section 3
in][]{Franco2020}, we searched for the optical-to-mid-IR counterparts of the
VLA 3 GHz sources within a matching radius of $0.\arcsec5$ and found 759 unique
objects.\footnote{As shown in Figure~\ref{fig:map}, the coverage of the 3D-HST
    catalog limited the footprint of the radio sources associated with 3D-HST
    objects, but it did not reduce the sensitivities { to detect possible AGNs}. In fact, almost all radio
    sources within the HST GOODS-S footprint have optical counterparts (besides
a few low-$z$ extended optical sources where the radio emission is broken up).
As a result, this sample can be considered complete by radio flux.} 
{ This radius is
adopted following \cite{Alberts2020}; it encompasses the offsets found between
the optical and radio flux peaks \citep{rujo2016}. In fact, all counterparts are found within
$0.\arcsec2$ and the median separation between radio and optical is $\sim0.\arcsec05$, consistent with the intra-source feature/variations at $\lesssim$0.5 kpc given the redshifts of our sources. With the source density of 3D-HST sources
and our $>5\sigma$ detection criteria, the predicted number of false detection is well below 0.5.} 
This
defines the sample of radio-detected { objects} where our SED analysis and other
AGN selection techniques will be applied and compared.

Following \cite{Alberts2020}, we identified multi-wavelength counterparts of
these objects. This includes: (1) the MIPS 24~$\mum$ measurements from the
Rainbow Cosmological Database \citep{Perez-Gonzalez2008, Barro2011a,
Barro2011b} (707/759), (2) the IRS 16~$\mum$ measurements from
\cite{Teplitz2011} (309/759), and (3) X-ray counterparts in the {\it Chandra} 7
Ms X-ray catalog \citep{Luo2017} within a 2$\arcsec$ radius (342/759).  We also
retrieved the object redshifts (for 756 of the 759 sources) from the 3D-HST
``zbest'' catalog \citep{momcheva2016} and use them for our SED analysis later in this paper.

\subsubsection{Radio-Undetected Sample}\label{sec:sample-norad}

Although our VLA images are very deep, we could miss a large number of AGNs in
sources not detected in the radio band. For example, the radio emission of a
radio-quiet AGN in a low SFR or quiescent galaxy could be too weak to be
detected even if the AGN is reasonably bright at other wavelengths.
Consequently, to complete the AGN census in terms of bolometric luminosity, we
also searched for AGNs in the following categories of radio-undetected { objects}
within the 3D-HST footprint:

\begin{itemize}
    \item 437 X-ray sources: We include all the other X-ray sources in the {\it
        Chandra} 7 Ms X-ray catalog \citep{Luo2017} within the 3D-HST footprint
        that do not have 3 GHz radio counterparts. We will utilize their X-ray
        properties and  various additional methods to identify AGNs (see
        Section~\ref{sec:x-ray-select}).
        
    \item 543 objects with 3-sigma detections in the MIPS 24 $\mum$ image
        and 154 objects with 3-sigma detections in the IRS peakup 16 $\mum$
        band: These objects have reasonable mid-IR wavelength coverage,
        allowing us to carry out SED analysis to search for AGNs.

    \item 135 objects with IRAC colors that satisfy the AGN selection criteria
        in \citealt{Kirkpatrick2017}: In total, we have 3979 3D-HST
        sources with 3-sigma detections in all IRAC bands\footnote{3D-HST
            spectra in GOODS-S are largely of galaxies with $0.5 < z < 3.0$, of
            which a total of $\sim$ 8000 have redshift measurements
        \citep{momcheva2016}.}, of which 671 have detailed SED analysis from
        their membership in other parts of the sample.  For the 3308 remaining
        objects we have applied IRAC color selections to pre-select AGNs and
        used  SED fitting to improve the selection credibility. See
        Section~\ref{sec:irac-select}. 

    \item 113 variable sources either in X-ray or optical bands, as reported in
        the literature. See Section~\ref{sec:var-select}.
\end{itemize}

%IDL> help, where(mips_det)
%<Expression>    LONG      = Array[543]
%IDL> help, where(irs_det)
%<Expression>    LONG      = Array[154]
%IDL> help, where(irac_col)
%<Expression>    LONG      = Array[135]
%IDL> help, where(xray_det)
%<Expression>    LONG      = Array[364]
%IDL> help, where(xagn)
%<Expression>    LONG      = Array[228]
%IDL> help, where(var)
%<Expression>    LONG      = Array[113]
%IDL> help, var
%VAR             FLOAT     = Array[1127]

However, we do not extend this study to all the sources in the 3D-HST catalog
but focus on a subset of the sample as defined above. These criteria ensure
that these objects have the relevant data to search for both obscured and
unobscured AGNs (the selection methods will be presented in
Section~\ref{sec:agn-select}). Meanwhile, the other 3D-HST sources only have
optical-to-near-IR SED constraints and the selection would result in undesired
bias against obscured AGNs in the final sample.

This radio-undetected sample contains 1127 unique { objects}.  Similarly to the
radio-detected sample, we also compile the multi-wavelength counterparts for
these objects to identify AGNs with all possible methods and to characterize
their physical properties later.

\starbreak

In summary, our whole parent sample has 1886 unique objects, which include all
the 3D-HST sources that are (1) detected on the VLA 3 GHz images in the radio
(referred to as ``radio-detected'' hereafter), (2) included in the {\it
Chandra} 7Ms Catalog in the X-ray (referred as ``X-ray detected'' hereafter),
(3) detected in the {\it Spitzer} MIPS $24~\mum$ or IRS 16~$\mum$ images or
showing AGN-like IRAC color in the mid-IR (referred to as ``mid-IR detected"
hereafter), or (4) time variable in the HST or {\it Chandra} images. Given the
selections and the multi-wavelength data available for these objects, we expect
reasonably unbiased detections of objects that host AGNs over a wide range of
obscuration level.  In Figure~\ref{fig:mission_sen}, we summarize the 5$\sigma$
sensitivities of various photometric bands used for our AGN identification.
Figure~\ref{fig:map} shows the footprints of the multi-wavelength data as well
as the various source locations in the GOODS-S field.

\section{AGN Selection Methods}\label{sec:agn-select}

We have developed a sophisticated SED analysis technique to identify AGNs and we also 
explore seven other AGN selection methods on our parent sample.  We first
introduce the SED analysis approach in Section \ref{sec:sed} and then describe
the other AGN selection techniques in Section~\ref{sec:other-method}.

\subsection{A Robust SED Analysis}\label{sec:sed}

To reveal the AGN population and constrain the various properties of its
members, we have utilized a new SED analysis tool with (semi-)empirical
templates calibrated against various observations.  Section~\ref{sec:sed-tool}
gives a brief description of this tool, Section~\ref{sec:application} discusses
how we have adapted it in the current study, and Section~\ref{sec:sed-result}
presents some example fittings and discusses how the AGNs are identified by SED
analysis. 

\subsubsection{The Fitting Tool}\label{sec:sed-tool}

\begin{figure*}[h!tp]
    \begin{center}
  \includegraphics[width=1.0\hsize]{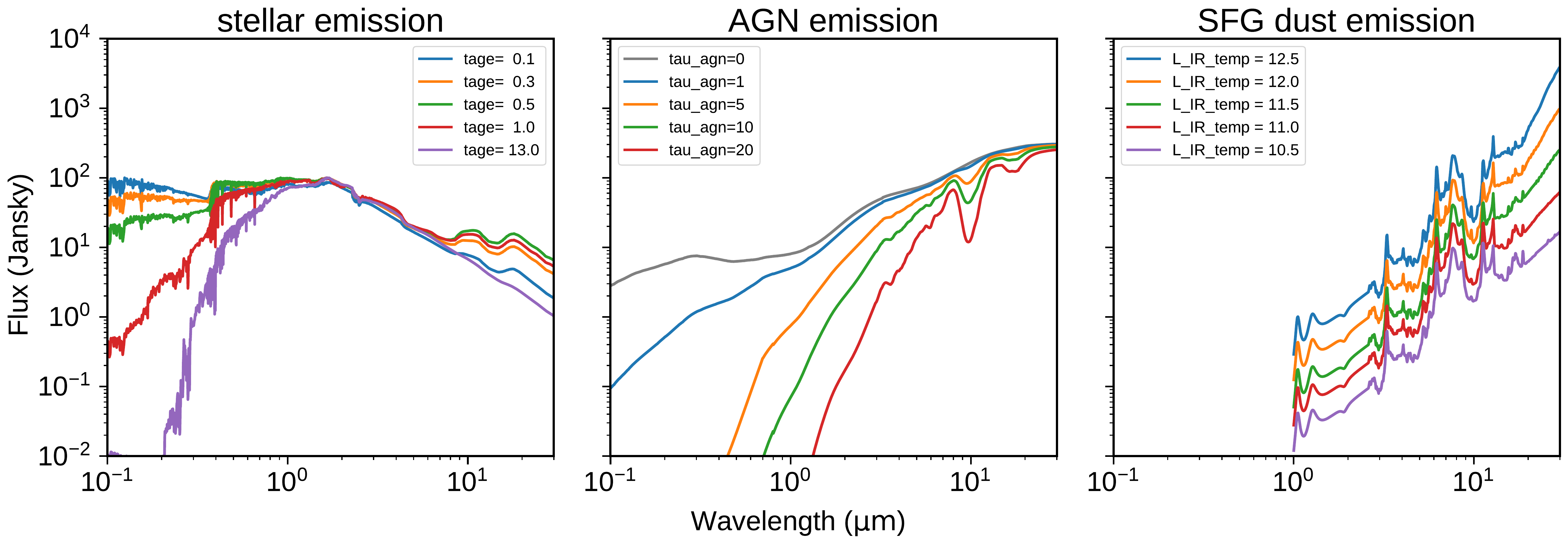}
    \caption{
     Examples of the galaxy stellar (left panel), AGN (middle panel), and SFG
     dust emission (right panel) templates used in our SED model.  We have
     shown the variations caused by the one single parameter that leads to the
     most significant changes of the SED model. See the text for more details.
     The stellar emission is dominated by photospheric output at wavelengths
     $<$ 10 $\mu$m and by reradiation by circumstellar dust at longer
     wavelengths.
    }
  \label{fig:sed_template}
    \end{center}
\end{figure*}

To select AGNs from SED analysis, we need (1) libraries of accurate
templates that capture the variations of AGN and galaxy SEDs without much
redundancy and (2) fitting algorithms that reveal and evaluate the possible
degeneracy of models when the data are limited. Considering the limited mid-IR
photometric bands accessible before JWST, selection of templates that provide
accurate fits with minimal free parameters is critical. Thus, we have
constructed semi-empirical templates that are calibrated against various
observations for the AGN optical to far-IR continuum emission and star-forming
galaxy (SFG)  dust emission, as detailed in Appendix A. In short, the AGN
component is based on an empirical AGN template library introduced in
\cite{Lyu2017a, Lyu2017b, Lyu2018} with its obscuration controlled by an
empirical AGN optical-to-IR attenuation curve. The SFG dust emission is based
on an improved version of the \cite{Rieke2009} templates with the stellar
emission subtracted and the wavelengths extended into the near-IR.
Figure~\ref{fig:sed_template} illustrates some examples of these SED models.

\begin{deluxetable*}{@{\extracolsep{4pt}}ccccc}
    %\tabletypesize{\scriptsize}
    \tabletypesize{\footnotesize}
    \tablewidth{1.0\hsize}
    \tablecolumns{5}
    \tablecaption{Parameter Setup of the Modified {\it Prospector} Code \label{tab:model-setup}
    }
    \tablehead{
  \colhead{Parameter} &
  \colhead{Free?} & %\tablenotemark{1}} &
  \colhead{Prior/Value} &
  \colhead{Unit} &
  \colhead{Comment}
}
\startdata
\multicolumn{5}{c}{Stellar Component (FSPS)} \\
\hline
  mass        & True  &  [1e8, 1e12], LogUniform      &  $M_\odot$   &   Solar masses formed   \\
  logzsol     & True  &  [-2, 0.19], TopHat           &   $\log (Z/Z_\odot)$  & Stellar metallicity \\
  dust2       & True  &  [0.0, 4.0], TopHat           &  dimensionless & stellar optical depth at 5500\AA   \\
  tage        & True  &  [0.001, 13.8], TopHat        &  Gyr   & stellar age \\
  tau         & True  &  [0.1, 30], LogUniform        &  Gyr$^{-1}$  & E-folding time of the SFH\\
  dust\_type  & N/A   &  2  &  N/A   & Calzetti attenuation curve is selected \\
  sfh         & N/A   &  4  &  N/A   &  Delay-tau SFH is selected \\
  \hline
\multicolumn{5}{c}{AGN Component} \\
\hline
  L\_AGN      & True  &  [1e-4, 1e6], LogUniform    &  $10^{10}L_\odot$  & AGN bolometric luminosity \\
  f\_hd       & False &  1.0, fixed      &  dimensionless  & relative strength of the AGN hot-dust component \\
  f\_wd       & False &  1.0, fixed      &  dimensionless  & relative strength of the AGN warm-dust component \\
  f\_pol      & False &  0.0, fixed      &  dimensionless  & relative strength of the AGN polar-dust component \\
  tau\_agn    & True  &  [0, 20], TopHat      &  dimensionless &  AGN continuum optical depth at 5500\AA \\
  \hline
\multicolumn{5}{c}{SFG Dust Component} \\
\hline
  L\_IR\_obs  & True  & [1e-4, 1e6], LogUniform & dimensionless &  Scaling factor of the SFG IR component \\
  L\_IR\_temp & False & 11.25, fixed &  $\log (L_{\rm IR, SF}/L_\odot)$  &  SFG template IR luminosity (8--1000~$\mu$m)\\
  f\_pah      & False &  1.0, fixed   & dimensionless & relative strength of the PAH component\\ 
\enddata
%\tablenotetext{1}{All these priors have been sampled linearly.}
%\tablenotetext{2}{The ``uncertainties'' of these median values are 2 $\sigma$ (i.e., 2.5\%, 50\% and 97.5\% quantiles).}
\end{deluxetable*}

We have integrated our empirical templates of AGN continuum emission and galaxy
dust emission with the FSPS stellar model within the SED fitting framework
provided by {\it Prospector} \citep{Johnson2021}. This code applies forward modeling
techniques to match the photometric data and infers the posterior parameter
distributions with Monte Carlo techniques. We have kept the FSPS stellar
component of the original {\it Prospector} code unchanged and replace both the
\cite{DL2007} dust model and the \cite{Nenkova2008} AGN torus model with our
(semi-)empirical SED model of SFG dust emission and AGN continuum emission
(see Appendix~\ref{app:sed_input}). { Compared to the original
theoretical models adopted in {\it Prospector}, ours are based directly on observations that include features that are missing in the other models\footnote{E.g., the \citealt{DL2007}
model does not include mid-IR silicate absorption features typically found for IR luminous galaxies; the
\citealt{Nenkova2008} model does not include AGN hot dust emission and the version adopted
in {\it Prospector} does not include AGN accretion disk emission in the  UV-optical from the type-1 AGN.}. Our  templates have been optimized to minimize the number of free parameters while preserving an adequate description of the modeled behavior. Further discussion of their applicability can be found in \cite{Lyu2022}.
} In addition, no energy balance between the
stellar extinction and galaxy IR emission is introduced, because of the strong
anisotropy of extinction in typical disk galaxies and the uncertain effects on the extinction behavior 
caused by the introduction of an AGN.  Table~\ref{tab:model-setup} provides a
summary of the various parameters in the SED model. 

In the fitting, we use Dynamic Nested Sampling to randomly sample the entire
parameter space given by a prior and to reveal any model degeneracy.  Unlike
the traditional Markov Chain Monte Carlo (MCMC) or similar methods, Nested
Sampling estimates both the Bayesian evidence and posterior of the fitting
parameters and has a number of appealing statistical properties, as described
in \cite{Speagle2020}.  In our SED analysis, the prior of the AGN bolometric
luminosity, {\it L\_AGN}, is assumed to be a log-uniform distribution from 10$^6$
$L_\odot$ to 10$^{16}~L_\odot$\footnote{ Fundamentally, {\it L\_AGN} describes the scaling factor
of the AGN component and we adopt the optical bolometric luminosity correction in \cite{Duras2020} to convert the fitted bolometric luminosity to the observed flux. See Section~\ref{sec:agn-measure}.} and the AGN obscuration level, tau\_agn, is
assumed to have a top-hat prior with the value ranging from 0 (unobscured) to
20 (heavily obscured).  Based on various tests to balance the running speed and
the quality of the fittings, we adopt the nested posterior threshold at 0.1;
the fitting solution is accepted when the fractional scatter in the
Kullback-Leibler (K-L) divergence reaches this value \citep{Skilling2004}. 

{ The validity of this SED modeling has been confirmed  with hundreds
of low-$z$ AGN and galaxies that have good IR SED constraints from 2MASS,
WISE, Spitzer, Herschel and AKARI, as described in previous publications \citep[e.g.,][]{Lyu2016,Lyu2017a,Lyu2017b,Lyu2018}. The upgrade of the fitting to utilize {\it Prospector} has also been tested on such samples (Lyu et al. in prep) and we will publish
more details of this tool and its performance in the near future.}

\subsubsection{Application to the Current Sample}\label{sec:application}

The full model described above is appropriate for studying galaxies with many
observations, e.g., spectra as well as detailed photometry.  Depending on the
wavelength coverage, the number of photometric constraints, and prior knowledge
of the source properties, we can increase or reduce the complexity of our SED
model by changing the setup of various parameter variations.  In this paper,
our goal is to increase the number of AGN candidates based on a sophisticated
analysis of existing relatively sparse photometry. We now describe how we have 
minimized the number of free parameters for this work, given that our fits are
based in the infrared on a modest number of photometric points. 

In the sources of interest for detecting AGN, the near infrared stellar
continuum is virtually  uniform in shape across a broad variety of galaxy types
\citep{willner1984,mannucci2001,mcintosh2006, brown2014}.  Thus, the five free
parameters for the stellar continuum used in {\it Prospector} basically collapse to a
single parameter (the normalization) for the infrared; the other parameters are
sensitive to the presence or absence of an excess in the ultraviolet, with
little influence on the search for excess in the infrared. Similarly, the
infrared excess spectra of star forming galaxies of the luminosities and
metallicities of interest are dominated by the aromatic bands, which over the
3--15 $\mu$m range important for our fits (given typical redshifts of the
sources) vary little in relative strength \citep{smith2007}. A single free
parameter (again the normalization) is sufficient to fit this component. A similar choice has been made in other fitting programs, e.g., MAGPHYS \citep{dacunha2008}. For
the AGNs,  { \citet{Lyu2017a} provide three templates that span the observed near-infrared behavior: normal, warm dust deficient (WDD), and hot dust deficient (HDD). HDD outputs are sufficiently weak in the near infrared that they are likely to be overlooked by our fitting. Normal and WDD cases are similar in the critical rest wavelength range (2--6 $\mu$m) for our fits (see Figure 3 in \citet{Lyu2017a}). We use the normal template that for our purposes should work equally well for both.}   We find
there is no need to include a polar dust component (see \citealt{Lyu2018})
since at the redshifts, wavelengths, and luminosities of interest, good fits
are obtained for the great majority of AGNs without one \citep[e.g.,][]{xu2015,
Lyu2017a, Alberts2020}. We account for obscuration (e.g., Type-2 vs. Type-1) with a single
obscuration parameter of continuously adjustable strength.  Thus, the AGN fits
require just two free parameters, normalization and degree of obscuration.
Further details are provided in Appendix~\ref{app:sed_input}. 

The simplifications we have introduced are very unlikely to result in false
identifications of AGNs. AGNs are apparent through excess over the photospheric
SED in the 3--6 $\mu$m range, where star forming galaxies have only a weak
excess continuum: the aromatic band at $\sim$ 3.4 $\mu$m is weak, and the
emission by AGN polar dust is generally insignificant. This identification can
be reinforced if the galaxy has a relatively weak output at 10 - 20 $\mu$m; an
AGN with strong emission by polar dust would therefore tend to have overly
strong emission in this latter range, { potentially resulting in classification 
as a star forming galaxy rather than an AGN.}
Similarly, the weak continuum in the 2--6 $\mu$m range for HDD AGNs could cause them to be overlooked. 

\subsubsection{AGN SED Identification}\label{sec:sed-result}

\begin{figure*}[htp]
    \begin{center}
  \includegraphics[width=0.49\hsize]{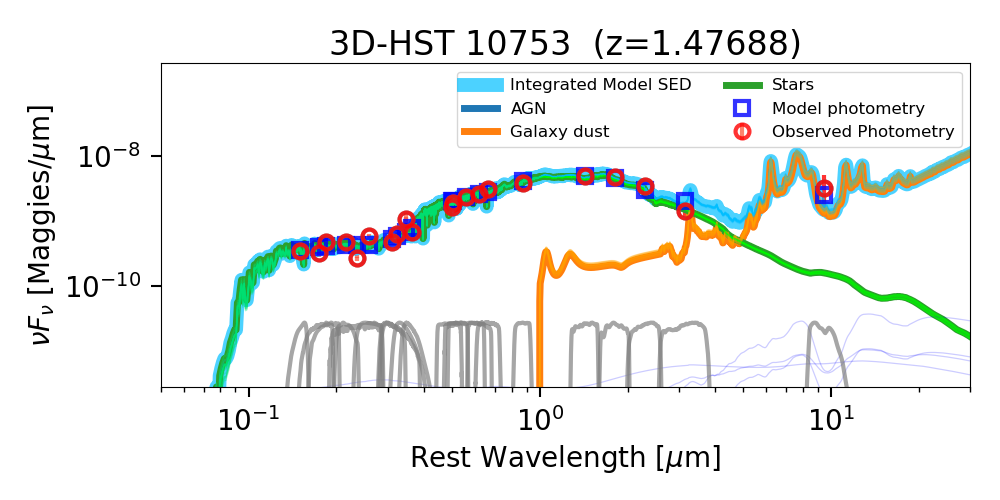}
  \includegraphics[width=0.49\hsize]{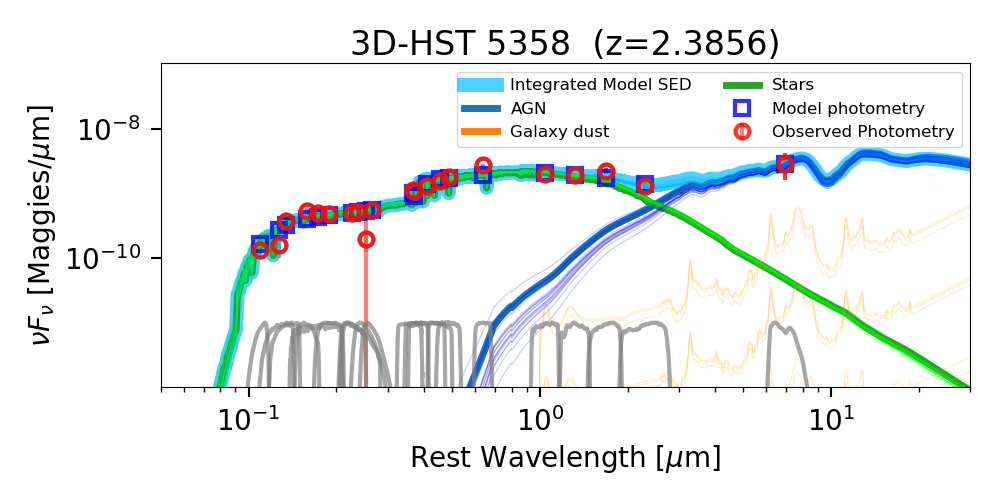}
  \includegraphics[width=0.49\hsize]{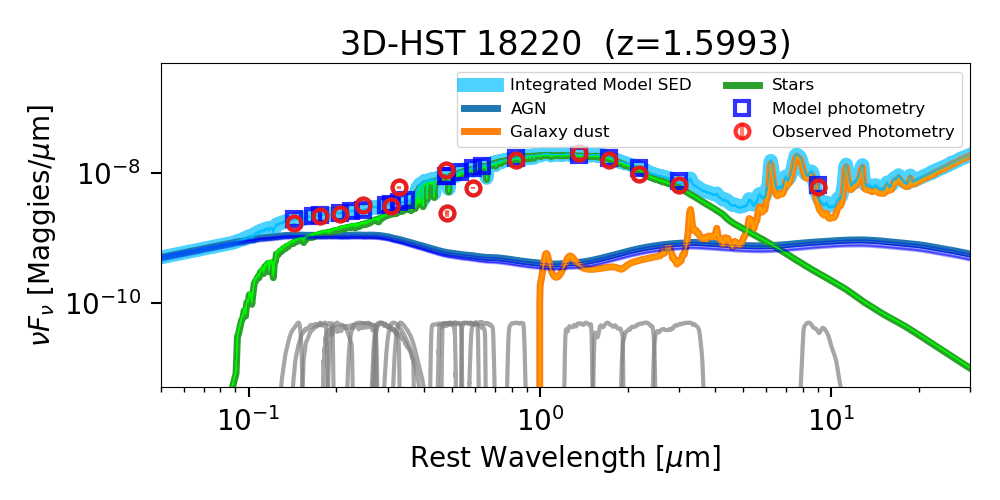}
  \includegraphics[width=0.49\hsize]{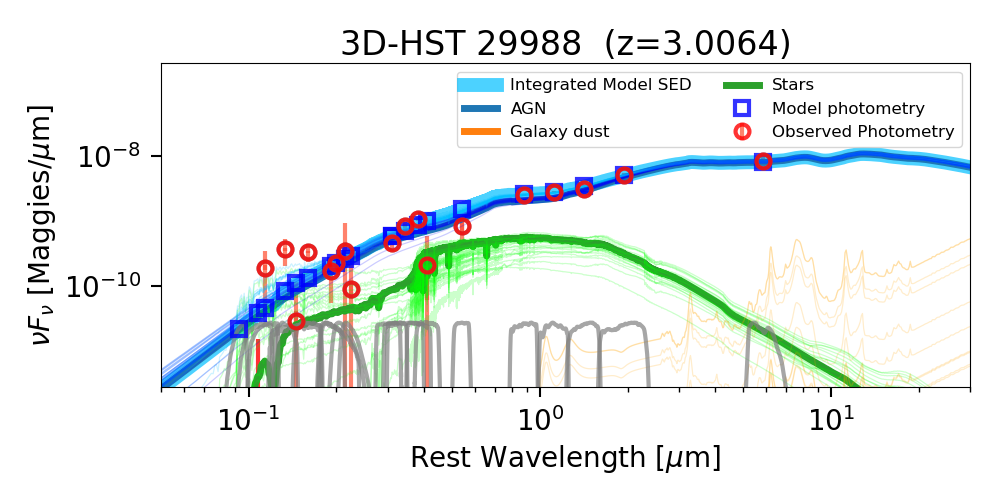}
  \includegraphics[width=0.49\hsize]{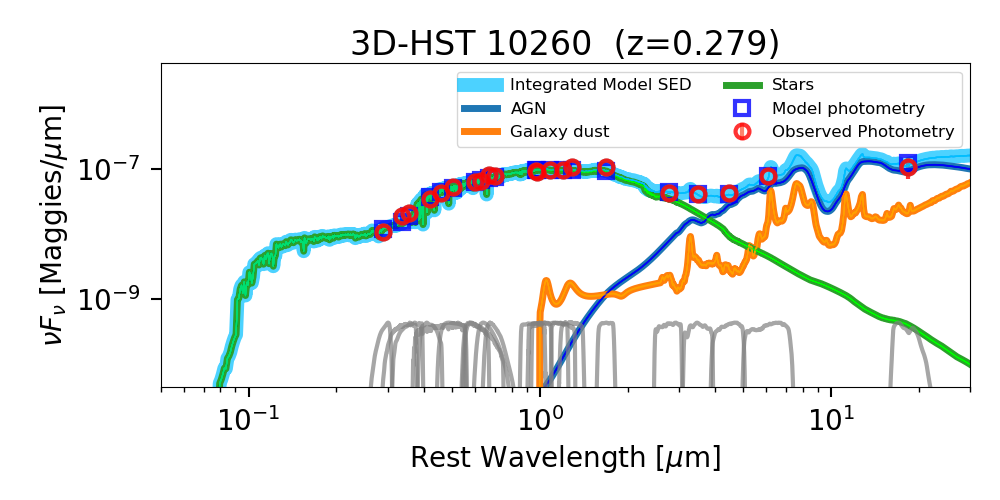}
  \includegraphics[width=0.49\hsize]{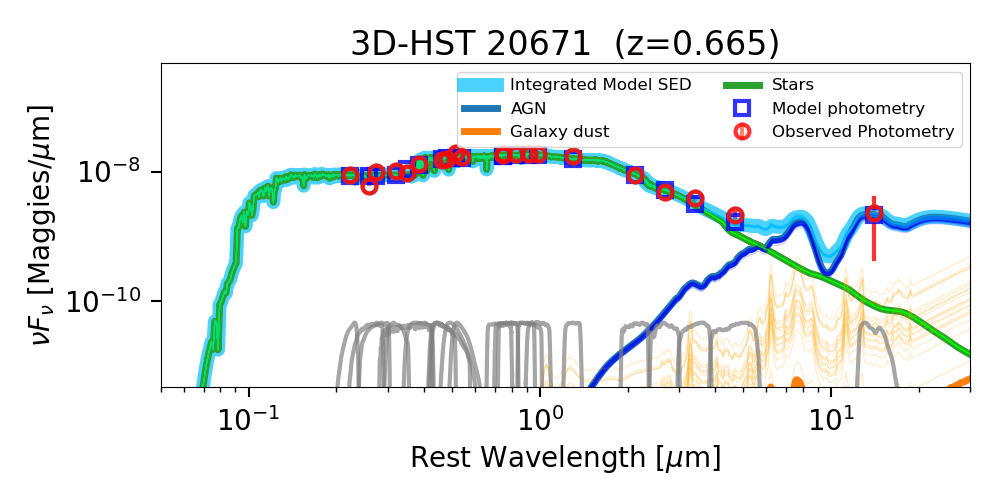}
  \includegraphics[width=0.49\hsize]{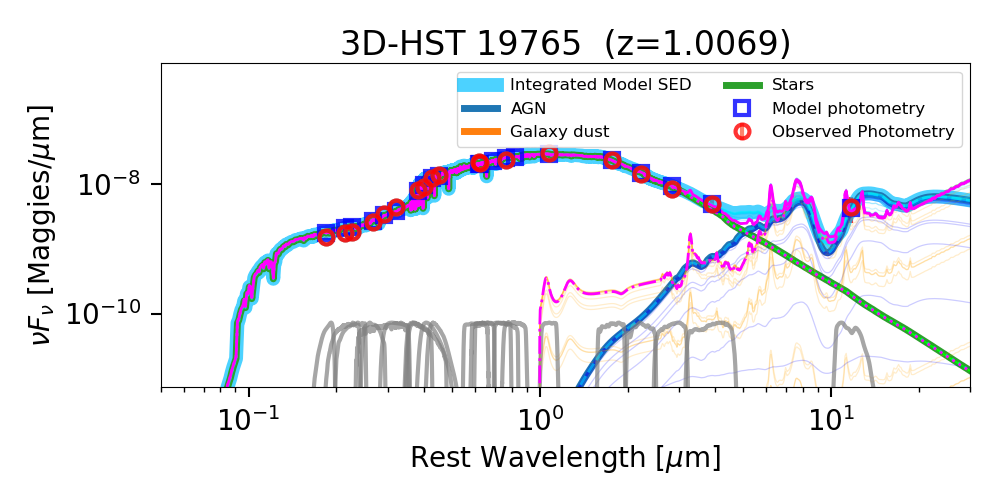}
  \includegraphics[width=0.49\hsize]{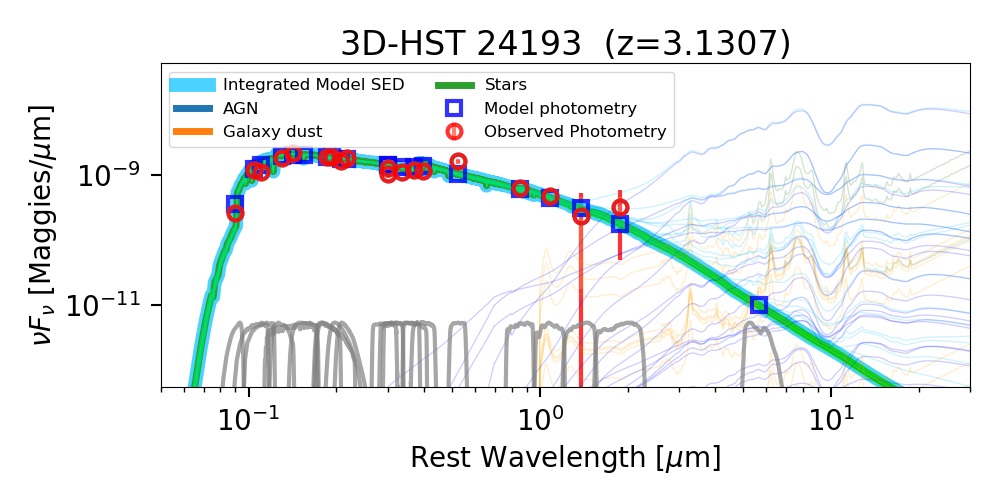}
    \caption{
    Example SED fitting of GOODS-S galaxies. The observed photometric data
    points (red circles) are fit with an SED model (cyan thick lines)
    composed of the FSPS stellar component (green thick lines), AGN component
    (blue thick lines) and SFG dust emission component (orange thick lines).
    The best-fit model photometry is shown as open blue squares. We also
    show 20 random SEDs drawn from the final MCMC chains to show the ranges of
    model degeneracy. Characteristics 
    of these examples are (from top to bottom and left to right): 10735 -- 
    pure star-forming galaxy; 5358 -- galaxy with a bright obscured AGN that
    dominates its near- to mid-IR emission; 18220 -- IR composite galaxy
    with UV-optical emission including a strong contribution from an unobscured
    AGN; 29988 -- galaxy whose UV to mid-IR emission is dominated by a
    moderately obscured AGN; 10260 -- IR composite galaxy with an obscured
    AGN; 20671 -- galaxy with a weak obscured AGN; 19765 -- galaxy where
    the existence of the AGN is highly ambiguous and we have shown the degenerate models in magenta; 24193 -- galaxy without
    enough data to detect any AGN component. The $y$-axis is the source flux ($\nu F_{\nu}$) described in the units
    of maggies/$\mu$m, where 1 maggie equals 1/3631 Jansky.
    }
  \label{fig:fitting_example}
    \end{center}
\end{figure*}

We have conducted SED fitting with the model introduced above for both
radio-detected and -undetected objects that have redshift constraints, either
photometric or spectroscopic, in the 3D-HST catalog.  As demonstrated in
Figure~\ref{fig:fitting_example}, our analysis has revealed a wide range of
variation in each component, from objects whose emission is totally dominated
by galaxy emission (e.g., 3D-HST 10753) or AGN (e.g., 29988) to composite
systems with a wide range of AGN strength and obscuration levels (e.g., 18220,
10260). Notably, the existence of AGNs in some sources can only be revealed by
the analysis of near- to mid-IR SEDs (e.g., 5358 and 20671) due to the
obscuration of the nuclei.

To confirm the existence of an AGN from SED analysis, we first require the
best-fit $L_{\rm AGN}>10^8~L_\odot$. For some objects, the fitting solutions
can be degenerate with multiple peaks of the $L_{\rm AGN}$ posterior
distribution. An example is shown in Figure~\ref{fig:fitting_example} (3D-HST
ID: { 19765}). Such cases are not counted as SED AGNs. There are also cases where
the photometric constraints are too limited to decide if the AGN component
exists (see 24193 in Figure~\ref{fig:fitting_example}), and we do not count
them as SED AGNs either.

Our SED-identified AGNs can be grouped into two classes:

\begin{itemize}
\item {\bf [mtype\_sed]} mid-IR SED AGNs: We visually inspect each fit and
    accept objects with a significant AGN component, using the criteria
    described above. For ambiguous or marginal cases (e.g., numbers 19765 and
    20671 in Figure~\ref{fig:fitting_example}), we run the SED fitting with and
    without the AGN component. If the chi-square values of the IRAC photometry
    data points have been reduced by at least a factor of 2.5 by introducing
    the AGN component, we classify the object as a mid-IR SED AGN.

\item {\bf [otype\_sed]} optical SED AGNs: Some objects are fitted with a
    significant UV/optical excess above the stellar component that is
    attributed to     an optically blue AGN component. We classify these
    objects as optical SED AGNs.
\end{itemize}

In total, 175 AGNs are identified with SED analysis in the radio-detected
sample; 52 of them are optical SED AGNs and 121 are mid-IR SED AGNs.  We have
also applied the SED analysis to the radio-undetected sample and found another
180 SED AGNs; 37 of them are optical SED AGNs and 167 are mid-IR SED AGNs.

Our SED selection criteria are quite conservative, as discussed above.  Many
additional AGNs are identified with other methods. We use the same SED analysis
to constrain the source properties (e.g., AGN luminosity, obscuration) in these
cases.

\subsection{Other Selection Techniques}\label{sec:other-method}

Besides SED analysis, there are other methods that can be used to identify AGN
within this field. { Besides variability identification (Section~\ref{sec:var-select})}, 
most of them have been well-documented in \cite{Alberts2020}
and here we only provide an overview and highlight some minor changes
introduced in this work.  For the reader's convenience,
Table~\ref{tab:agnsel-summary} provides a high-level summary of all the selection
techniques explored.

%\movetabledown=3.25in
%\begin{rotatetable*}
\begin{deluxetable*}{p{40pt}p{110pt}p{100pt}p{130pt}p{10pt}p{10pt}p{20pt}}
    %\tabletypesize{\scriptsize}
    \tabletypesize{\footnotesize}
    \tablewidth{1.0\hsize}
    \tablecolumns{7}
    \tablecaption{Overview of AGN Selection Methods \label{tab:agnsel-summary}
    }
    \tablehead{
  \colhead{key} &
  \colhead{band} & %\tablenotemark{1}} &
  \colhead{description} &
  \colhead{criteria} &
  \colhead{$f_{\rm acc}^a$} &
  \colhead{$f_{\rm yes}^b$} &
  \colhead{Ref.}
}
\startdata
\hline
  $[$xtype\_lum$]$     & X-ray (0.5--7 keV)          &  intrinsic X-ray luminosity higher than expected for stellar processes in a galaxy  &  \makecell{$L_\textrm{X,int}>10^{42.5} \textrm{erg~s}^{-1}$}    & 73.3\%  & 35.6\% & \S\ref{sec:x-ray-select}  \\
  $[$xtype\_x2r$]$     & X-ray (0.5--7 keV) + radio (3GHz) & X-ray to radio luminosity ratio higher than expected for stellar processes in a galaxy  &  \makecell{$L_\textrm{X,int} [\textrm{erg/s}]/L_\textrm{3GHz} [\textrm{W/Hz}]>8\times10^{18}$}   & 73.4\% & 65.3\% & \S\ref{sec:x-ray-select}    \\
  $[$otype\_spec$]$    & optical         & optical spectra showing hydrogen broad emission lines or hard line ratios representing  AGN-driven gas ionization & \makecell{\\ BPT (narrow-line AGN), \\ FWHM($H\alpha$)$>$1000--2000 km/s \\ (broad-line AGN)} & 50.8\% & 5.4\% & \S\ref{sec:optical-spec-select} \\
  $[$otype\_sed$]$     & optical to mid-IR (0.3--24$\mu$m)          & UV-optical excess emission above the stellar continuum, as judged from SED fitting & SED inference, notable AGN UV-optical emission similar or stronger than galaxy emission  & 100\% & 11.0\% & \S\ref{sec:sed-result} \\ 
  $[$mtype\_sed$]$     & optical to mid-IR (0.3--24$\mum$)         & near- to mid-IR excess emission from AGN hot and warm dust, as judged from SED fitting &SED inference, notable hot- to warm-dust excess at $\lambda\sim$2--8~$\mum$ & 100\% &   32.0\% &  \S\ref{sec:sed-result}\\
  $[$mtype\_color$]$   & mid-IR (3--8 $\mum$)         & mid-IR colors that are dominated by AGN warm dust emission & 
  \makecell{$\log(S_{5.8}/S_{3.6})>0.08$, \\ $\log(S_{8.0}/S_{4.5})>0.15$}  & 76.9\% & 11.3\% & \S\ref{sec:irac-select} \\
  $[$rtype\_rl$]$      & radio (3GHz) + mid-IR ($24\mum$)   & radio-loud AGNs with excess emission in the radio band compared with the prediction of templates of normal SFGs & \makecell{ $q_{24}=\log(S_\textrm{24$\mu$m}/S_\textrm{1.4GHz})$, \\ $q_\textrm{24, obs} < q_\textrm{24, temp}$} & 32.9\% & 4.8\% & \S\ref{sec:radio-select}\\
  $[$rtype\_fss$]$     & radio (3--6GHz)             &  radio slope similar to flat spectral radio sources      &  \makecell{radio SED: $f_\nu\propto\nu^\alpha$, \\  $\alpha>-0.5$ } & 13.1\% & 2.0\% & \S\ref{sec:radio-select} \\
  $[$var$]$            & X-ray , optical  &  significant photometric variation in any wavelength  & \makecell{e.g., Median Absolute Deviation \\ (see \citealt{Pouliasis2019})} & 100\%  & 12.4\% &  \S\ref{sec:var-select}
\enddata
\tablenotetext{a}{the fraction of AGN identified with this method when it can be applied;}
\tablenotetext{b}{the fraction of AGN to which this method can be applied;}
\end{deluxetable*}
%\end{rotatetable*}

\subsubsection{X-ray properties}\label{sec:x-ray-select}

{\bf [xtype\_lum]} We regard the sources as X-ray identified AGNs once the
intrinsic X-ray luminosity $L_{\rm X-ray}>10^{42.5}$~erg/s following
\cite{Luo2017}. If the SED fitting indicated low levels of star formation { ($\lesssim10M_{\odot}/$yr)}, we
also accepted $L_{\rm X-ray}>10^{42}$ erg/s as AGNs \citep[e.g.,][]{Pereira-Santaella2011,mineo2014, symeonidis2014, algera2020}. Unlike \cite{Luo2017}, we
do not treat the X-ray spectral index as an AGN tracer as massive X-ray
binaries can also have hard X-ray spectra
\citep[e.g.,][]{Pereira-Santaella2011} (however, after our other selections
were completed we checked and found that at most only one more AGN would have
been identified via X-ray spectral index).  In total, 171 X-ray bright AGNs are
found in the radio-detected sample and another 150 X-ray bright AGNs are found
in the radio-undetected sample. 

\begin{figure}[htp]
    \begin{center}
  \includegraphics[width=1.0\hsize]{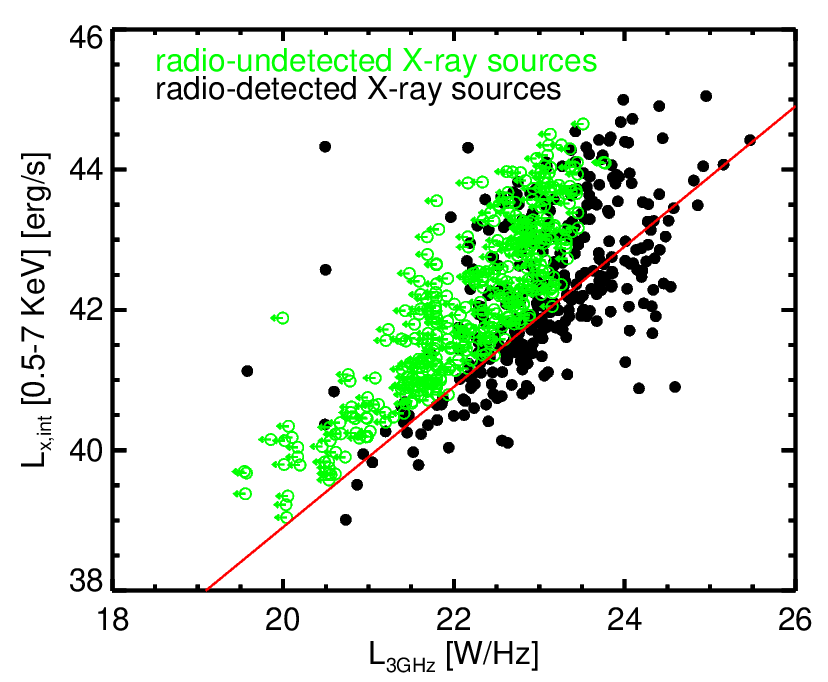}
    \caption{
    The X-ray 0.5--7 keV intrinsic luminosity versus the radio 3~GHz
luminosity of the VLA detected and un-detected X-ray sources. The red line
is the AGN identification criterion, above which the source is considered as an
[xtype\_x2r] AGN.}
  \label{fig:x2r_3ghz}
    \end{center}
\end{figure}

{\bf [xtype\_x2r]} Following \cite{Alberts2020}, we have also used the $L_{\rm
X-ray}-L_{\rm radio}$ relation for SFGs to pick out AGNs. Details of this
selection applied to 6 GHz radio data can be found in that reference. We have
corrected the threshold to apply to the 3 GHz radio data and identify AGNs if the objects
satisfy $L_{\rm X-ray} [{\rm erg/s}]/L_{\rm 3GHz} [{\rm W/Hz}]>8\times10^{18}$
with a 2-sigma significance. \footnote{As shown in our previous work (\citealt{Alberts2020}; Section 4.4.3), the average 
radio spectral slope of our 3 and 6 GHz detected sources is $\alpha=-1$ (assuming $S_\nu\propto\nu^\alpha$), which
makes the monochromatic luminosity ($\nu L_\nu$) independent on the frequency. As a result, we do not need to worry 
about the redshift corrections.} There are 214 objects picked out this way. In
Figure~\ref{fig:x2r_3ghz}, we present the distribution of the X-ray intrinsic
and radio luminosities of all the X-ray sources.

For X-ray sources without radio detections, we have used the VLA 5-sigma upper
limits to derive upper limits on the radio luminosity and compute a lower limit
of $L_{\rm X-ray} [{\rm erg/s}]/L_{\rm 3GHz} [{\rm W/Hz}]$. There are 374 AGN
identified by this criterion.

\subsubsection{Optical spectra}\label{sec:optical-spec-select}

{\bf [otype\_spec]} \cite{Santini2009} and \cite{Silverman2010} have identified
AGNs in GOODS-S through optical spectra. The first work collected all the
available spectroscopic information from public surveys  and cataloged a total
of 59 AGNs.  The second work obtained deep optical spectra of X-ray sources and
reported 112 broad-line AGNs with any emission line FWHM$>$2000 km/s.  After
cross-matching, we found 22 radio-detected and 26 radio-undetected AGNs in our
parent sample through their optical spectra.

\subsubsection{Mid-IR Colors}\label{sec:irac-select}

{\bf [mtype\_color]} Selections through mid-IR color-color diagnostic diagrams
are widely used to look for obscured (and unobscured) AGNs (e.g., see the
summary by \citealt{Padovani2017}), especially during the {\it Spitzer} era
when deep IRAC 3--8 $\mu$m photometry was accessible \citep[e.g.,][]{lacy2004,
stern2005, Donley2012}.  Such techniques are effective owing to the distinct IR
SED features of AGN compared with other types of objects. However, the mid-IR
color-color diagnostics are limited to picking out relatively strong AGN
components with a limited redshift range ($z \lesssim$ 2--3).

\begin{figure}[htp]
    \begin{center}
  \includegraphics[width=1.0\hsize]{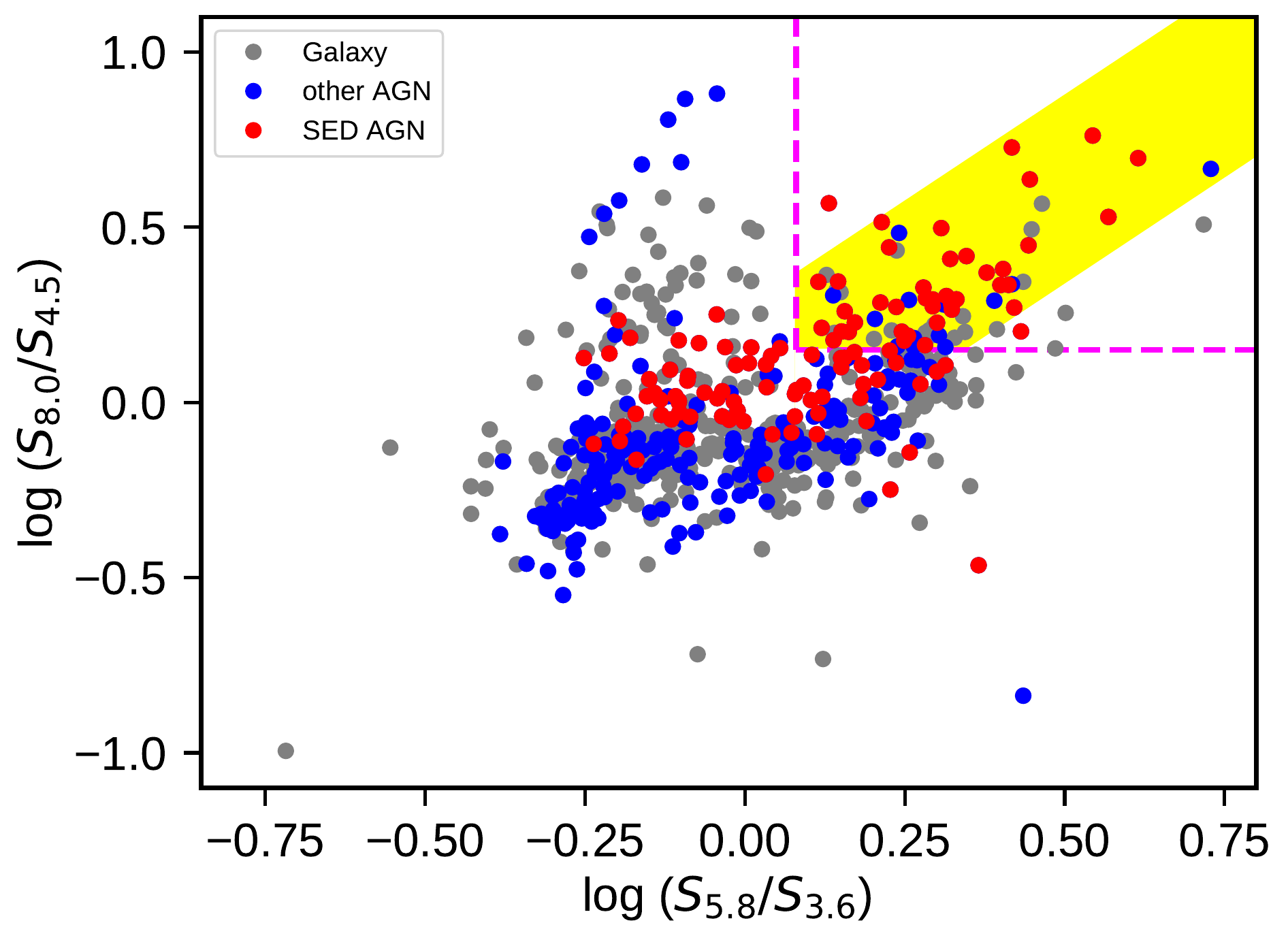}
    \caption{
     IRAC color-color diagram of the VLA-detected AGNs. We have denoted the
     selection criteria proposed by \cite{Kirkpatrick2017} (magenta dashed
     lines) and \cite{Donley2012} (yellow shaded region).
    }
  \label{fig:irac_color}
    \end{center}
\end{figure}

To compare with other AGN selection techniques, especially SED identification,
we revisit the mid-IR AGN color-color selections in this field (e.g.,
\citealt{Donley2008}). Figure~\ref{fig:irac_color} presents the IRAC
color-color diagram of all the radio-detected { objects} with 3-sigma detections
in all four IRAC bands. We have identified 65 AGN candidates with the criteria
in \cite{Kirkpatrick2017}. 

In the same figure, we also color-code objects selected to be optical or mid-IR
SED AGNs. Some sources fall into the \cite{Kirkpatrick2017} AGN color-color
region but are not picked out by our SED analysis. Some of these objects are
high-redshift where the stellar continua can mimic the AGN color and others
have SED constraints that are too noisy to confirm the presence of an AGN. 

There are many AGNs identified by our SED fitting, obscured and unobscured,
that are not identified in IRAC color-color selection.  This is a frequent
issue with IRAC color-color selection: e.g.,  \citet{cardamone2008} find that
42\% of their X-ray selected AGN fall outside the Lacy color-color AGN zone,
which is significantly more liberal than the Kirkpatrick/Donley zone. A serious
issue for our study is that the color-color selections are relatively
insensitive to obscured AGNs at modest redshift. Thus, we adopt the SED
analysis as being superior to simple color cuts and use it to pick out AGN through 
the mid-IR continuum. We also apply it to confirm AGN candidates from color-color methods.

Across the 3D-HST footprint, there are 135 objects with AGN-like IRAC colors with \citet{Kirkpatrick2017} criteria
and we confirm 104 are AGNs with SED analysis as well as other selection methods.

\subsubsection{Radio properties}\label{sec:radio-select}

\begin{figure}[htp]
    \begin{center}
        \includegraphics[width=1.0\hsize]{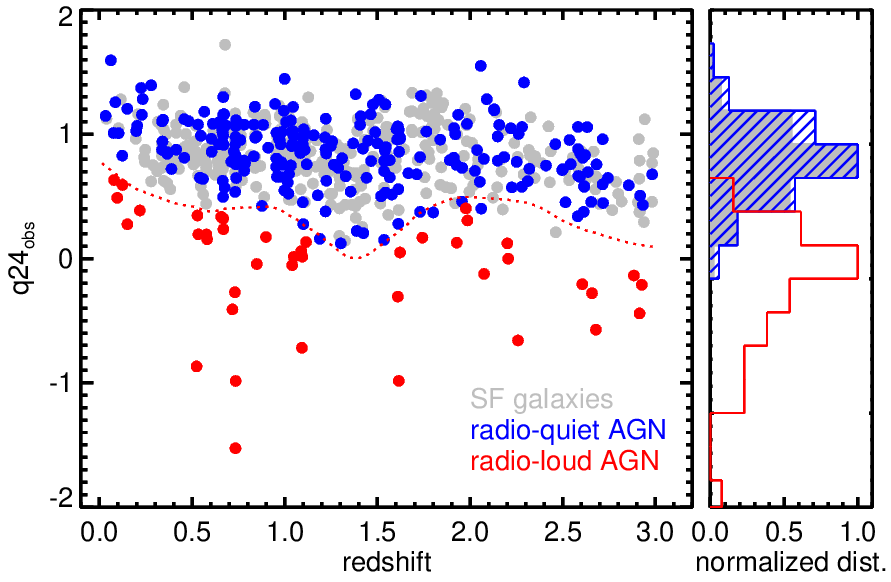}
    \caption{
    The observed IR-to-radio flux $q_\textrm{obs, 24}$ as a function of
    redshift of all the sources. The red dotted line is 0.5 dex below the
    mid-point of the radio-infrared correlations in the \cite{Rieke2009} SFG
    templates; we define the source as a radio-loud AGN if it falls below this
    line at $\ge$ 2-$\sigma$ significance. We color-code different sources on
    the main plot and show their relative distribution of $q_\textrm{obs, 24}$
on its right side.}
  \label{fig:q24_redshift}
    \end{center}
\end{figure}

{\bf [rtype\_rl]} For the radio-detected objects, we have picked out radio-loud
AGNs if the sources deviate from the radio-infrared correlation for SFGs.
Similar to \cite{Alberts2020}, we compute $q_{\rm 24, obs}=\log(S_{\rm 24,
obs}/S_{\rm 1.4 GHz, obs})$, where $S_{\rm 24, obs}$ is the observed MIPS
24~$\mu$m flux and $S_{\rm 1.4 GHz, obs}$ is the radio flux density at 1.4 GHz
derived by extrapolating the 3 GHz flux assuming the radio spectrum can be
described as a power-law with $\alpha=-0.7$, and compare them with the
radio-infrared correlations described by the \cite{Rieke2009} SFG templates at
the appropriate redshifts. If an object is 0.5 dex below the mid-point of the
radio-infrared relation at $>$2-$\sigma$ significance, we classify it as an
AGN. In total, 43 AGNs are selected. In Figure~\ref{fig:q24_redshift}, we show
the distribution of $q_{\rm 24, obs}$ values as a function of redshift. 

{\bf [rtype\_fss]} We have computed the radio slope assuming a power-law
spectrum $f_\nu\propto\nu^\alpha$ between the observed 3 and 6 GHz bands. Following
\cite{Alberts2020}, if the slope index $\alpha$ is larger than $-$0.5 with a
2-sigma significance, we classify this object as an AGN as it is indicated to
be a flat-spectrum radio source.  18 AGNs are identified.

\subsubsection{Time Variability}\label{sec:var-select}

{\bf [var]} Several authors used HST data to identify AGNs in the GOODS-S field
by their time variability   \citep[e.g.,][]{Villforth2010, Sarajedini2011,
Pouliasis2019}. Although using much of the same data, the third reference was
unable to confirm the variability of the majority of the candidate variables
from the first two references. 

{ 
One possibility for this discrepancy is the effects of cosmic ray hits. With
allowance for the statistical distribution of ionization yields from cosmic
rays, there is a very wide distribution extending at low but not negligible
probability to small values in CCDs \citep{garcia2018} and infrared detectors \citep{hagan2021}. 
In these types of detectors, which accumulate charge in an analog fashion, the electrons from a cosmic ray event will just be
added to the signal and hence can produce a tail toward large fluctuations in
the distribution of signal sizes, i.e., can mimic the presence of variability
in some of the sources. 
As a result, it is possible that studies of variability in the
optical and infrared have been affected by incomplete removal of cosmic rays.
\citet{Pouliasis2019,Pouliasis2020}
introduce improved methods to eliminate cosmic ray signals. They confirm $<$ 20\% of the variable sources from previous studies of
the same field, but identify 113 variable sources using their more demanding
selection. We include these latter sources in our compilation.

In detectors that count each hit in a pulse-counting fashion, e.g. X-ray CCDs,
the effect of a hit is just one pulse regardless of its size and cosmic rays
will be less likely to be confused with variability. For this reason, we accept
studies of X-ray variability without further review. In the X-ray, variable AGN
candidates have been identified in \cite{Young2012, Ding2018}. 
}

After matching these variable sources with our radio-detected parent sample, we
found 30 variable sources (19 in the optical and 12 in the X-ray).  For the
radio-undetected 3D-HST { objects}, we found another 113 variable sources (93 in the
optical and 24 in X-ray) from these works.  Since the radio
emission is typically dominated by star formation \citep{Alberts2020}, young
stars may be diluting the AGN optical/UV emission with a non-variable component 
and hence reducing the overall variability in the radio-detected galaxies. After investigating
other information (e.g., redshifts, SEDs) of these variable sources, we conclude two objects
are stars and the other 111 variable objects are likely to be AGNs.

\section{The Pre-JWST GOODS-S AGN Catalog}\label{sec:agn-sample}

With the selections above, we have identified 323 AGNs from the
radio-detected { objects} and another 578 AGNs from the radio-undetected { objects}. Given
the size of the 3D-HST GOODS-S footprint ($\sim$171 arcmin$^2$), these numbers correspond
to lower limits of the AGN number density at $\sim$1.9 { arcmin$^{-2}$} with radio detections (about twice our previous value in \citealt{Alberts2020}) 
down to 2.55 $\mu$Jy~beam$^{-1}$ 
and a total AGN number density of $\sim$5.3 { arcmin$^{-2}$} within 
all the sensitivity limits of relevant X-ray, mid-IR and radio surveys in GOODS-S. 
{ Given the small field of GOODS-S, cosmic variance is an important source of uncertainty on the source density \citep[e.g.,][]{somerville2004} and these numbers may not apply generally to other surveys.}

In the following, we describe
some measurements of basic AGN properties and present a value-added
catalog of AGNs identified in GOODS-S with currently available data 
(Table~\ref{tab:agn-sample}). 

\begin{deluxetable*}{cccc}
    %\tabletypesize{\scriptsize}
    \tabletypesize{\footnotesize}
    \tablewidth{1.0\hsize}
    \tablecolumns{4}
    \tablecaption{Table Format of the Pre-JWST GOODS-S AGN Catalog\label{tab:agn-sample}}
    \tablehead{
  \colhead{Column} &
  \colhead{Type} & %\tablenotemark{1}} &
  \colhead{Unit} &
  \colhead{Description} \\
}
\startdata
id\_3dhst      &    int64   &                & ID in the 3D-HST Catalog   \\
id\_vla        &    int64   &                & ID in the VLA Catalog $^a$   \\
id\_chandra    &    int64   &                & ID in the CDF-S Catalog $^a$   \\
redshift      &  float64   &                & object redshift   \\
ra            &  float64   &   deg          & 3D-HST RA with an offset $\Delta$RA=$+$96 mas $^b$   \\
dec           &  float64   &   deg          & 3D-HST Dec with an offset $\Delta$Dec=$-$252 mas $^b$   \\
ztype         &   int64   &                 &  redshift type: 1 -- spectroscopic; 2 -- grism ;  3 -- photometric \\  
mtype\_sed     &   string   &                & classification tag for [mtype\_sed] $^c$   \\
mtype\_color   &   string   &                & classification tag for [mtype\_color] $^c$   \\
xtype\_lumcut  &   string   &                & classification tag for [xtype\_lum] $^c$   \\
xtype\_x2r     &   string   &                & classification tag for [xtype\_x2r] $^c$   \\
rtype\_rl      &   string   &                & classification tag for [rtype\_rl]  $^c$  \\
rtype\_fss     &   string   &                & classification tag for [rtype\_fss]  $^c$  \\
otype\_sp      &   string   &                & classification tag for [otype\_sp]  $^c$  \\
otype\_sed     &   string   &                & classification tag for [otype\_sed]  $^c$  \\
var           &   string   &                 & classification tag for [var]  $^c$ \\
lbol\_int      &  float64   &  log(erg/s)    & final adopted $L_{\rm AGN, bol}$   \\
lbol\_tag      &    int64   &                & tag of final $L_{\rm AGN, bol}$ source: 1 --- from SED fitting, 2 --- from X-ray luminosity   \\
lbol\_sed      &  float64   &  log(erg/s)    & $L_{\rm AGN, bol}$ estimated from SED fitting   \\
lbol\_xray     &  float64   &  log(erg/s)    & $L_{\rm AGN, bol}$ estimated from the X-ray   \\
m\_bh          &  float64   &  $M_\odot$     & $M_{\rm BH}$ based on Maggorian relation   \\
m\_star        &  float64   &  $M_\odot$     & $M_{\rm star}$ based on SED fitting   \\
r\_tag         &    int64   &                & radio sub-sample tag: 1 --- the source is detected in radio   \\
x\_tag         &    int64   &                & X-ray sub-sample tag: 1 --- the source is detected in X-ray   \\
m\_tag         &    int64   &                & mid-IR sub-sample tag: 1 --- the source is detected in mid-IR   \\
o\_tag         &    int64   &                & other sub-sample tag: 1 --- the source is not detected in any of the above   \\
N\_H           &  float64   & log(cm$^{-2}$) & gas column density from \citet{Luo2017}   \\
l\_x           &  float64   &  log(erg/s)    & observed X-ray luminosity from \citet{Luo2017}   \\
lx\_int        &  float64   &  log(erg/s)    & absorption corrected X-ray luminosity from \citet{Luo2017}   \\
gamma         &  float64   &                & X-ray spectral slope from \citet{Luo2017}   \\
gamma\_err     &  float64   &                & error of the X-ray spectral slope from \cite{Luo2017}   \\
chandra\_type  &   string   &                & object classification from \citet{Luo2017}   \\
L\_3GHz        &  float64   &  log(erg/s)    & Radio luminosity at the observed 3GHz, $L_{\rm 3GHz}$   \\
L\_6GHz        &  float64   &  log(erg/s)    & Radio luminosity at the observed 6GHz, $L_{\rm 6GHz}$   \\
L3\_tag        &    int64   &                & quality tag of $L_{\rm 3GHz}$: -1 -- no detection, 0 -- detection, 1 --- upper limit   \\
L6\_tag        &    int64   &                & quality tag of $L_{\rm 6GHz}$: -1 -- no detection, 0 -- detection, 1 --- upper limit   \\
q24\_obs       &  float64   &                & mid-IR to radio flux flux ratio, $q_{\rm 24, obs}$   \\
q24\_obs\_err  &  float64   &               & error of $q_{\rm 24, obs}$   \\
q24\_crt       &  float64   &                & SFG template $q_{\rm 24, temp}$   \\
alpha          &  float64   &                & radio SED slope, $\alpha$   \\
alpha\_err     &  float64   &                & error of $\alpha$   \\
tau\_agn       &  float64   &                & AGN optical depth from SED fitting, tau\_agn   \\
L5100\_obs     &  float64   &  log(erg/s)    & observed AGN luminosity at 5100\AA, $L_{\rm AGN, 5100\AA}$   \\
L3um          &  float64   &  log(erg/s)    & observed AGN luminosity at rest-frame 3 $\mu$m, $L_{\rm AGN, 3\mu m}$   \\
L6um          &  float64   &  log(erg/s)    & observed AGN luminosity at rest-frame 6 $\mu$m, $L_{\rm AGN, 6\mu m}$   \\
l8um          &  float64   &  log(erg/s)    & observed AGN luminosity at rest-frame 8 $\mu$m, $L_{\rm AGN, 8\mu m}$   \\
L12um         &  float64   &  log(erg/s)    & observed AGN luminosity at rest-frame 12 $\mu$m, $L_{\rm AGN, 12\mu m}$   \\
L20um         &  float64   &  log(erg/s)    & observed AGN luminosity at rest-frame 20 $\mu$m, $L_{\rm AGN, 20\mu m}$  
\enddata
%\tablenotetext{1}{All these priors have been sampled linearly.}
%\tablenotetext{2}{The ``uncertainties'' of these median values are 2 $\sigma$ (i.e., 2.5\%, 50\% and 97.5\% quantiles).}
\tablecomments{ $^a$ --- if the object is not listed in the catalog, we assign
    an ID = $-1$; $^b$ --- this follows the coordinate corrections in
    \citet{Franco2020}; $^c$ --- four values exist: ``agn'' --- the source is
    identified AGN, ``gal'' --- the source behaves as a galaxy, ``none'' ---
    the situation is unclear, ``N/A'' --- there is not enough data for the
    technique.
\\
(This table is published in its entirety in the machine-readable format.)}
\end{deluxetable*}

\subsection{Measurements of Basic AGN Properties}\label{sec:agn-measure}

The rich multi-wavelength dataset in GOODS-S allows us to estimate the AGN
properties at the X-ray, optical, infrared, and radio.  Our SED analysis
constrains the AGN and galaxy emission across the UV into the IR wavelength
range, and it can be used to estimate the AGN luminosity and obscuration as
well as the properties of the host galaxies. We adopt X-ray properties as
presented in \cite{Luo2017} and \cite{Liu2017}. The former reference estimated
the gas column density $N_{\rm H}$ and intrinsic 0.5--7 KeV luminosity by
comparing the X-ray slope to standard spectral models.  \cite{Liu2017} made a
more robust spectral analysis for objects with enough counts. We adopt the
latter results if available. In the radio band, we measure the radio spectral
slope and radio loudness with the two-band radio data and comparisons with the
infrared, respectively. 

\begin{figure}[htp]
    \begin{center}
  \includegraphics[width=1.0\hsize]{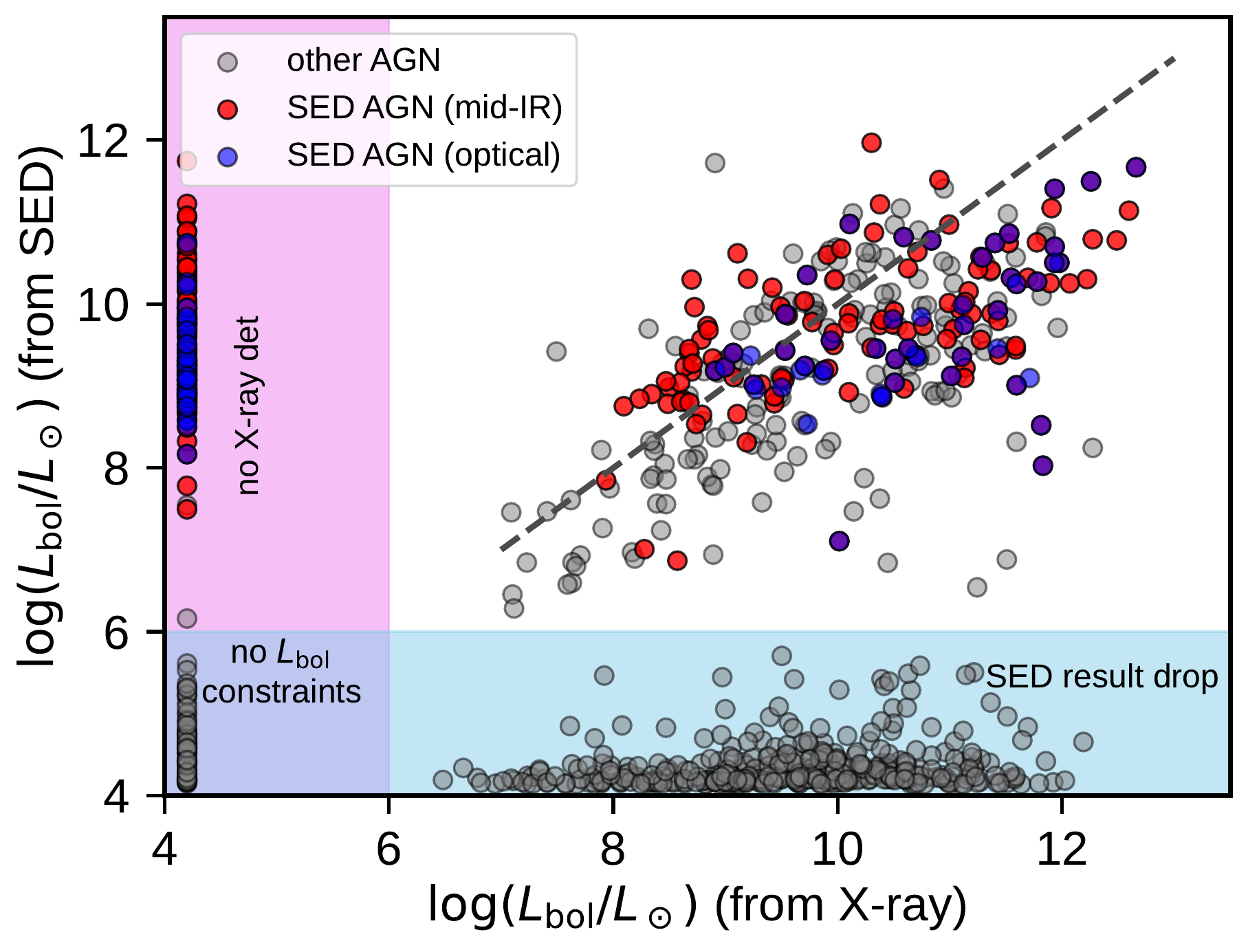}
    \caption{
        Comparison of AGN bolometric luminosity measurements from the X-ray and
        SED analysis. We denote [otype\_sed], [mtype\_sed] and other AGNs in
        blue, red and gray dots, respectively. The dashed black line represents
        the 1:1 relation. The blue and magenta shaded regions indicate where
        the X-ray or SED $L_{\rm bol}$ estimations are not useful.
    }
  \label{fig:agn_lum_measure}
    \end{center}
\end{figure}

One of the most critical parameters is the AGN luminosity $L_{\rm AGN, bol}$.
However, not all of our objects have enough data to constrain the full SED
model robustly. In accordance with the results in \cite{Duras2020}, we adopt a
hybrid approach and estimate $L_{\rm AGN, bol}$ either from (1) the
un-attenuated rest-frame 4400 $\rm \AA$ AGN luminosity inferred from SED
analysis, or (2) the X-ray 0.5-7 KeV intrinsic luminosity from the {\it Chandra}
observations \citep{Luo2017, Liu2017}. We apply the corresponding optical and
X-ray bolometric corrections in \cite{Duras2020} to make the measurements
consistent within the same framework.  Figure~\ref{fig:agn_lum_measure}
compares the $L_{\rm AGN, bol}$ estimations from these two methods. Within a
large scatter, there is general agreement for most objects. However, there are
objects either with very large $L_{\rm AGN, bol}$ from the X-ray but not from
the SED analysis, or with very large $L_{\rm AGN, bol}$ from the SED analysis
but not from the X-ray. For sources with both luminosity indicators, we select
the larger one for $L_{\rm AGN, bol}$; this accommodates possible causes such
as strong obscuration in the X-ray, or a HDD continuum in the infrared.
Otherwise we adopt whichever is available. There are also 65 AGNs in the "no
$L_{\rm AGN}$ constraints" region for which the mutual constraints indicate
$L_{\rm AGN, bol}<10^6~L_\odot$.  Since the mid-IR SED constraints for these
objects are very poor, we use the upper limits of the $L_{\rm AGN, bol}$
estimated from the SED fittings as their bolometric luminosities. 

Another key parameter of an AGN is the black hole mass. As most of our objects
do not have spectral observations, we rely on empirical relations to estimate
the black hole masses from the host galaxy stellar masses assuming a standard
relation of black hole to stellar mass.  We estimate the host galaxy stellar
masses with a color-dependent mass-to-light ratio from \cite{Zibetti2009}
\begin{equation}
\log(M/L_i) = 1.032 (g-i) -0.963 ~~,
\end{equation}
where $g$ and $i$ are the AB magnitudes of the stellar model from SED fitting,
$L_i$ is the galaxy $i$ band luminosity (in units of $L_\odot$) and $M$ is the
galaxy stellar mass (in units of $M_\odot$). We then estimate the black hole
mass assuming the local $M_{\rm BH}/M_{\rm star}=0.025\%$, following
\cite{Reines2015} where the same galaxy stellar mass estimation method is
adopted.

\begin{figure}[htp]
    \begin{center}
  \includegraphics[width=1.0\hsize]{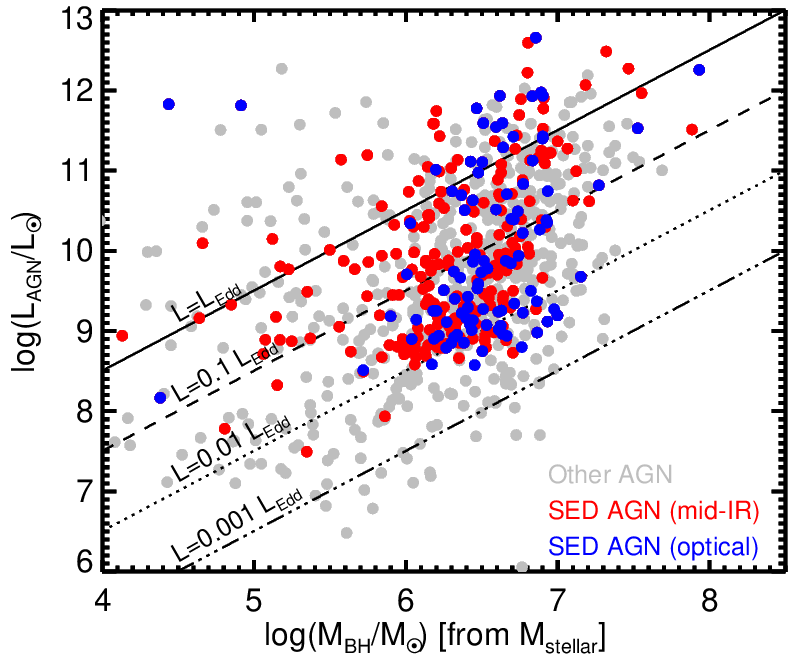}
    \caption{
       The distribution of AGN luminosity and BH mass for AGNs identified in this work. 
    }
  \label{fig:agn_lum_mbh}
    \end{center}
\end{figure}

Figure~\ref{fig:agn_lum_mbh} presents the distribution of the GOODS-S AGNs in
the $L_\textrm{bol}$-$M_\textrm{BH}$ diagram. Although there is a large scatter
due to various uncertainties in the measurements, most of the AGNs have
$M_\textrm{BH}\sim10^6$--$10^7 M_\odot$,
$L_\textrm{bol}\sim10^8$--$10^{12}~L_\odot$ and $L_\textrm{bol}/L_\textrm{Edd}\sim0.01$--1.\footnote{ We define AGN Eddington luminosity $L_{\rm Edd}=3.2\times10^4 (M_\textrm{BH}/M_\odot)L_\odot$.}

\begin{figure*}[htp]
    \begin{center}
        \includegraphics[width=1.0\hsize]{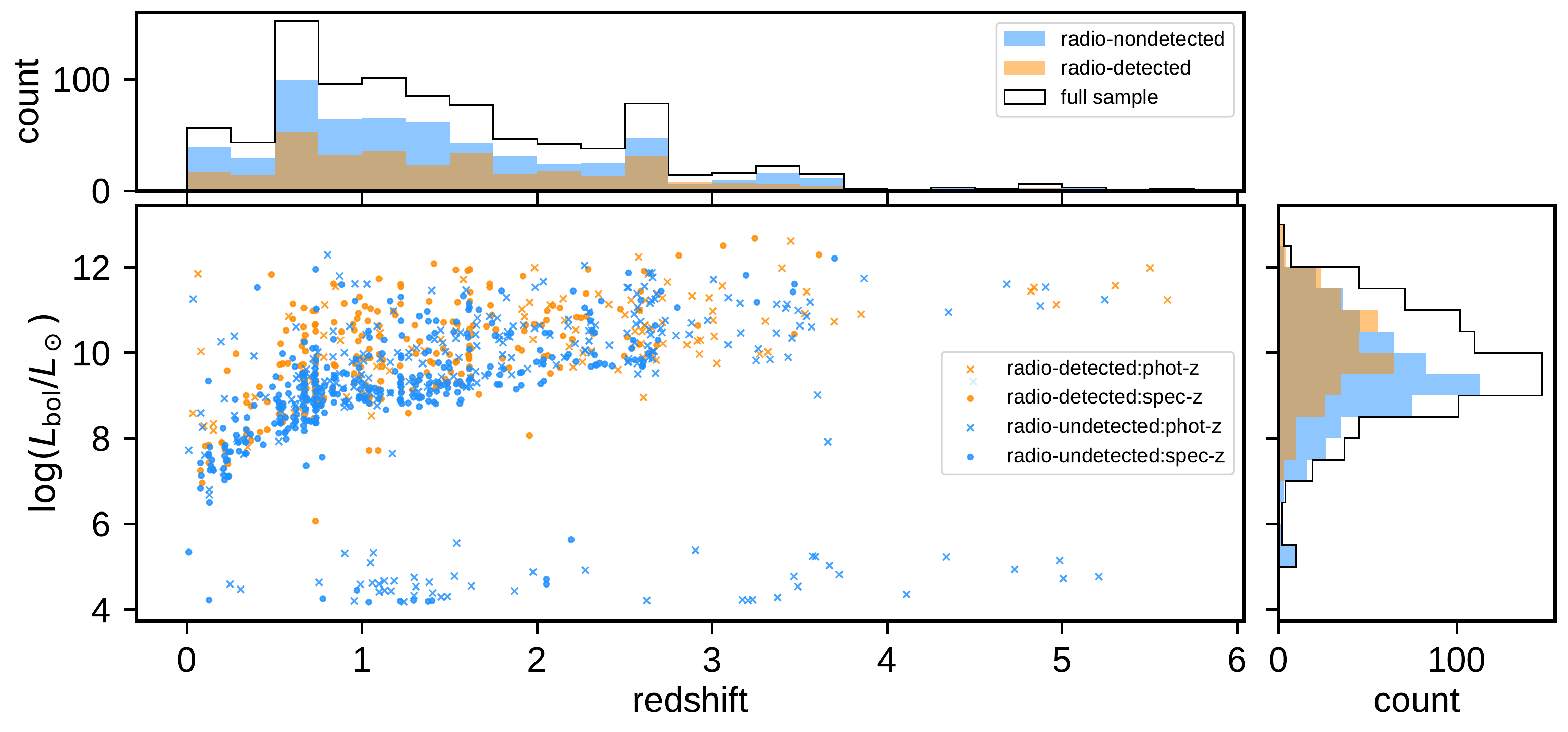}
        \caption{ The redshift and luminosity distribution of GOODS-S AGNs identified in this work. { We have denoted
        radio-detected and radio-undetected AGNs in orange and blue. Sources with spectroscopic redshifts and photometric redshifts are shown as dots and crosses respectively. }
    }
  \label{fig:agn_lum_z}
    \end{center}
\end{figure*}

In Figure~\ref{fig:agn_lum_z}, we show the GOODS-S AGNs on the
luminosity-redshift plane as well as their distributions. The AGN bolometric
luminosity ranges from $\sim10^7~L_\odot$ to $\sim10^{13}~L_\odot$ with the
peak distributed around $10^9$--$10^{11}~L_\odot$, and the redshift ranges from
$\sim0$ to $\sim$5.5 with the peak around 1--3.  { The key properties of the radio-detected and -non-detected samples are similar: (1) their redshift distributions are largely 
indistinguishable with a Kolmogorov-Smirnov (K-S) probability 
that they are drawn from the sample sample of 0.27; (2) the bolometric luminosities
for the radio-detected AGN sample are slightly larger than for the radio-undetected AGN sample. This is 
understandable as the requirement of radio detection will bias towards galaxies with higher 
SFRs (see Section~\ref{sec:survey_limit}), which corresponds to more luminous AGNs given the correlations
between galaxy SFR, stellar mass, supermassive black hole mass, and average AGN luminosity over a wide range of redshift \citep[e.g.,][]{xu2015b}.}

\subsection{Value-added AGN Catalog}

Table~\ref{tab:agn-sample} provides a value-added catalog of various AGN properties, with the following content:

\begin{itemize}
    \item Basic source information: object IDs in the 3D-HST catalog, VLA source catalog and {\it Chandra} 7Ms source catalog, 
    if available, and the source coordinates and redshift;
%    \item Source detection information: whether the AGN is detected in radio (3GHz or 6 GHz), MIPS 24$\mu$m, IRS 16 $\mu$m or X-ray;
    \item AGN identification methods: the status of each selection method: agn -- confirmed AGN, gal -- galaxy, none -- ambiguous, N/A -- data/measurements unavailable;
    \item AGN subsample tag: whether this AGN is "radio-detected", "x-ray-detected" or "mid-IR detected";
    \item AGN bolometric luminosity and BH mass: the final adopted AGN bolometric luminosity, bolometric luminosity estimated from 
    X-ray luminosity and that from intrinsic 4400 ${\rm \AA}$ luminosity from SED decomposition. BH mass inferred from galaxy stellar mass following the Magorrian relation;
    \item SED related information: this includes the AGN luminosities at optical 4400 $\rm \AA$ and mid-IR 3, 6, 8, 12 and 20 $\mu$m, and the AGN extinction level
    tau\_agn from our SED fitting;
    \item X-ray properties: these measurements are from the {\it Chandra} catalog in \cite{Luo2017}, which includes gas column density, the observed and absorption-corrected X-ray luminosity, the X-ray spectral slope and its uncertainty;
    \item radio properties: this includes the source radio luminosity at 3 and 6 GHz, radio slope and its uncertainty;
    \item other properties: this includes the ratio of X-ray to radio luminosity, the observed ratio of 24 $\mum$ to radio luminosity $q_{\rm 24,obs}$ and the corresponding value of star forming templates $q_{\rm 24, temp}$.
\end{itemize}

%{ Some more details on this sample would be helpful, like describing the FOV for the selection (center and diameter, assuming all the IDs are in the footprint of the 3 GHz primary beam), the density on the sky, with a breakdown for types (e.g., unobscured, obscured, etc.). You give these numbers later, but a quick summary in a sentence or two with a reference to later in the paper would be good. }

\section{Results and Discussion}\label{sec:discuss}

With the AGN sample constructed above, now we discuss the relative performance
of all the selection techniques in Section~\ref{sec:agn-demo-sel} and various
issues in building a complete AGN sample in Section~\ref{sec:missed_agn}.
Section~\ref{sec:obscured_pop} focuses on the obscured AGN population and
presents investigations of the relation between AGN gas and dust obscuration,
and the dependence of obscuration fractions on redshift and luminosity.
Section~\ref{sec:agn_lf} is a preliminary inspection of the AGN number density
and luminosity function from our study. Finally, we briefly discuss the outlook
for upcoming JWST surveys in terms of AGN study in
Section~\ref{sec:jwst-predict}.

\subsection{Performance of Different AGN Selection Methods}\label{sec:agn-demo-sel}

\begin{figure*}[htb]
    \begin{center}
  \includegraphics[width=1.0\hsize]{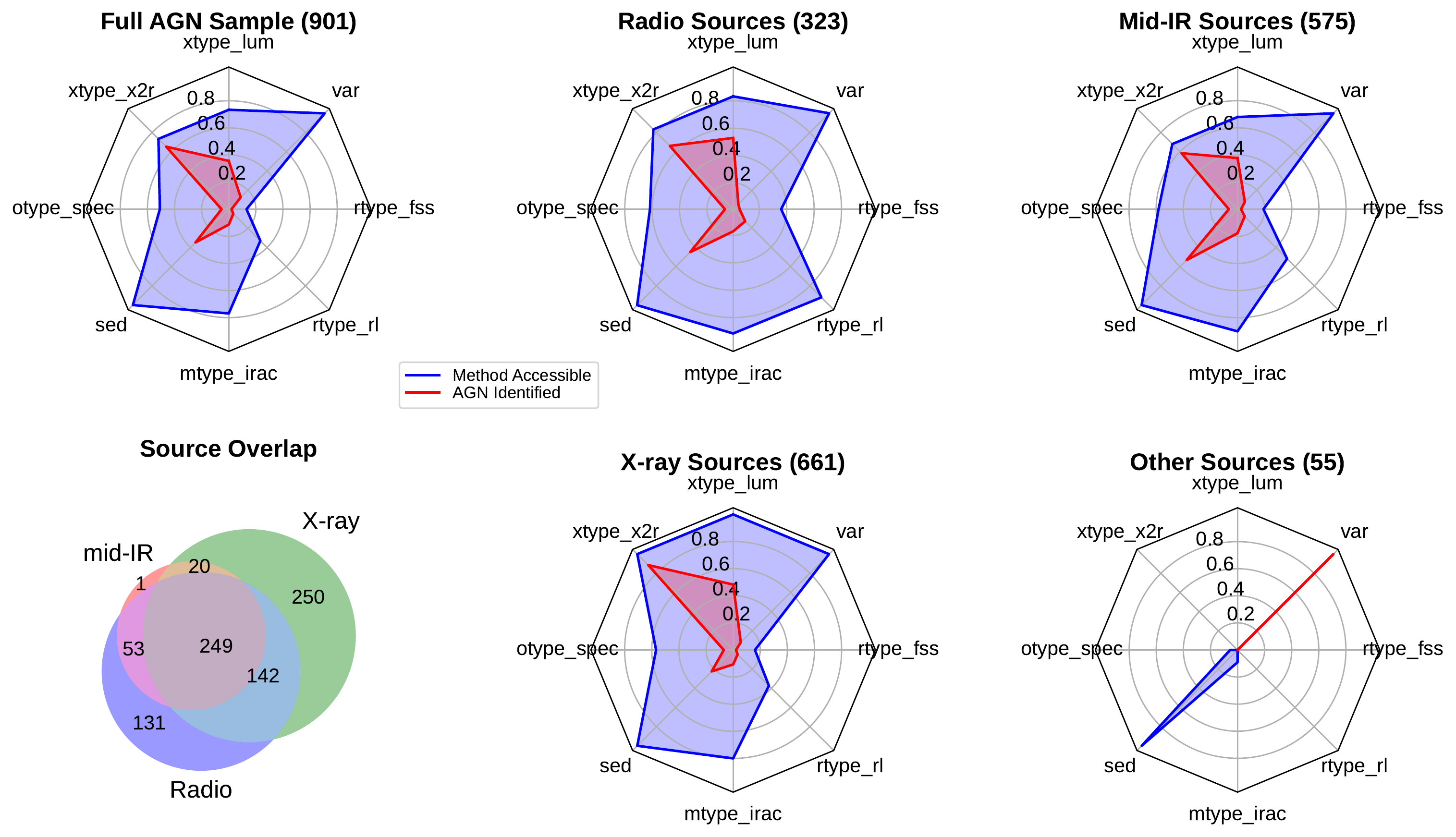}
    \caption{
     Radar diagrams showing the fraction of sources to which each AGN selection
     can be applied to and the fraction of AGN that each selection identifies.
     These numbers are calculated for the whole AGN sample and the AGN samples
     detected in the radio, X-ray, mid-IR, or none of the above. On the bottom
     left panel, we show the AGN sample overlaps for sources in our sample that
     are detected in radio, X-ray or mid-IR.
    }
  \label{fig:selection_radar}
    \end{center}
\end{figure*}

Eight different selection techniques, covering a broad wavelength range from
the X-ray to the radio bands, have been explored to identify AGNs in GOODS-S in
this work. { To visualize the accessibility and productivity of each AGN selection method,
Figure~\ref{fig:selection_radar} presents a few radar diagrams of the AGN
fractions where the criteria of each method can be applied (blue line) and
where they can confirm the AGN existence (red line) for the AGN sample defined in
different ways. As seen from the moderate differences among
the full AGN sample and those detected in the radio, X-ray or mid-IR, the effectiveness of any 
one AGN selection technique clearly depends on how the AGN sample is defined and whether the accessible data 
are good enough to apply the selection. It is also obvious than no single method comes close to selecting all
the AGNs.} 

\begin{figure*}[htb]
    \begin{center}
  \includegraphics[width=1.0\hsize]{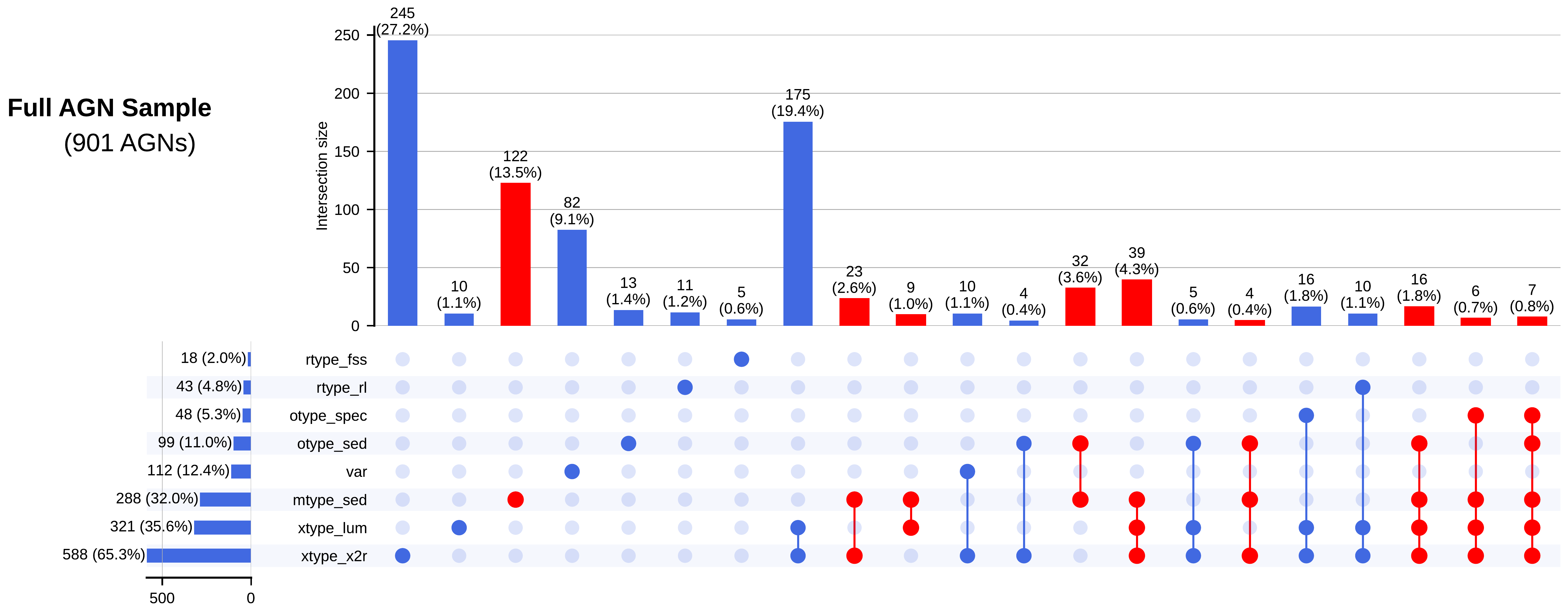}
    \caption{
        { Sizes of the AGN samples} selected with different methods and the source
        intersections in the full AGN sample, { visualized with the UpSet technique. The sample intersections of AGNs selected with different methods are plotted as a matrix. Each row corresponds to one selection method and the bar charts on the left show the size of the AGN identified with this method. Each column corresponds to a possible intersection: the filled-in cells indicate which selection method is part of the intersection. The lines connecting the filled-in cells show in which direction the plots should be read. The bar charts on the top give the size of the AGN sample identified with the corresponding intersection of various selection techniques.} We only show subsets with sample
        size greater than 4 ($\gtrsim4\%$ of the whole sample). We highlight
        the subsets relevant to mid-IR SED selections in red, where we expect
        that the numbers will be increased significantly with the upcoming JWST
        surveys of this field.
    }
  \label{fig:agn_counts}
    \end{center}
\end{figure*}

{ To show the relative performance and possible overlaps of different selections}, Figure~\ref{fig:agn_counts}
summarizes the { sizes of AGN samples} selected from each method and the sample
intersections among different methods for all the AGNs identified, using the
\textit{UpSet} visualization technique \citep{Lex2014}.\footnote{ The traditional Venn Diagram cannot 
effectively visualize the intersections of many ($>3$) sets of data while \textit{UpSet} is an alternative that can be applied 
more generally. See details in \url{https://upset.app/}}

The first conclusion is that no single method has identified more than two
thirds of the sample.  The top three methods that selected most of the AGNs
are: (1) the X-ray to radio luminosity ratio ([xtype\_x2r]), identifying a
fraction of { 65.3\%}, (2) SED analysis ([mtype\_sed] + [otype\_sed]), finding a
fraction of { 39.4\%}, and (3) the X-ray luminosity cut ([xtype\_lum]), yielding  a
fraction of { 35.6\%} (there is significant overlap with [xtype\_x2r] -- { 19.4\%}
are indicated by both methods).  Variability also makes a significant contribution, { 12.4\%}, but the remaining methods were less prominent: { 5.3\%} for [otype\_sp], { 4.8\%} for [rtype\_rl] and { 2.0\%} for
[rtype\_fss]. Considering the selection overlaps, $\sim88.7$\% of the AGN sample
was identified with the first three methods while the other methods contributed
an additional 12.3\%. Notably, only one AGN has been uniquely identified
by its optical spectrum ([otype\_sp]), indicating that the additional
discovery space for spectral selection is limited once the AGN identifications
have been completed using all the other multi-wavelength approaches.
(Nevertheless, spectral follow-up remains, of course, an important means to
    confirm and expand on the nature of a source.)

Among the final 906 AGNs, 45.8\% of the sample are identified with at least two
methods. Besides [xtype\_lum] $\cap$ [xtype\_x2r] with a $\sim$ 19.4\% fraction
of the whole sample, none of the other selection intersections contribute
$>$5\% of the whole sample. { This reveals the wide diversity of the AGN behaviors
and the complicated nature of AGN selections.}  54.2\% of the AGNs are only identified by
one method, with 27.2\% for [xtype\_x2r], 13.5\% for [mtype\_sed], 9.1\% for
[var], 1.4\% for [otype\_sed], 1.2\% for [rtype\_rl], 1.1\% for
[xtype\_lum], 0.5\% for [rtype\_fss] and 0.1\% for [otype\_spec]. Despite their
relatively small sample sizes, (nearly) each selection method identifies some
unique AGNs that would be missed by other selection techniques.

In Figure~\ref{fig:agn_counts}, we also highlight (in red) the subsets relevant
to mid-IR SED selection.  Despite the limited wavelength coverage in the mid-IR
(only {\it Spitzer} IRAC 3--8 $\mu$m, { MIPS 24 $\mu$m, { plus IRS 16 $\mu$m for part of the sample}}) 
and the conservative selection criteria adopted in this work, we have demonstrated that SED analysis
can select obscured (and unobscured) AGNs effectively. The upcoming JWST GTO
surveys (program 1207) of this field have a much denser wavelength sampling in
the mid-IR, and we expect a significant increase in AGN numbers and an
improvement in reliability from SED selection. This discussion will be expanded
in Section~\ref{sec:jwst-predict}.

In Appendix~\ref{app:agn-demo-more}, we present similar
\textit{UpSet} diagrams for AGNs but detected in the radio, X-ray or mid-IR.
The exact values of the AGN yield fraction varies. For example, for AGNs
detected in X-ray, the [xtype\_x2r] identified 89.0\% of the sample while this
value changed to 58.4\% for AGN selected from the mid-IR parent sample. Due to
our conservative SED selection, [mtype\_{sed}] AGNs account for only 48.9\% of
the mid-IR Sources, 37.5\% of the radio sources and 20.6\% of the X-ray
sources. Nevertheless, the general conclusion is that X-ray to radio luminosity
ratio, optical to mid-IR SED analysis and X-ray luminosity are the top three
most effective selections (for $\sim$30--90\% of the sample for which each
selection can be applied) while the other methods have more limited success
($\sim<15$\%).

\subsection{Issues in building a complete AGN sample}\label{sec:missed_agn}

As shown above, no single selection technique can identify all the kinds of
AGNs (with the most efficient ones yielding no more than 2/3 of the whole
sample) and the sample overlaps between different methods are generally small
($\sim$1--20\% of the whole sample). Understanding the reasons { why many AGNs are missed}
in these selections is critical to complete the AGN census.

The AGN itself has both intrinsic (determined by the AGN physical properties)
and apparent (caused by various obscuration sources) SED variations that change
its multi-wavelength behavior significantly. As a result, it is a challenge to
define a complete AGN sample cleanly. For most observers selecting AGNs in a
particular band, ``complete'' means the sample is flux-limited at some
wavelength. However, AGN samples identified at one wavelength will be
seriously incomplete and, worse, strongly biased. Thus, multi-wavelength data have
to be invoked to reduce sample bias. Ideally, one would like a selection that
can be traced to bolometric AGN luminosity.  

{ Assuming that the relevant data types are 
all available over the survey area or sample, there are two remaining issues
that undermine a complete AGN census. First, the data used for AGN identification 
may introduce biases to the population that is found (e.g., the requirements for source detection yield a biased sample).
Second, the selection technique does not work or is incomplete for some types of objects (e.g., the galaxy contamination may be too strong).
The first point will be elaborated upon in Section~\ref{sec:survey_limit}, and is illustrated with a check of the sample biases among radio-detected AGNs in Section~\ref{sec:bias_detection}. For the second point, we focus on some AGNs with abnormal X-ray to mid-IR SED behavior that complicates their selections in Section~\ref{sec:irxray}, and also we briefly mention incompleteness caused by the source time variability in Section~\ref{sec:bias-var}.
}
%In this work, the issues raised by the first possibility are minimized since
%all our AGN identifications have been limited to the HST GOODS-S footprint and
%nearly all the data we use (besides the 6 GHz VLA image, which only influences
%    the [rtype\_fss] AGN detections) have covered this area fully (see
%    Figure~\ref{fig:map}).  Perhaps the largest remaining need is very deep
%    multi-color photometry in the mid-infrared, 10--25 $\mu$m, which should
%    soon become available through JWST.  We now focus on the other two issues. 

\subsubsection{Survey Limits of $L_{\rm AGN, bol}$}\label{sec:survey_limit}

\begin{figure*}[htp]
    \begin{center}
  \includegraphics[width=1.0\hsize]{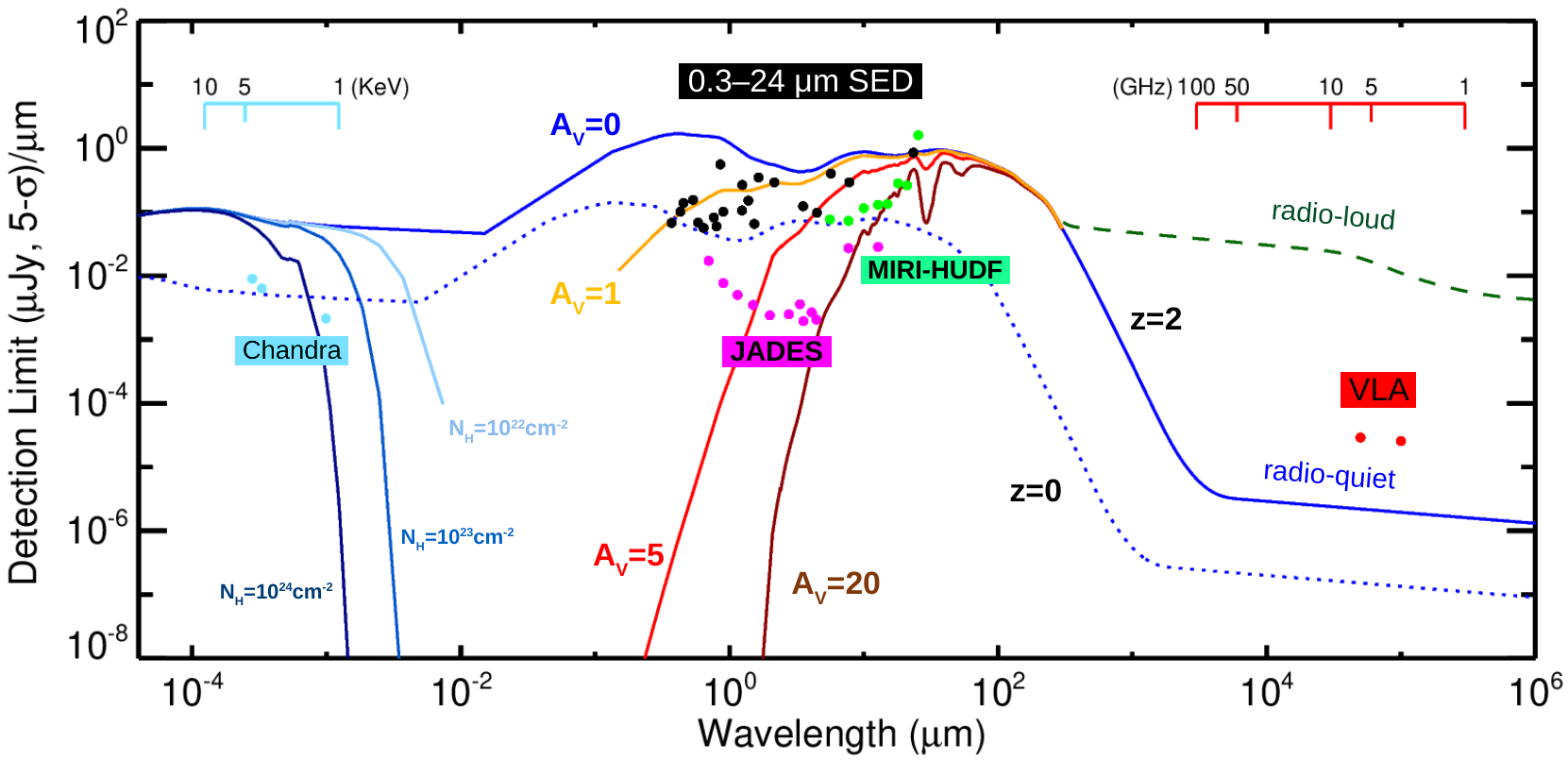}
    \caption{
      The standard AGN SED model at $z=2$ (solid blue line, normalized by the
      MIPS 24$\mu$m detection limit and plotted in $\lambda F_{\lambda}$ units).
      We show examples of variations with the
      different obscuration levels in the X-ray ($N_{\rm H}=10^{22}, 10^{23},
      10^{24}~{\rm cm}^{-2}$) or in the UV-to-MIR ($A_{\rm V}=1, 5, 20$). The
      default AGN model is assumed to be radio-quiet (solid lines) and we also
      show the far-IR to radio SED of the radio-loud case as a dashed green line.  A
      radio-quiet AGN template without obscuration at $z=0$ is also presented
      for comparison (dotted blue line). The median 5-$\sigma$ detection limits
      of various surveys are also denoted as solid dots (X-ray data from
          {\it Chandra} in light blue, optical to mid-IR SED photometry data used in this
      work in black, radio data from VLA in red and planned JADES and MIRI-HUDF
  observations in magenta and green). See text for details.
    }
  \label{fig:sed_sens}
    \end{center}
\end{figure*}

To compare detection limits of the AGN brightness in different bands, we have
developed a simple SED model to predict the typical AGN flux at the X-ray to
the radio wavelengths with different luminosity and obscuration, as described
in Appendix~\ref{app:agn-sed-model}. { This model lets us compute the lower limits of the AGN brightness that could be
detectable in each band given its flux sensitivity. Figure~\ref{fig:sed_sens} shows some key
features of this model: in the X-ray, with the increase of gas column density
$N_{\rm H}$, the AGN soft band emission drops quickly while the hard band
emission changes little. In the optical to mid-IR, with an increase of the dust
attenuation level, the AGN optical to near-IR emission drops quickly while the
mid-IR emission changes only slightly. In the radio band, the AGN emission is
not influenced by obscuration but presents a large range of intrinsic
variations. Although the relative strength of the AGN radio emission is a
continuous distribution in reality, we show the SEDs of two representative
cases --- one radio-loud and another radio-quiet.

In the X-ray, the emission is dominated
by the AGN in most cases, so we can transfer the {\it Chandra} measurements to
estimates of the intrinsic AGN flux and luminosity (as functions of redshift and gas column
density).  However, this approach can fail for AGNs with high $N_H$ columns,
whose output can be strongly suppressed even in the {\it Chandra} hard band
\citep[e.g.][]{buchner2021}.

In the optical to mid-IR the AGN portion of the observed flux
can be contaminated or even dominated by the host galaxy emission, which depends on a large set of  parameters (stellar mass,
stellar age, star formation history, stellar dust extinction, AGN obscuration,
etc). The
minimum AGN light fraction that the SED analysis could reveal is not
decided by one or two photometric bands but by an integrated analysis of all
the data points where the SED analysis can be conducted.  A full analysis is clearly beyond the scope of this
paper.  Therefore, we take a very simple approach to estimate the AGN
fractional detection limits in our current UV-to-IR SED selection.}

\begin{figure}[htp]
    \begin{center}
  \includegraphics[width=1.0\hsize]{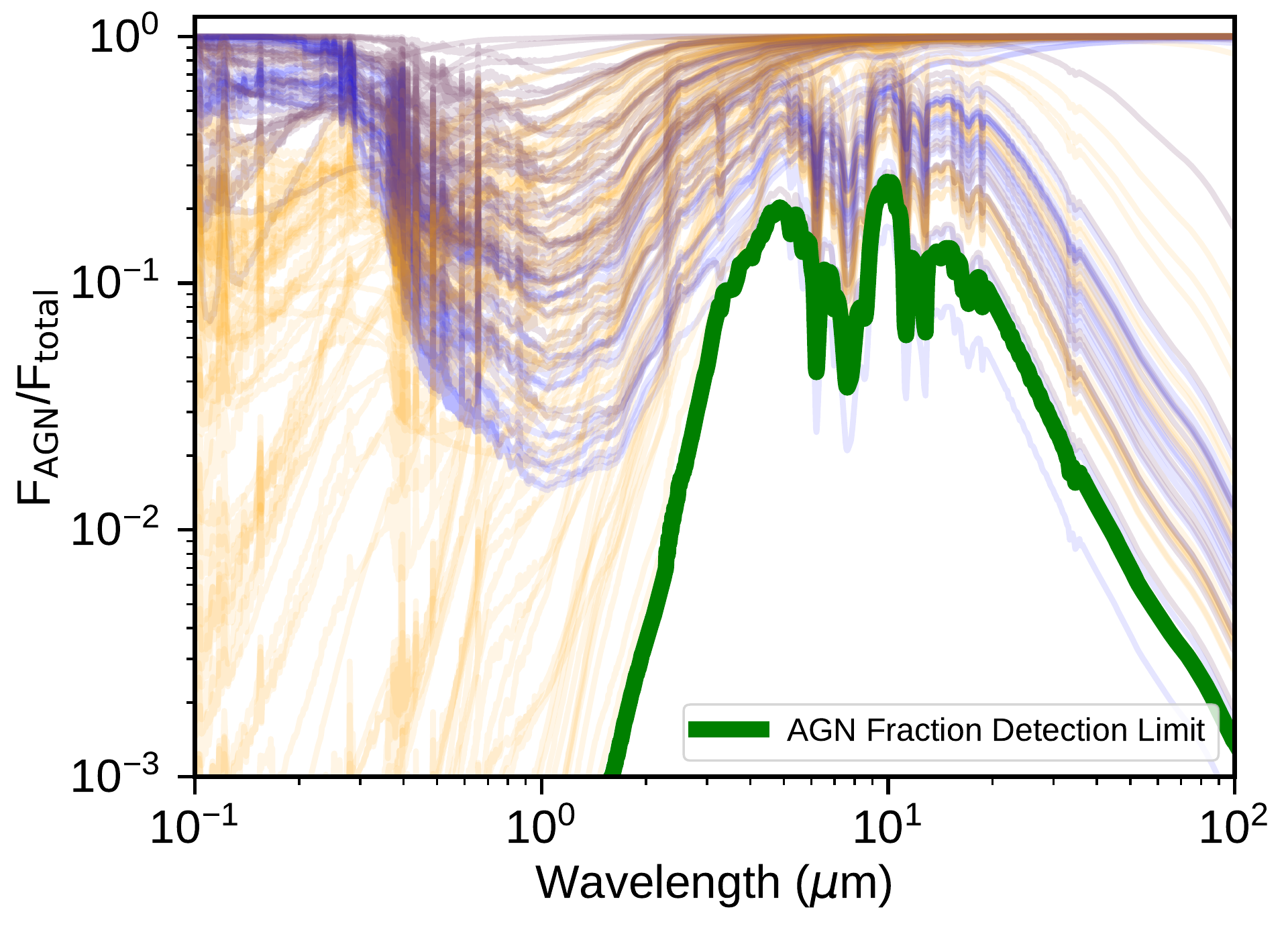}
    \caption{
AGN light fraction as a function of { rest-frame wavelength} in our SED-identified AGN. Optical SED AGNs are 
shown in blue and mid-IR SED AGNs are shown in orange. We denote the AGN fractional 
detection limit with a thick green line, above which the existence of an AGN can be 
confirmed by our selection criteria once the wavelength coverage is sufficient. 
See text for details.
    }
  \label{fig:sed_curve}
    \end{center}
\end{figure}

Figure~\ref{fig:sed_curve} shows the relative AGN light fraction of the galaxy
total emission ($F_{\rm AGN}/F_{\rm total}$) as a function of rest-frame
wavelength for both optical and mid-IR SED-identified AGNs.  Despite their huge
differences at shorter wavelengths where the [otype\_sed] AGNs have higher
$F_{\rm AGN}/F_{\rm total}$ values with a limited range of variations but
[mtype\_sed] AGNs show a broad range of variations, the patterns in the mid-IR
for these two populations are similar. We calculate the minimum $F_{\rm
AGN}/F_{\rm total}$ values of all the SED-identified AGNs as a function of
rest-frame wavelength ({ dark green} line in Figure~\ref{fig:sed_curve}) and treat
this curve as the AGN fraction detection limit for all redshifts and different
AGN/galaxy SED types. A local peak is found near $\lambda\sim5~\mum$ with
$F_{\rm AGN}/F_{\rm total}\sim0.2$.  If the data cover this peak and the
indicated AGN contribution is larger than 20\%, we can consider the AGN to be
identifiable. (In fact, the MIPS $24~\mum$ band covers the rest-frame 5 $\mum$
at a maximum redshift at 3.8 and our SED AGN numbers drop quickly above this
redshift.) This could be a very useful result given the strong correlation
between the X-ray and mid-IR emission  \citep[e.g.,][]{Lutz2004,
Gandhi2009,asmus2015, Stern2015}, indicating that both bands are indicators of
bolometric luminosity.  In fact, they may be complementary; for example, NGC 424 and the
Circinus galaxy, heavily obscured in the X-ray \citep{buchner2021}, have a
prominent AGN contribution in the 3--6 $\mu$m range
\citep{gallimore2010,prieto2004}.  %The correlation is discussed further in
%Section~\ref{sec:irxray}, where we show that there are exceptions that need to
%be understood better to clarify the relation of the two bands to the bolometric
%luminosity.

In the radio, the story is more complicated and bolometric results are
unlikely.  For a radio-loud AGN, the emission is essentially all from a
relativistic jet launched by the AGN.  In comparison, the radio emission in
radio-quiet AGNs comes from multiple sources including star formation
within the galaxy \citep[see the review by][]{Panessa2019}. Given this range, we cannot use radio measurements to predict $L_{\rm AGN, bol}$. In our model, two
extreme cases, ``radio-loud'' and ``radio-quiet,'' are considered;
their relative radio brightness differs by a factor of $\sim10^5$--$10^6$.

\begin{figure}[htp]
    \begin{center}
   \includegraphics[width=1.0\hsize]{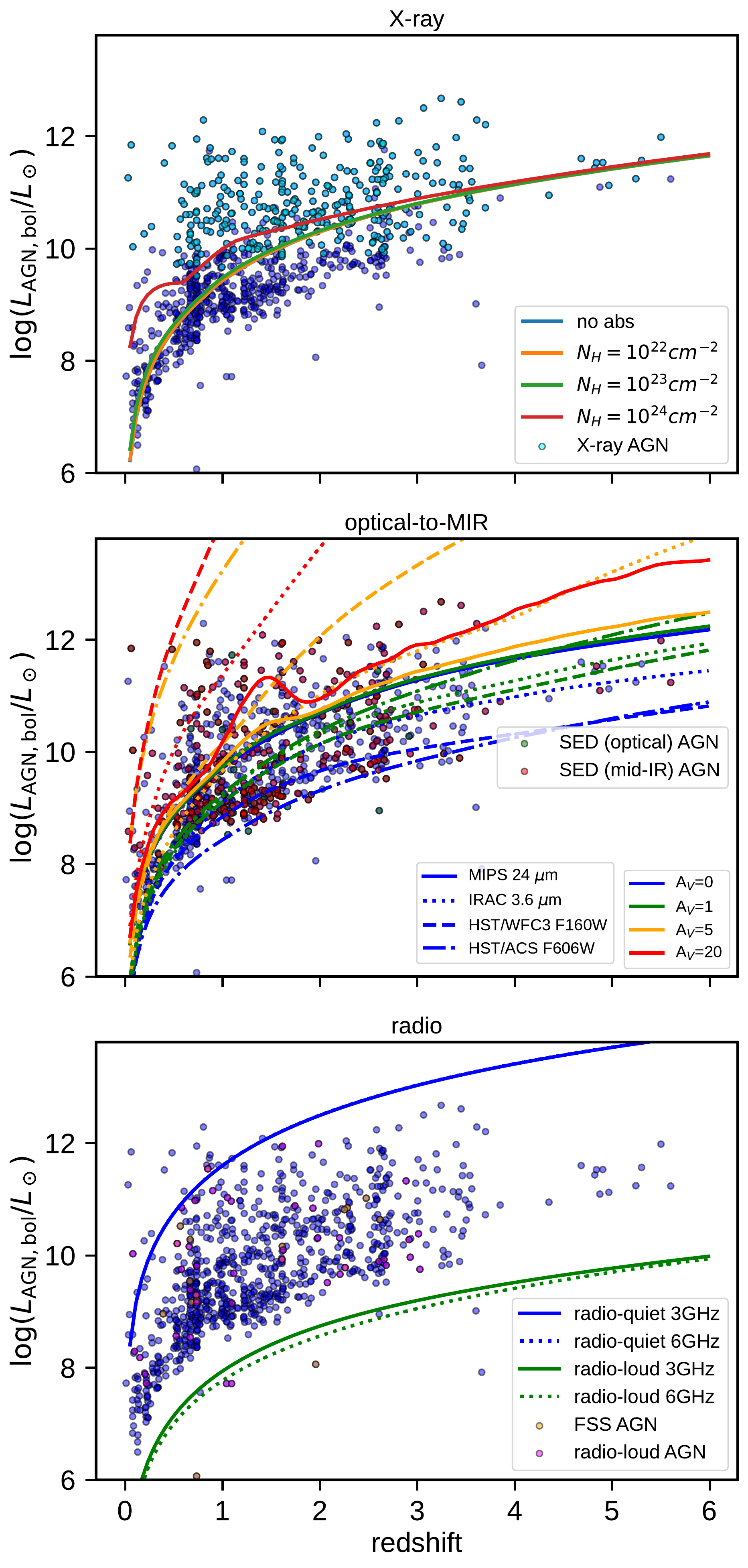}
    \caption{
      $L_{\rm AGN, bol}$ detection limits as a function of redshift for
      different photometric bands in X-ray (top), optical-to-MIR (middle) and
      radio (bottom), under the assumption that the emission is coming purely
      from the AGN.
    }
  \label{fig:curve}
    \end{center}
\end{figure}

Figure~\ref{fig:curve} presents the $L_{\rm AGN, bol}$ detection limits
as a function of redshift for various photometric bands, according to
the AGN model.  In the X-ray, except for Compton-thick AGNs
($N_H\gtrsim10^{24}$~cm$^{-2}$), the detection limits are similar among AGNs with a
large range of obscuration levels. In the optical to the mid-IR, the
optical-to-near-IR photometry is strongly biased against AGNs with high obscuration ($A_V\gtrsim5$) while the detection limits in the mid-IR
bands change little, showing the necessity of mid-IR observations to detect the
obscured AGN population.  In the radio, we expect to see all the ``radio-loud''
AGNs and fail to detect the AGN emission in the ``radio-quiet'' population. In
fact,  the radio emission in most AGNs comes
from the host galaxy (see discussion below).  In the bottom panel of
Figure~\ref{fig:curve}, we plot the host SFR detection limits in the radio
bands; the detections are biased toward galaxies with active star formation at
high-redshift.

\subsubsection{Selection Bias Imposed by Single-band Source Detection}\label{sec:bias_detection}

From the summary of AGN demographics in Section~\ref{sec:agn-demo-sel}, we
concluded that no single method, or even a few methods, can pick out all the
AGNs and each method can identify some unique AGNs. As shown in
Figure~\ref{fig:selection_radar}, the accessibility of each AGN selection
varies with the definition of the parent sample, { i.e., if the sample is flux-complete at some particular band(s).} Thus, selection bias is
inevitable. In this section, we explore the sample bias of our radio-detected
AGN sample as an example.

\begin{figure}[htp]
    \begin{center}
  \includegraphics[width=1.0\hsize]{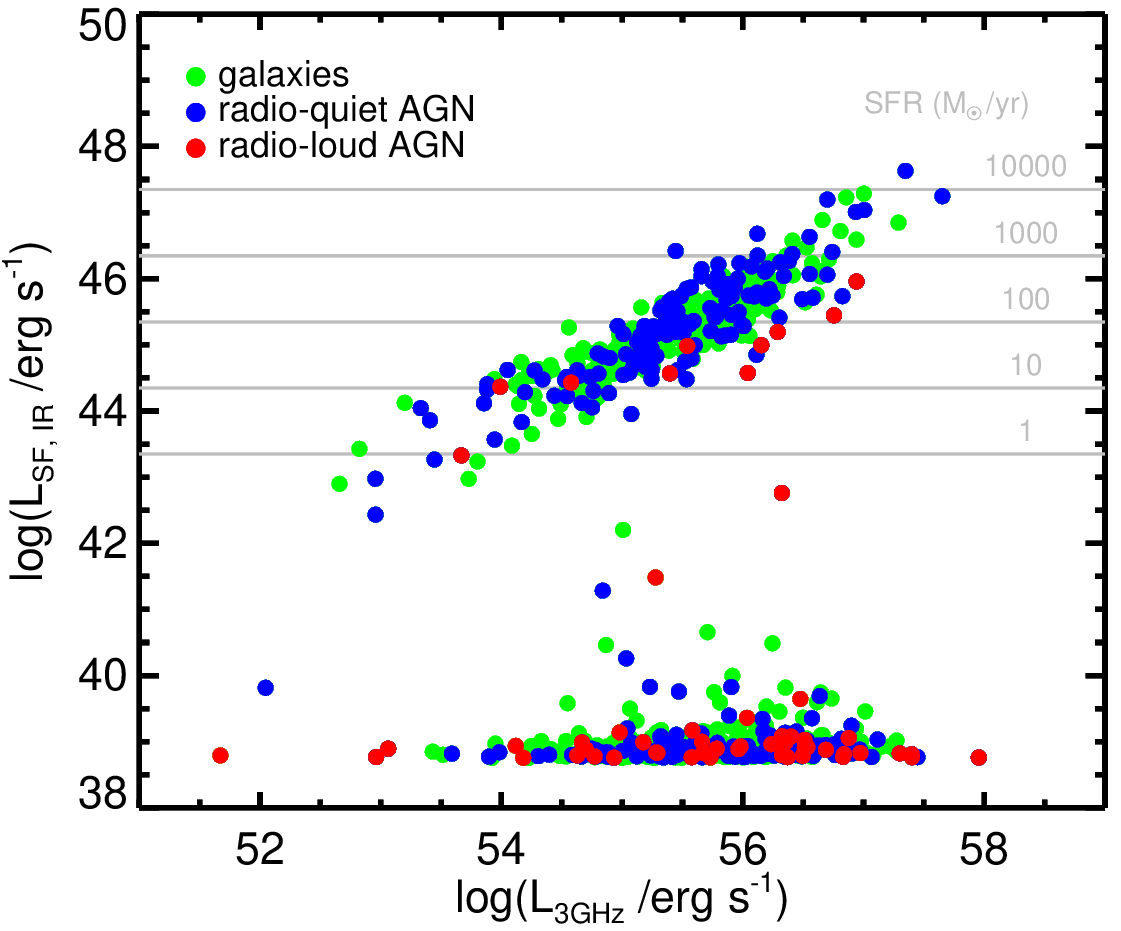}
    \caption{
        SFG IR luminosity, $L_{\rm SF, IR}$, derived from our SED fitting { versus}
        the 3 GHz radio luminosity, $L_{\rm 3GHz}$, for our radio sample. We
        show radio-quiet AGNs, radio-loud AGNs and galaxies in blue, red and
        green dots. Adopting the \citet{kennicutt1998} star formation law, we
        converted the $L_{\rm SF,IR}$ to SFR and denote the corresponding
        values with grey horizontal lines. { (Note: For some objects with $L_{\rm SF, IR}\lesssim10^{44}$~erg~s$^{-1}$, the 
        SED constraints are not good enough to constrain the galaxy IR luminosity.) }
    }
  \label{fig:radio-sf}
    \end{center}
\end{figure}

Our AGN selections in the radio-detected parent sample require the sources to
be detected in the radio 3 GHz image. However, this does not mean that our
sample would be biased towards radio-loud AGNs. Our VLA sensitivity { at the beam center} is deep
enough to match {\it Spitzer} MIPS 24~$\mu$m data in terms of detecting SFGs
during cosmic noon { with 5-$\sigma$ SFR limits from $\sim$25 to $\sim$85~$M_\odot$/yr at $z\sim$2--3 assuming 
the \citealt{Rieke2009} $L_{\rm IR}=10^{11.25}~L_\odot$ SFG template). Most of the MIPS 
sources fall in the 3~GHz HPBW and the radio data is within
$\sim$2$\times$ the MIPS SFR limit near the survey footprint edges}. 
Thus, the radio signals detected for most AGNs come from
the host galaxies. As shown in Figure~\ref{fig:q24_redshift}, the observed
$q_{\rm obs, 24}$ distributions of radio-quiet AGN and star-forming galaxies
appear to be similar, with a K-S probability of 0.17 { for drawing these two types of objects
from the same sample.} In Figure~\ref{fig:radio-sf}, we also compare the galaxy
IR luminosity derived from our SED fitting code with the radio luminosity of
all the VLA sources. A strong correlation is seen between these two parameters,
further indicating that star formation activity dominates the integrated radio
emission in most AGN/host galaxy systems (see also \citealt{Alberts2020}). As
discussed in the last section, the requirement of a radio detection introduces
biases towards detecting galaxies with higher SFR at higher redshifts, thus our
AGN sample with radio detections (of the host galaxy signals in most cases)
would miss AGNs in quiescent or relatively lower SFR galaxies.

We can test the relative AGN sample bias of the radio sources to those in the
whole X-ray detected source catalog.  Among the 1008 X-ray sources in the {\it
Chandra} 7 Ms catalog \citep{Luo2017}, 721 are matched in the 3D-HST catalog
(255 are in the final AGN sample and 170 are identified by X-ray luminosity)
and 5 X-ray sources do not have 3D-HST counterparts (they could be relatively
low-luminosity AGNs missed by HST or {\it Spitzer}) { but are within the GOODS-S 
footprint}; the other 282 X-ray sources are out
of the GOODS-S footprint. In the radio band, 438 of these 1008 X-ray sources
have 3 GHz counterparts. In total, there are 283 X-ray sources without
radio-detections but 3D-HST counterparts; 49 of them are [xtype\_lumcut] AGNs
and 230 are [xtype\_x2r] AGNs. Given the discussion at the end of
Section~\ref{sec:survey_limit}, these sources are likely those AGNs in galaxies
with relatively low levels of star formation and therefore with radio output
below the detection limit of the JVLA data.

\begin{figure}[htp]
    \begin{center}
        \includegraphics[width=1.0\hsize]{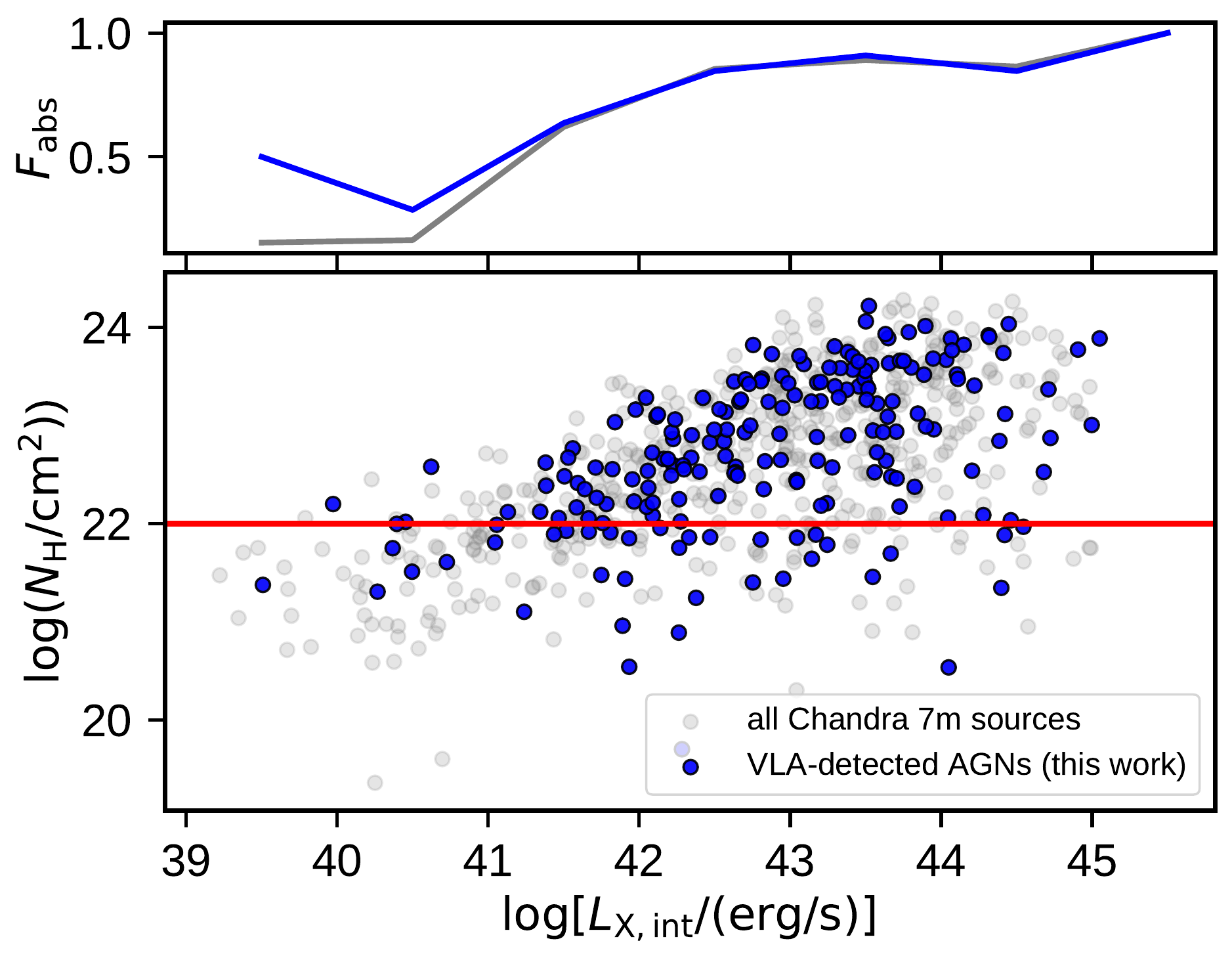}
    \caption{
        Gas column density, $N_{\rm H}$ and X-ray intrinsic luminosity, $L_{\rm X,int}$ of
        all Chandra 7Ms sources (grey dots) and VLA-detected AGNs (blue dots). The red horizonal
        line indicates $\log (N_{\rm H}/cm^2)=22$, above which the source is classified as an obscured 
        AGN. The top panel shows the fraction of objects that are obscured as a function of $L_{\rm X, int}$ (blue for VLA-detected AGNs
        and green for all Chandra 7Ms sources).
    }
  \label{fig:xray_nh_lx}
    \end{center}
\end{figure}

To explore how the requirement of a radio-detection changes the X-ray source
properties, Figure~\ref{fig:xray_nh_lx} shows the gas column densities and
X-ray intrinsic luminosities of radio-detected and -undetected X-ray sources;
the distributions are very similar. In fact, a K-S test
of the X-ray properties shows no significant differences in X-ray (0.5--7 KeV)
column density ($N_H$) ($D$=0.070, $p$=0.414); apparent luminosity ($L_{\rm X,
app}$) ($D$=0.085, $p$=0.205), intrinsic luminosity ($L_{\rm X, int}$)
($D$=0.074, $p$=0.349), or redshift ($z$) ($D$=0.068, $p$=0.463). As a result,
we conclude that the X-ray properties are not biased by the requirement for
radio detection.  

{ Since the radio-band is dominated by star formation activity
in the host galaxies, we will miss AGNs in galaxies with low SFRs towards higher
redshifts given the survey limit, which may indirectly introduce a bias against relatively low-luminosity AGNs if the AGN luminosity correlates with the host SFR (see Section~\ref{sec:agn-measure}). However, other AGN 
properties such as obscuration do not show a significant dependence on the level of  star
formation and thus are not biased by the requirement for radio detections.  } 

%In addition, the detection of 326 AGNs among the 759 radio sources
%also indicates that there are  reasonably luminous AGNs in the half of the galaxy
%population with active star formation.

%\subsection{Concerns about selection completeness}
%\label{concerns}

\subsubsection{Objects with Non-standard SEDs}\label{sec:irxray}

%31240          45910         8094        11403
%38.5943      38.6283     -99.9000      38.4714
%42.4223      43.2580      44.2209      42.9539
% 23.281       ----          23.40       21.4377

A strong correlation between the X-ray and mid-IR emission from AGNs has been
well-established in the literature \citep[e.g.,][]{Lutz2004,
Gandhi2009,asmus2015, Stern2015}  and has been adopted as evidence for the
similarities of AGN intrinsic SEDs, since the obscuration in the X-ray and
mid-IR is minimal compared to other wavelengths. This linkage is useful for searches where there are detections in only one of the spectral ranges. 

\begin{figure}[htp]
    \begin{center}
        \includegraphics[width=1.0\hsize]{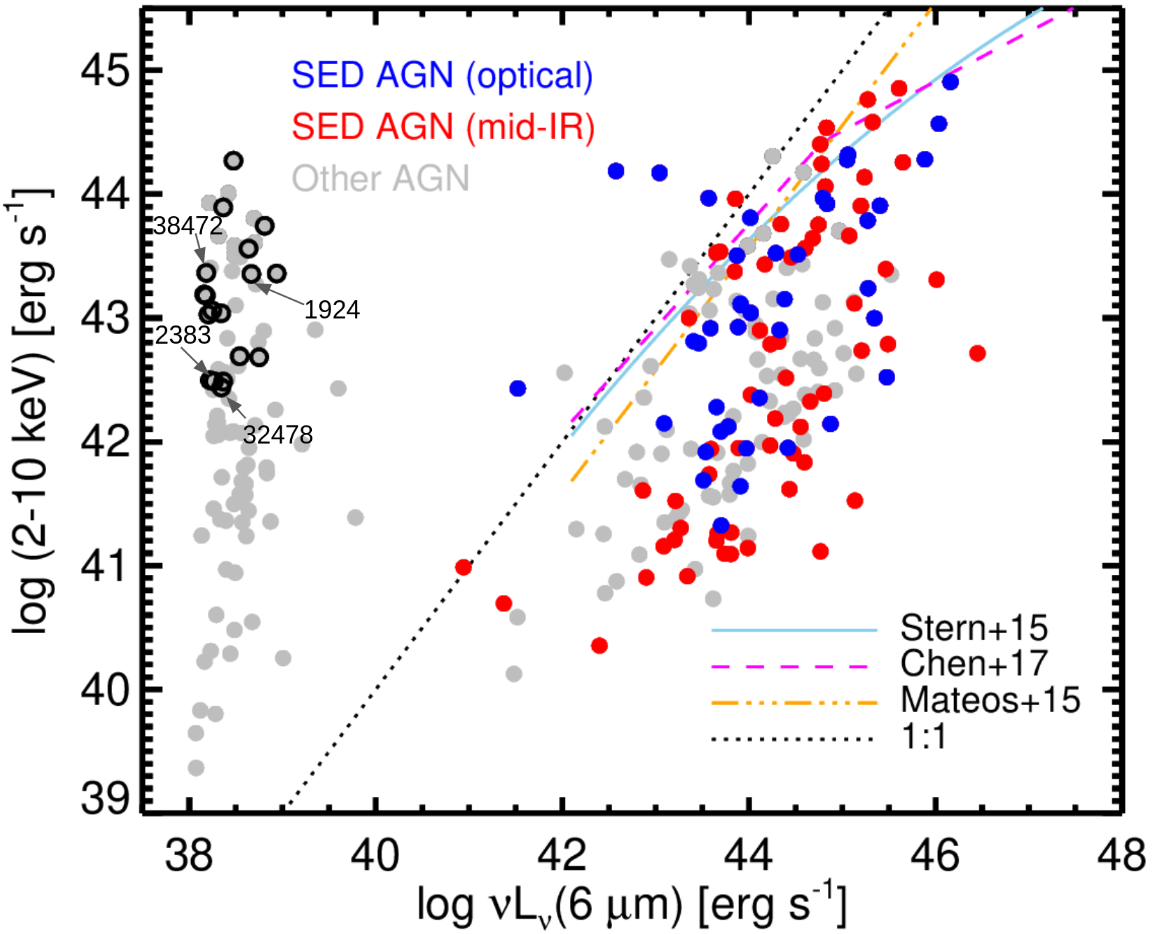}
    \caption{
        The relation between the rest-frame 6 $\mu$m luminosity from AGNs and
        the intrinsic 2--10 keV luminosity for our radio-flux-limited AGN
        sample. { We have shown the SED-identified AGNs in blue and red, corresponding to [otype\_sed] and [mtype\_sed], and other AGNs in
        gray. The open black circles indicate the locations of some AGNs with very low mid-IR emission relative to their X-ray luminosity after detailed visual SED inspections ({ Four objects whose SEDs will be shown in Figure~\ref{fig:fitting_example_irweak} are highlighted with their 3D-HST IDs}). We have also plotted some empirical relations of these quantities for bright AGNs ($\nu L_\nu(6~\mum)\gtrsim10^{42}$--$10^{43}$ erg~s$^{-1}$) reported in the literature.}
    }
  \label{fig:xray_ir_relation}
    \end{center}
\end{figure}

With our SED analysis and the
deep X-ray observations of the AGN in this field, we have revisited this relation. { Since \citet{Luo2017} only reported
the absorption-corrected (i.e., intrinsic) X-ray luminosity at 0.5--7 keV, we have converted these values to the rest-frame 2--10 keV
energy band, assuming a power-law spectrum with a fixed photon index of 1.8 using the Portable, Interactive, Multi-Mission
Simulator (PIMMS)\footnote{\url{http://cxc.harvard.edu/toolkit/pimms.jsp}}, to make them directly comparable to other literature studies.}
In Figure~\ref{fig:xray_ir_relation}, we present the AGN rest-frame mid-IR
6~$\mum$ luminosity from our SED fitting and X-ray 2--10 keV for our radio-flux-limited AGN. Previous
correlations from \cite{Stern2015}, \cite{Mateos2015} and \cite{Chen2017} are
also included in the figure as a comparison.

We confirm the existence of a correlation between the AGN absorption-corrected
X-ray luminosity at 0.5--7 keV and the AGN luminosity at 6 $\mu$m for
most AGNs. { At high AGN luminosity, the trend is similar to those reported
in the literature \citep[e.g.,][]{Stern2015, Mateos2015, Chen2017}. However, there seems to be an offset of the trend at relatively low luminosity compared to
the extrapolations of previous work and the scatter is also larger.}  \cite{Lambrides2020} also
studied the mid-IR and X-ray luminosity relation of AGNs in this field and
found similar trends (see their Figure 4). These discrepancies can be caused by either the uncertainties in the X-ray
absorption correction, or the limited data { constraints} in the mid-IR to measure
a robust $L_{\rm AGN, 6\mum}$. 
%The latter issue can be resolved with the
%multi-band JWST MIRI observations of these AGNs.

As shown in Figure~\ref{fig:xray_ir_relation}, our SED fitting approach has
failed to detect strong 6 $\mum$ AGN emission ($L_{6\mum}>10^{40}$erg s$^{-1}$)
for { 88} out of 269 X-ray detected AGN (33\% of the sample). If we limit the
objects to the { 145} X-ray bright AGNs ($L_{\rm X-ray, int}>10^{42.5}~{\rm
    erg~s}^{-1}$), there are still { 42} objects that do not have appreciable
    6~$\mum$ AGN emission (29\% of the sample). Interestingly, based on a study
    of bright AGNs in the SDSS Stripe 82 field, \cite{LaMassa2019} found that
    61\% of their X-ray bright AGNs do not have sufficiently strong mid-IR
    emission to be identified as AGNs,  39\% due to  failure of the mid-IR
    color method and 22\% that are undetected by WISE. As a result, it seems
    that X-ray bright but mid-IR weak AGNs do exist in a general sense.

\begin{figure*}[htp]
    \begin{center}
  \includegraphics[width=0.49\hsize]{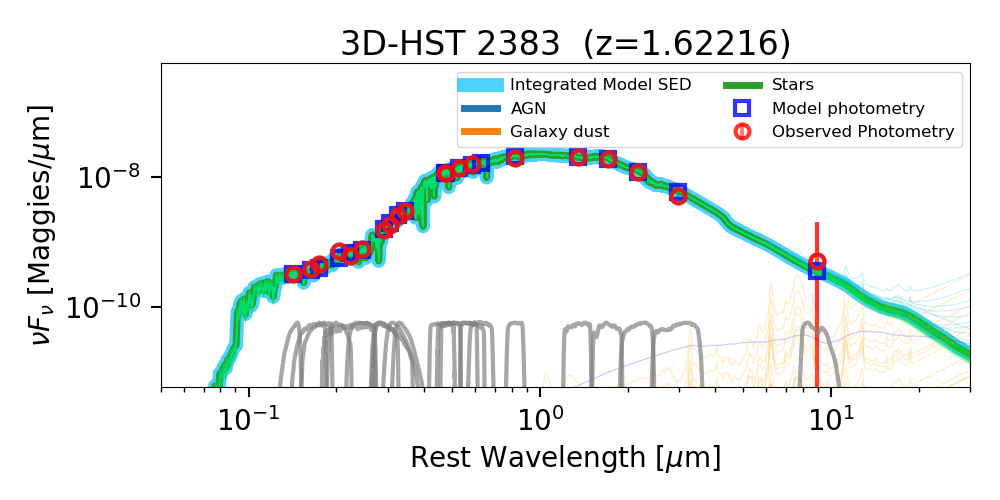}
  \includegraphics[width=0.49\hsize]{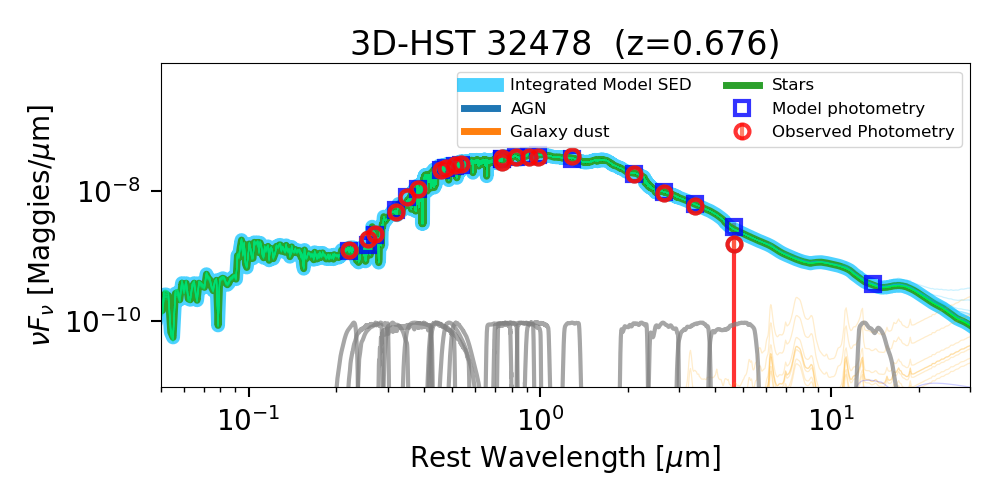}
  \includegraphics[width=0.49\hsize]{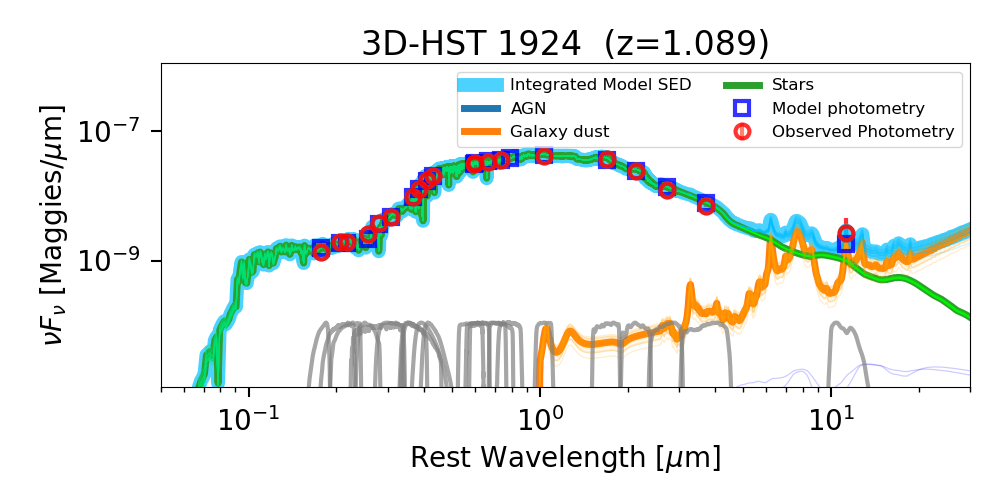}
  \includegraphics[width=0.49\hsize]{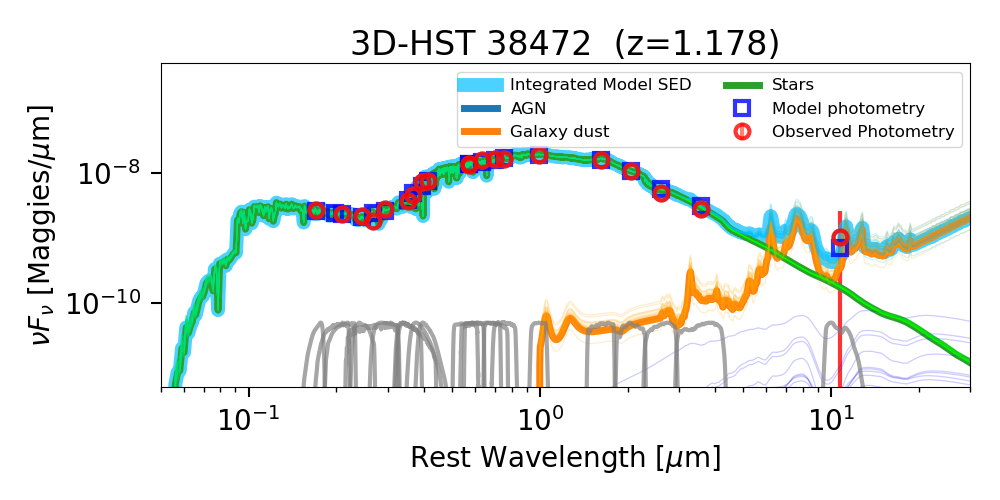}
    \caption{
    Example SED fittings of X-ray bright but IR weak AGNs (these sources are identified in Figure~\ref{fig:xray_ir_relation}). The meanings
    of the lines and the symbols are the same as Figure~\ref{fig:fitting_example}. 
    %The logarithms of the absorption-corrected X-ray luminosities (erg/s) between 0.5 and 7.0 keV are respectively 42.64, 42.58, 43.50 and 43.50 for sources with ID 3D-HST 2383, 32478, 1924, and 38472 \citep{Luo2017}. 
    The logarithms of the absorption-corrected X-ray luminosities (erg/s) between 2 and 10 keV are respectively 42.50, 42.44, 43.36 and 43.36 for sources with ID 3D-HST 2383, 32478, 1924, and 38472. 
    }
  \label{fig:fitting_example_irweak}
    \end{center}
\end{figure*}

It is possible that some of these seemingly abnormal AGNs do not show strong
mid-IR emission simply because of the poor SED constraints.  However,
convincing cases that deviate significantly from the X-ray and mid-IR
correlation do exist. { There are  26 AGNs with the non-stellar
emission at 6 $\mum$ notably below  the 1:1 relation with the X-ray, even if we assume all the mid-IR emission is
from the nuclei; that is, with X-ray output significantly greater than
predicted by the X-ray/infrared relation.} Thus, such objects constitute at least $\sim$10\% of the sample.
Again, if we limit the sample to the bright X-ray AGNs, 18 objects are convincing cases for this behavior { after visual inspections  (open
circles in the figure)}, and they contribute also about
$\sim$12\% of the X-ray bright AGN sample.  { Figure~\ref{fig:fitting_example_irweak} presents some examples of such objects.}

In summary,  $\sim$10\%--40\% of the X-ray bright AGNs may 
not have strong mid-IR emission. 
\cite{LaMassa2019} has suggested that they could be dust-deficient AGNs
\citep[e.g., see][]{Lyu2017a}. However, our study reveals a more complicated
picture as some of these AGNs { have a very weak central
engine component inferred from optical to mid-IR SED fittings despite the strong X-ray AGN emission. Those in Figure~\ref{fig:fitting_example_irweak} are examples, with no excess emission from an AGN apparent from the far UV through the mid-infrared. }

\subsubsection{Incompleteness due to variability}\label{sec:bias-var}

There is a chance that some AGNs could be missed if their time variability
dropped their flux below the detection/selection threshold at the time of the
observations. An investigation of this issue requires both very extensive monitoring over an extended period and knowledge of the time evolution of the AGN multi-wavelength properties. Systematic conclusions are not feasible due to the
lack of sufficient observational constraints even in GOODS-S. 

\subsection{Properties of the Obscured AGN Population}\label{sec:obscured_pop}

The identification of obscured AGNs is known to be very challenging. Our
multi-wavelength search campaign has greatly increased the AGN numbers compared
to any single band study and offers a chance to characterize the statistical
properties of the obscured AGN population with the deepest data available up to
date. As the radio wavelengths are least influenced by gas and dust
obscuration, we focus on the 326 AGNs detected in the { radio band} to
avoid notable bias on AGN obscuration properties.

Generally speaking, AGN obscuration can be divided into two categories. In the
UV-to-mid-IR band, dust obscuration introduces wavelength dependent reddening
and reduces the AGN continuum emission. By fitting the observed SED, our model
provides preliminary constraints on the AGN dust attenuation level. About 2/3
of the sample can be defined as obscured AGNs with $A_V\gtrsim1.0$.  This
number is consistent with the dust covering factor estimated based on the
intrinsic AGN template \citep[see e.g.,][]{Lyu2017b}. This level of attenuation significantly reduces the probability of finding the AGN in the optical, increasing the importance of X-ray identification.

In the X-ray, it is the optically-thick gas that dominates
AGN obscuration.  We have estimated the intrinsic  absorption column density 
by comparing the observed X-ray spectral slope with an assumed intrinsic
power-law spectrum with photon index of 1.8 using the PIMMS in
\cite{Luo2017}. Adopting $N_H=10^{22} \text{cm}^{-2}$ as the criterion, we have
37 X-ray unobscured AGNs and 161 X-ray obscured AGNs (and 51 not detected in
X-ray), yielding an X-ray obscured fraction of 81-88\% { among the radio-detected AGN sample in this work}.

In the following, we first discuss if there are correlations between the
SED-inferred obscuration and the X-ray gas absorption and then explore how the
fraction of obscured AGNs changes with the luminosity and redshift.

\subsubsection{Gas Absorption vs. Dust Extinction}\label{sec:extinction_relation}

\begin{figure}[htp]
    \begin{center}
  \includegraphics[width=1.0\hsize]{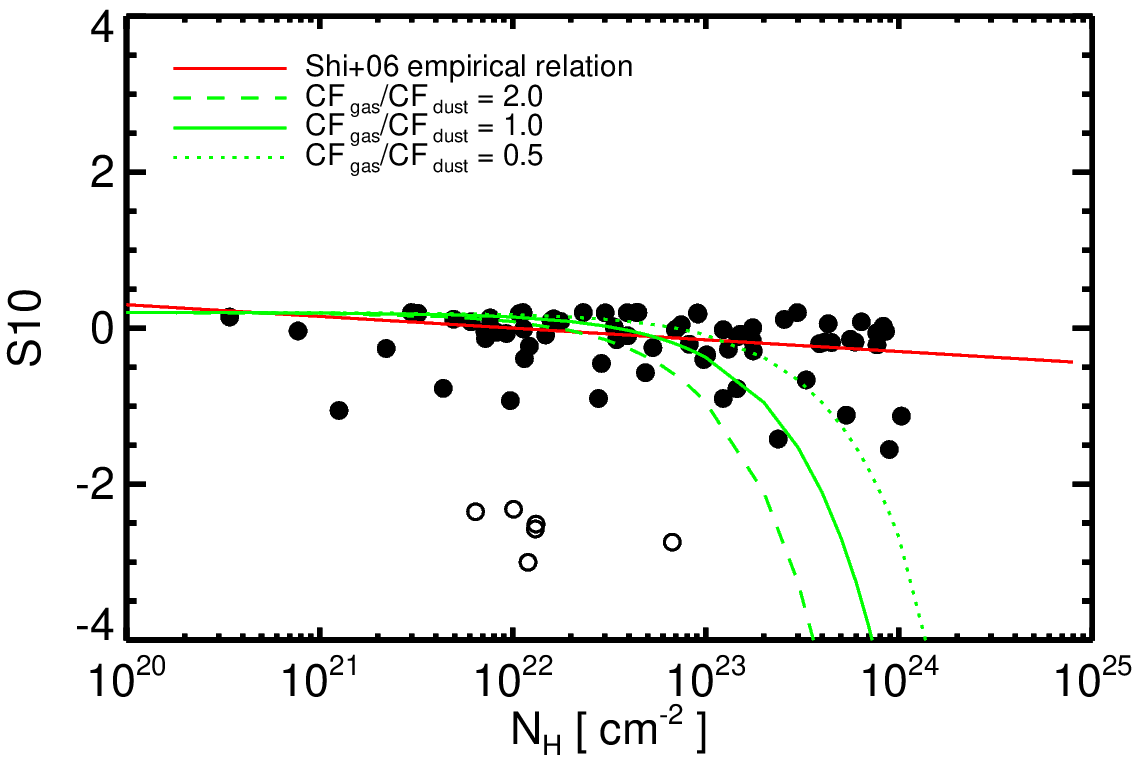}
    \caption{
       The distribution of the X-ray gas column density and SED-inferred
       10~$\mu$m silicate strength of SED AGNs. { After dropping seven AGNs with very high mid-IR extinction ($S10>-2$, open circles), the other objects (solid dots) show a weak trend. } The red line is the correlation
       established in \cite{Shi2006}. The green lines are the analytical
       relations described in this work.
    }
  \label{fig:nh_extinction}
    \end{center}
\end{figure}

{ The relation between obscuration by dust (in the optical and near infrared) and by gas (in the X-ray) is complex, complicating the identification of obscured AGNs. } To explore it, we
selected 76 SED-identified AGNs that have X-ray column density measurements
from \cite{Luo2017}. As the host galaxy would dominate the UV-to-near-IR
emission, our constraints on the AGN SED obscuration level is mainly based on
the near-to-mid-IR bands. Thus, we have converted the optical attenuation $A_V$
to the mid-IR silicate strength $S10$ assuming $S10=-A_V/5.5+0.2$ (
    \citealt{Lyu2014}; an offset of 0.2 is introduced to be consistent with the
    average silicate values for unobscured quasars, see \citealt{Hao2007}). We
    present the distribution of our AGN sample on the $N_H$--$S10$ plane in
    Figure~\ref{fig:nh_extinction}. 

\cite{Shi2006} first reported a possible correlation between $\log N_H$ and
$S10$ for a sample of 97 low-$z$ AGNs, despite the large dispersion.  After
ignoring the seven AGNs with $S10<-2$ (open circles in
Figure~\ref{fig:nh_extinction}), the linear Pearson correlation coefficient of
our sample is $-0.26$ with the significance of its deviation from zero at
0.026, indicating a weak negative correlation that is roughly consistent with
the result of \cite{Shi2006}.

However, we should not expect a well-defined linear correlation. In fact, the
relation between gas column density $\log N_H$ and the X-ray flux correction
$L_{X, app}/L_{X, int}$ is highly non-linear (see e.g., \citealt{Gilli2007}).
If we define the fraction between gas and dust covering factors as
$CF_\text{gas}/CF_\text{dust}= (L_\text{X, app}/L_\text{X, int})/(L_\text{opt,
app}/L_\text{opt, int})$ and note $A_V=-2.5\times\log(L_\text{opt,
app}/L_\text{opt, int})$, the theoretical correlation between $S10$ and $N_H$
would be described by the green lines in Figure~\ref{fig:nh_extinction}: for
low values of $N_H$, the dust attenuation varies little despite the large
changes of the gas column density. The correlation becomes stronger after $N_H$
reaches some threshold. It seems our data cannot validate or rule out either
case. In other words, we do not find a strong correlation between the dust
extinction and gas absorption in our AGN sample.

There are several issues that could undermine any conclusions.  We first note
the measurements of both $N_H$ and $S10$ (or $A_V$) are highly uncertain. Due
to the limited number of X-ray photons,  rather than carrying out a robust
spectral analysis the gas column density $N_H$ is estimated by comparing the
observed X-ray spectral slope with the assumed intrinsic spectrum under several
assumptions.  The AGN dust attenuation level is also uncertain due to the
limited near- to mid-IR data coverage.  We expect that upcoming near- to mid-IR
surveys of this field by JWST NIRCam and MIRI will help constrain the latter
parameters more accurately. Better measurements of the X-ray properties of AGNs
in this field are also expected with future X-ray telescopes such as Athena and
Lynx.

The study of the relation of AGN X-ray gas absorption and optical-to-mid-IR
dust attenuation in the literature also yields very complicated results.  As
shown in \cite{Shi2006, Xu2020}, the reported weak correlation between $S10$
and $N_H$ at low-$z$ always is associated with very large scatter.
\cite{Goulding2012} demonstrated the possibility of a host galaxy contribution
to the  X-ray extinction in some sub-groups of the AGN population, which may
further increase the scatter of the relation. { The lack of strong correlation
between the X-ray gas absorption and optical-to-IR dust extinction indicates the complicated
nature of AGN obscuration, which in turn influences
the AGN identification across different wavelengths.}

\subsubsection{Obscured AGN Fraction and its Possible Evolution}

To explore how the obscured AGN fraction changes with redshift and luminosity,
we need a complete sample that does not have e.g., Malmquist bias. After
checking the luminosity and redshift distribution, we find the objects at
$\log(L_{\rm bol}/L_\odot)=[9.5, 12.5]$ and $z=[0.5, 3.5]$ are mostly complete.
There are 216 such AGNs and we have binned them at $\log(L_{\rm bol}/L_\odot)$=[9.5,
10.5], [10.5, 11.5] and [11.5, 12.5], and $z$=[0.5, 1.5], [1.5, 2.5] and [2.5,
3.5].

\begin{figure*}[htp]
    \begin{center}
        \includegraphics[width=0.48\hsize]{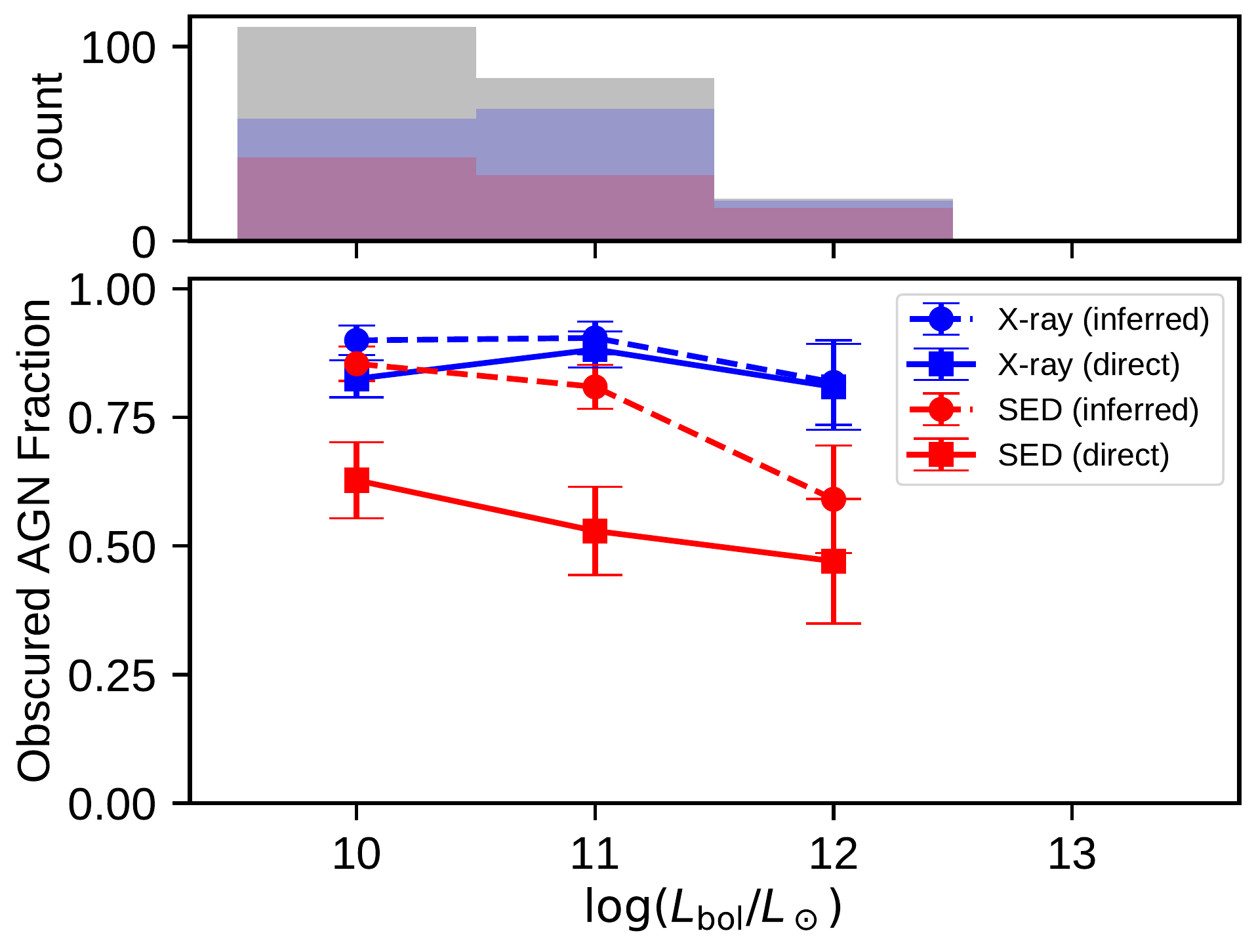}
        \includegraphics[width=0.48\hsize]{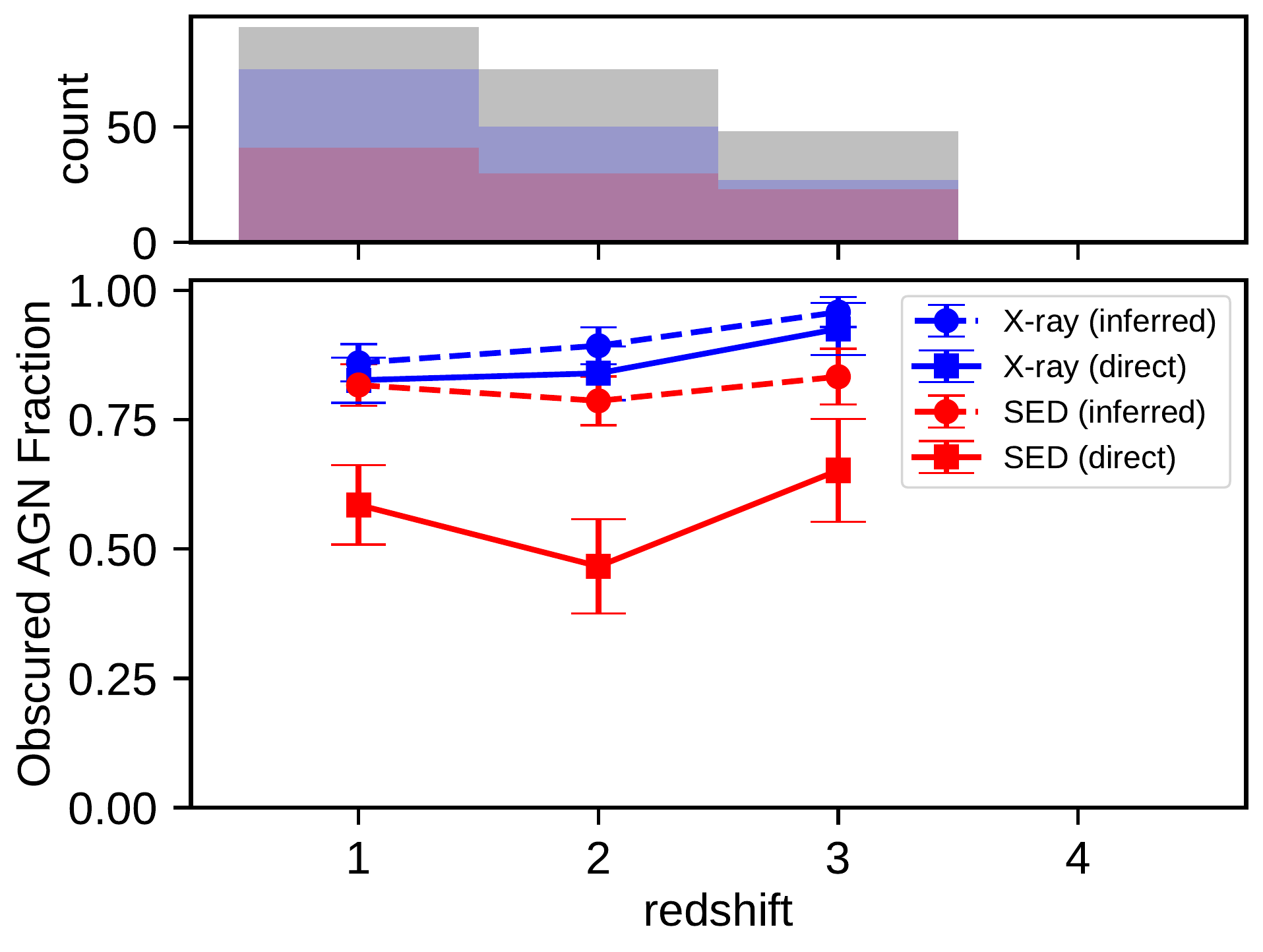}
    \caption{
        { AGN number counts (top panels) and obscured fraction (bottom panels) as a function of luminosity (left) and redshift
        (right). We have limited the sample to 216 AGNs with $\log(L_{\rm bol}/L_\odot)=[9.5, 12.5]$ and $z=[0.5,
        3.5]$.}  The SED obscuration is defined by the number of
        optical SED AGNs relative to the total sample and the X-ray obscuration
        is defined by the number of objects with $N_N<10^{22} cm^{-2}$ to the
        total sample. We define two groups of obscuration: direct -- the sample
        is limited to objects with the relevant measurements, inferred -- the
        whole AGN sample is considered and objects without the relevant
        measurements are considered as obscured AGNs. For example, if an AGN identified in other ways is not detected at the expected X-ray luminosity, we attribute this to obscuration in the X-ray. 
    }
  \label{fig:obs_dependent}
    \end{center}
\end{figure*}

In Figure~\ref{fig:obs_dependent}, we show the luminosity and redshift
dependence of the X-ray and UV-optical obscured fraction. There is a luminosity
dependence of the AGN optical obscuration but not X-ray obscuration. There seems to 
be no strong redshift evolution of the obscured fraction. 

{ As mentioned above, the robustness of our SED analysis is largely limited by the mid-IR wavelength coverage 
of current data, thus the results presented here are preliminary. We will conduct 
a detailed study with upcoming JWST NIRCam and MIRI observations 
of this field (see Section~\ref{sec:jwst-predict}) and discuss  
this topic in-depth in another paper.}

\subsection{Constraints on AGN Statistics}\label{sec:agn_lf}

{ Now we discuss some statistics that can be derived from our AGN sample. It should 
be noted that all these quantities are subject to the uncertainties caused by cosmic variance, as already
mentioned in Section~\ref{sec:agn-sample}.}

\subsubsection{AGN Sky Surface Density near the HUDF}

\begin{figure}[htp]
    \begin{center}
        \includegraphics[width=1.0\hsize]{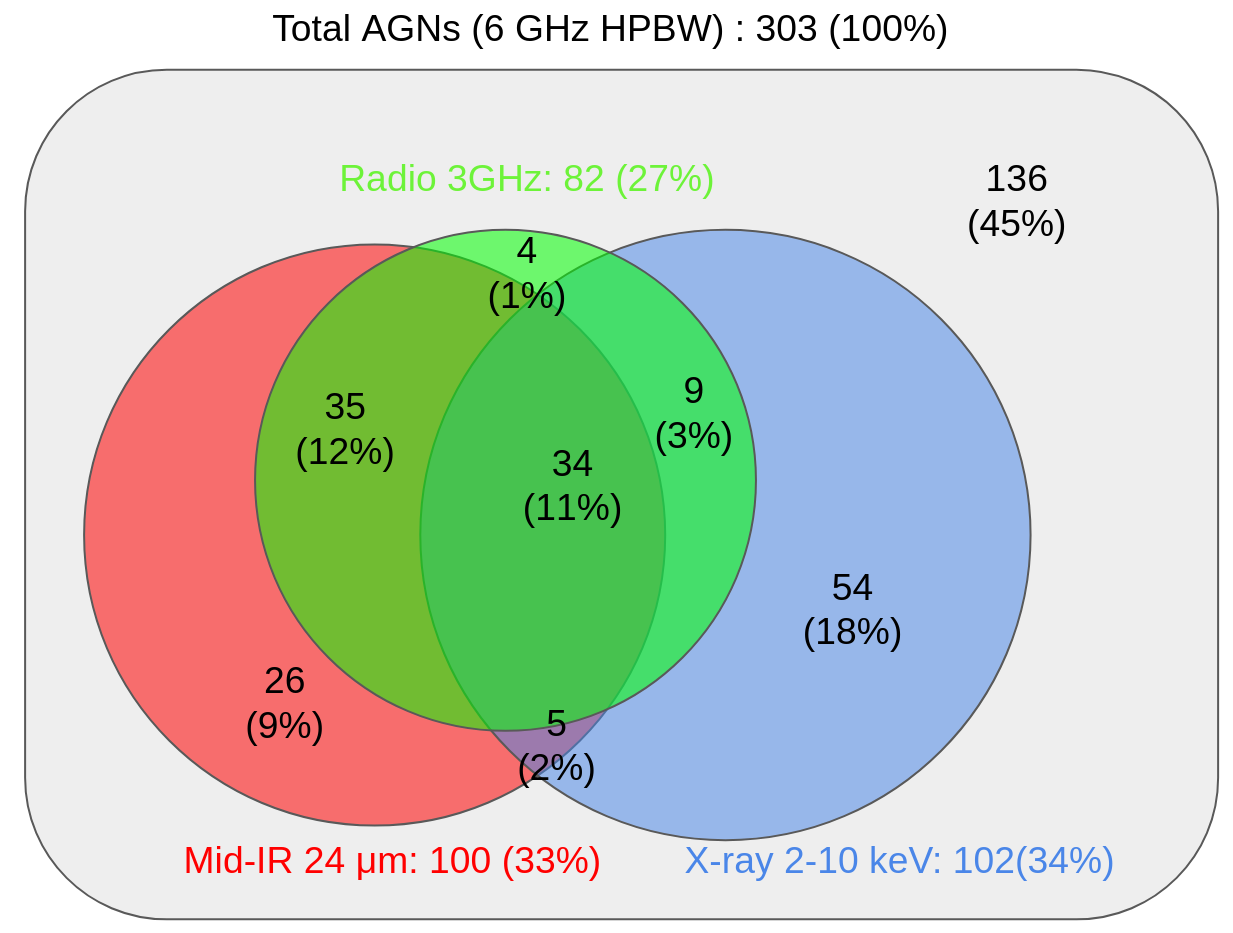}
    \caption{
   Venn diagram of AGNs in the VLA 6GHz HPBW that are detected in X-ray 0.5--7.0 keV, mid-IR 24$\mum$ and radio 3GHz.
    }
  \label{fig:6ghz_agn_venn}
    \end{center}
\end{figure}

To get a robust estimation of AGN number density, we have limited the sample to
the VLA 6 GHz primary beam HPBW with a radius of 220\arcsec. This small field
is centered around the HUDF and has the deepest observations of all the
multi-wavelength data. There are 221 radio-undetected AGNs (162 have  $L_{\text
bol}>10^6L_{\odot}$) and 82 radio-detected AGNs, yielding a total AGN count of
303. Given the corresponding sky area of 42.237 arcmin$^2$, the total AGN
surface density is about 7.2 per arcmin$^2$. {
Figure~\ref{fig:6ghz_agn_venn} shows the Venn diagram of the AGN detections at different
bands in this sky area and Table~\ref{tab:agn-density} presents the corresponding
flux-limited AGN surface densities.} We can conclude again that no single
band can detect all the AGNs despite their similar AGN bolometric luminosity
ranges.

\begin{deluxetable}{ccccc}
    %\tabletypesize{\scriptsize}
    \tabletypesize{\footnotesize}
    \tablewidth{1.0\hsize}
    \tablecolumns{4}
    \tablecaption{AGN sky surface density in 6GHz HPBW\label{tab:agn-density}}
    \tablehead{
  \colhead{band} &
  \colhead{5$\sigma$ limit} &
  \colhead{AGN count} &
  \colhead{number density}  \\
   \colhead{} &
   \colhead{$\mu$Jy} &
   \colhead{} &
   \colhead{arcmin$^{-2}$} 
}
\startdata
  radio 3GHz       &  2.55 &  82 &  1.94  \\
  mid-IR 24$\mum$  &   30 &  100 &  2.37 \\
  X-ray 0.5--7 keV &  2.09e-6 & 102 &  2.41 \\
  integrated$^*$   & $\cdots$       & 303 &  7.17 
\enddata
%\tablenotetext{1}{All these priors have been sampled linearly.}
\tablenotetext{*}{The integrated results also include sources that are not detected in X-ray, mid-IR nor radio.}
\end{deluxetable}

\subsubsection{Luminosity Function}

For each AGN, we have assigned the luminosity of the AGN in
Section~\ref{sec:agn-measure}.  We compute the AGN bolometric luminosity function with
the $1/V_a$ method \citep{Page2000}.  For each AGN, the co-moving volume within
which it has been observed is 
\begin{equation}
    V_a = \int_{\Delta z} p(L_{\rm bol}, z)\frac{dV}{dz}dz,
\end{equation}
where $p(L_{\rm bol},z)$ is the selection function, defined as the probability
that the AGN with a given $L_{\rm bol}$ and z can be identified.  The total
{ volume} density and its uncertainty are
\begin{equation}
    \rho = \sum_{i} \frac{1}{V_a^i}~~~, 
\end{equation}
\begin{equation}
    \sigma(\rho) = \left[\sum_{i} \left(\frac{1}{V^i_a}\right)^2\right]^{1/2}  ~~~.
\end{equation}
The selection function of each individual method is quite complicated since
host galaxy contamination, AGN obscuration and intrinsic SED variations all
play some roles. However, we can limit the sample to $z=0.5$--3.5 and
$log(L_{\rm bol}/L_\odot)=$9.5--12.5 and assume $p(L_{\rm bol},z)\sim1$
considering all possible selection techniques have been utilized. In contrast
with previous studies, we do not assume some pre-defined AGN model to make any
corrections to the selection but rely on the integrated results of an extensive
list of multi-wavelength AGN identifications. 

\begin{figure}[htp]
    \begin{center}
  \includegraphics[width=1.0\hsize]{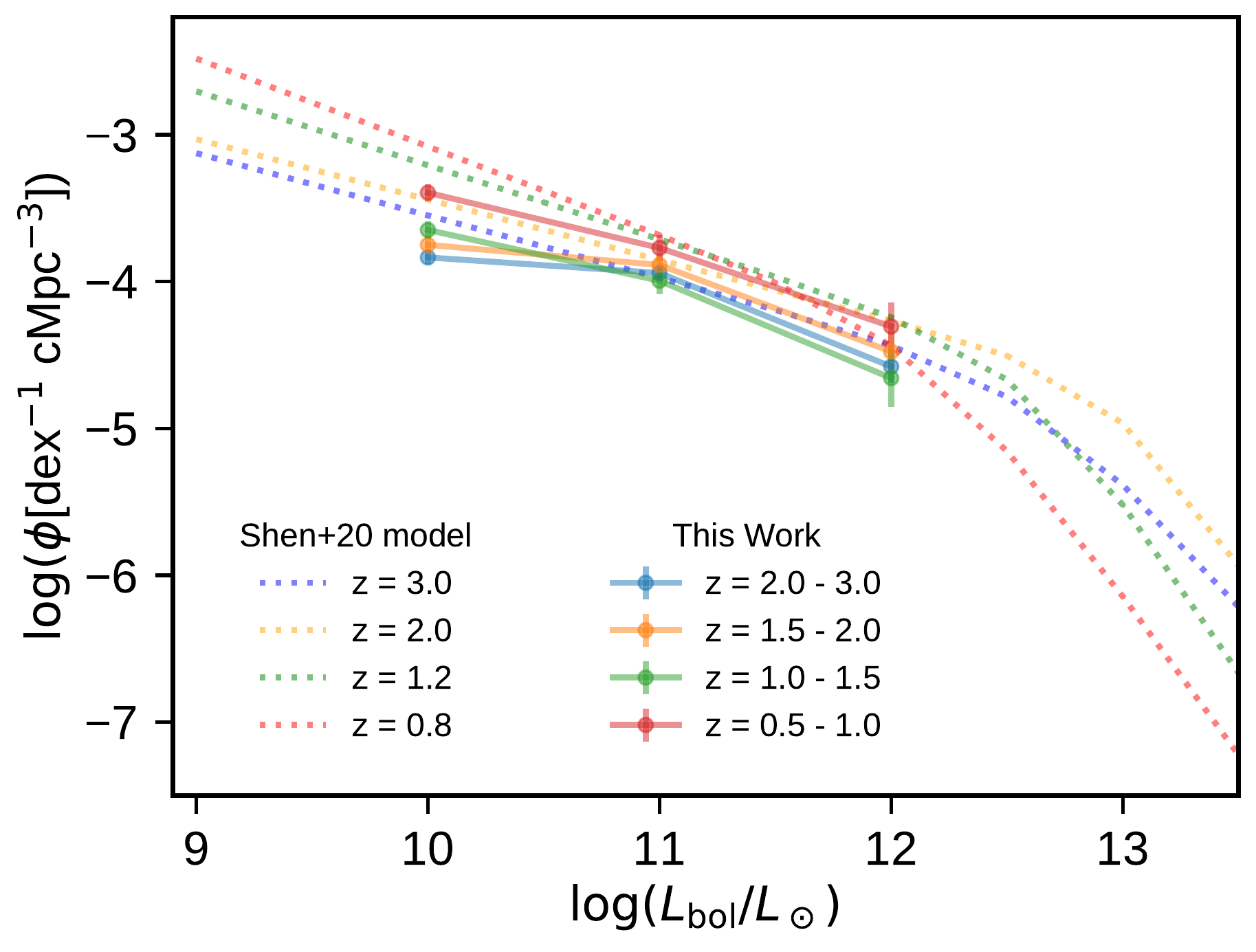}
    \caption{
       The apparent AGN luminosity function from this work and a comparison
       with the modeling work by \cite{Shen2020}.
    }
  \label{fig:agn_lum_function}
    \end{center}
\end{figure}

We calculate the luminosity functions over the whole 3D-HST GOODS-S footprint 
at $z=$0.5-1, 1--1.5, 1.5--2, 2--3 and present the results in Figure~\ref{fig:agn_lum_function}.  
As a comparison, we
also show the luminosity functions derived in \cite{Shen2020}.  With various
model assumptions on the AGN SED, bolometric correction, and gas and dust
extinction, \cite{Shen2020} have proposed a bolometric luminosity function
model and fitted the observed AGN luminosity functions in the optical, UV, IR
and X-ray from different surveys. Interestingly, our model-independent
empirical AGN bolometric luminosity  functions in the same sky area agree with their
results in general. Nevertheless, this agreement should not be
over-interpreted. For example, our study has found a significant number of
obscured AGN not included in \citet{Shen2020}, not a surprising result given
that they comment that there is `` an apparent deficiency in IR observations
compared with other wavelengths ...  the total number of IR data points are
limited and thus they have low statistical significance in the fit of the
bolometric quasar luminosity function." However, the relative number, $\sim$
20\% of the total, is too small to show in Figure~\ref{fig:agn_lum_function}
outside the errors. In addition, both studies are likely to miss-estimate the
incidence of Compton-thick AGNs. A better comparison with theoretical estimates
will be possible with characterization of the obscured population using JWST.

\subsubsection{Comparison with Other Work}

{ There have been many previous efforts to identify complete samples of AGNs. 
A comprehensive review can be found in \citet{Lyu2022}. Our approach represents an advance 
because of the depth of the observational database and the breadth of the methods employed. 
Here we discuss some of the leading previous efforts toward very thorough AGN identification, 
using them as a guide to the next generation of searches.} 

\cite{Luo2017} have identified AGNs with various methods among  X-ray sources
in the Chandra survey of GOODS-S and reported 464 AGNs from 721 X-ray sources
within the 3D-HST footprint. In comparison, we have identified 906 AGNs in total, 684 of which are X-ray sources listed by \cite{Luo2017}.\footnote{Our AGN
detections require the sources to be  cataloged in 3D-HST so the X-ray source
number in our catalog (684) is smaller compared with the value (721) in
\cite{Luo2017}} In other words, with more extensive use of ancillary data, we have
doubled the AGN number compared to the results in \cite{Luo2017}. As shown in Section~\ref{sec:agn-demo-sel}, X-ray luminosity alone only identifies $\sim$ 35\% of the total sample of AGNs we have found; deep X-rays are most effective when combined with other indicators. The increase of X-ray detected AGNs is
largely driven by those selected through the ratio of X-ray to radio luminosity, thanks to our ultra-deep VLA radio data. Roughly 
26\% of our AGN sample is not even detected in this deepest {\it Chandra}
image, possibly indicating a large number of heavily obscured or Compton-thick
AGNs missed in the X-ray survey. The  AGNs not detected in the X-ray { must be} 
identified through various approaches such as SED analysis, radio features and variability.

%342 radio-undetected AGN are X-ray sources, 269 radio-detected AGN are X-ray sources.   ===> 611 AGNs detected in X-ray
%174 MIPS AGN not detected in VLA,  298 MIPS AGN detected ===> 472 AGNs detected in MIPS
%      326 AGNs detected in radio 

\cite{Alberts2020} studied the AGN census in the radio sample of GOODS-S/HUDF
and reported an AGN number density of 1 per armin$^2$ in their radio-detected
sample and a similar density of radio-undetected AGNs identified via X-ray
properties. The radio-detected AGN number density obtained in this work is
about 1.9 arcmin$^{-2}$, almost double their value. { This comparison emphasizes 
the importance of comprehensive search methods, such as the ability to find heavily 
obscured AGNs as introduced in our work. }  This gain is also reflected in the
radio-undetected AGN number density
obtained in this work, about 3.5 arcmin$^{-2}$, also much higher than the
value in \cite{Alberts2020}.

{ There have been a number of AGN search programs in the COSMOS field \citep[e.g.,][]{delvecchio2017}. 
The deepest is based on the JVLA COSMOS-XS survey \citep{vander2021}, at similar frequencies and 
comparable rms sensitivities (0.53 $\mu$Jy beam$^{-1}$ at 3 GHz and 0.41 $\mu$Jy beam$^{-1}$ at 10 GHz) 
as our study (0.51 $\mu$Jy beam$^{-1}$ at 3 GHz and 0.29 $\mu$Jy beam$^{-1}$ at 6 GHz). Taking advantage 
of the multi-wavelength data available for this field, \citet{algera2020} have carried out an AGN census
among the radio sources. Over the 3GHz area of $\sim$350 arcmin$^2$, they reported an AGN fraction of $\sim$23\% 
among the 1437 radio sources with redshift constraints, corresponding to
an AGN number density $\sim$0.95 arcmin$^{-2}$. We find virtually an identical number of radio-emitting AGNs (323) 
as were found by \citet{algera2020} (334), but over a field half as large, i.e., our AGN number density for the 
GOODS-S radio-detected sample is higher by a factor of $\sim$2. 

Although cosmic variance might play a role, there are  several additional reasons for this difference. 
First, the X-ray coverage of GOODS-S is far deeper than in the COSMOS field, { allowing us to select
more AGNs per unit area with tracers associated with AGN X-ray emission}. \citet{algera2020} 
identified AGNs through X-ray fluxes relative to 
expected star formation luminosities in the far infrared, whereas we identified them relative to 
luminosities at 24 $\mu$m and 3--6 GHz, both of which are significantly deeper and thus supported a 
higher identification rate. The Spitzer IRAC data are also much deeper, particularly in bands 3 and 4. 
The IRAC band 4 at 8~$\mu$m is critical for infrared SED analysis; the deep detection limits for the 
majority of sources in GOODS-S contributed to higher identification rates through our SED fitting. 
However, for $z\gtrsim$1.5, this band moves out of the most diagnostic region, $\sim$5~$\mu$m 
(see Section~\ref{sec:survey_limit}); photometric bands to 12 or 15~$\mu$m would be even more powerful. 
We  also used  selection techniques that are not included in \citet{algera2020}, such as radio loudness 
([rtype\_rl]), radio slope ([rtype\_fss]), and time variability ([var]), and our fitting procedure provides 
a more flexible and complete approach to AGN obscuration levels. 

{ Last but not least, \citet{algera2020} 
have limited their AGN sample to be radio-detected while our study included AGNs detected in X-ray, mid-IR or radio. As
a result, our total AGN number density is
larger than the value reported in \citet{algera2020} by more than five times.}

{ These comparisons emphasize} how extremely deep observations at all of radio, mid-infrared, visible, and 
X-ray are required for a complete AGN census.

\subsection{Outlook for Upcoming JWST GTO Surveys}\label{sec:jwst-predict}

We now turn to the improvements in finding obscured AGN with JWST data. The
largest shortcoming in our current study is the lack of deep data between 8 and
24 $\mu$m. The result is potential degeneracy between star-forming galaxy
aromatic band emission and the smoother continua of obscured AGNs. The most
prominent aromatic band feature is the blended set of bands with central
wavelengths between 7.4 and 8.6 $\mu$m, which has a width $\Delta
\lambda$/$\lambda$ $\sim$ 18\%. The JWST/MIRI imaging bands from 10 through 21
$\mu$m have widths of  $\Delta \lambda$/$\lambda$ $\sim$ 17--23\%, well
matched to identifying SEDs with strong aromatic emission. Another strong
indication of such emission comes from the lack of strong features at $\lambda
<$ 6.2 $\mu$m, resulting in a sharp drop in the SED, where the broad continua
expected from embedded AGN have a distinctly different behavior. 

We estimate the number of suitable AGN detections per square arcmin from our
results for the 6 GHz field, where there are 69 radio-detected examples and 27
radio undetected ones all with 24 $\mu$m fluxes detected at $\ge$ 5 $\sigma$
(corresponding to $\sim$ 100 $\mu$Jy). The density of these AGNs is 2.27
arcmin$^{-2}$. 

There are two programs in the Cycle 1 observations with JWST that obtain  MIRI
data in multiple bands and to sufficient depth: ``The Cosmic Evolution Early
Release Science (CEERS) Survey'' (ERS 1345; \citealt{ceers}) and ``MIRI in the Hubble Ultra-Deep
Field'' (GTO 1207; \citealt{miri-hudf}).  Program ERS 1345 surveys three MIRI fields of view, or
about 7 arcmin$^2$, so the predicted number of AGN with such data is
$\sim$16, and without deep NIRCam photometric coverage. This program probably has too
small a sample for this type of study, particularly without the NIRCam
measurements. Program GTO 1207 surveys about 30 arcmin$^2$, with a predicted
return of about 70 sufficiently bright AGNs { detected in all the MIRI bands}, adequate for an initial study.
This program has the further advantage to be coordinated with the deep NIRCam
data from the James Webb Space Telescope Advanced Deep Extragalactic Survey
(JADES; \citealt{jades}) for the entire area.     

We therefore discuss this latter program (GTO 1207 + JADES) in more detail, as a prototype for the
more extensive surveys we anticipate will be conducted with JWST over its
lifetime. In Figure~\ref{fig:jwst_curve}, we show the predicted detection limits for the
NIRCam and MIRI photometric bands from unobscured (left panels), heavily-obscured (middle-panels) and
extremely-obscured (right panels) AGNs as a function of redshift.  For lightly
obscured AGNs ($A_V\sim0$), strong detections will be obtained over a large range of
luminosity and to z $\gtrsim$ 3 in all the NIRCam and MIRI bands (except
F2550W). { For heavily-obscured AGNs ($A_V=10$, corresponding to a $\sim10~\mum$ silicate absorption depth $\tau_{10}\sim1.5$), 
good detections will be possible for the longer-wavelength NIRCam bands (F335W, F410W and F444W) and all the MIRI bands. 
A representative sample of objects as extremely obscured as $A_V = 20$ (corresponding to $\tau_{10}\sim$2.6) 
will be detected over a reasonable range of luminosity and redshift 
with the MIRI bands (excepting F560W and F2550W at z $>$ 1) } and the longest wavelength NIRCam ones out to z $\sim$
1.5--2.  At a redshift of 2, the F2100W band is at a rest wavelength of
7~$\mu$m, and thus at higher redshifts the sensitivity to heavily embedded AGNs
will start to drop. { These multi-band measurements will suffice for a thorough search for 
obscured AGNs, similar but much more thorough than the one reported here.} Given the number of AGNs 
at $z\le$2, the result should be { a rigorous determination of the
fraction of embedded objects among the AGN population.} 

\begin{figure*}[htp]
    \begin{center}
  \includegraphics[width=1.0\hsize]{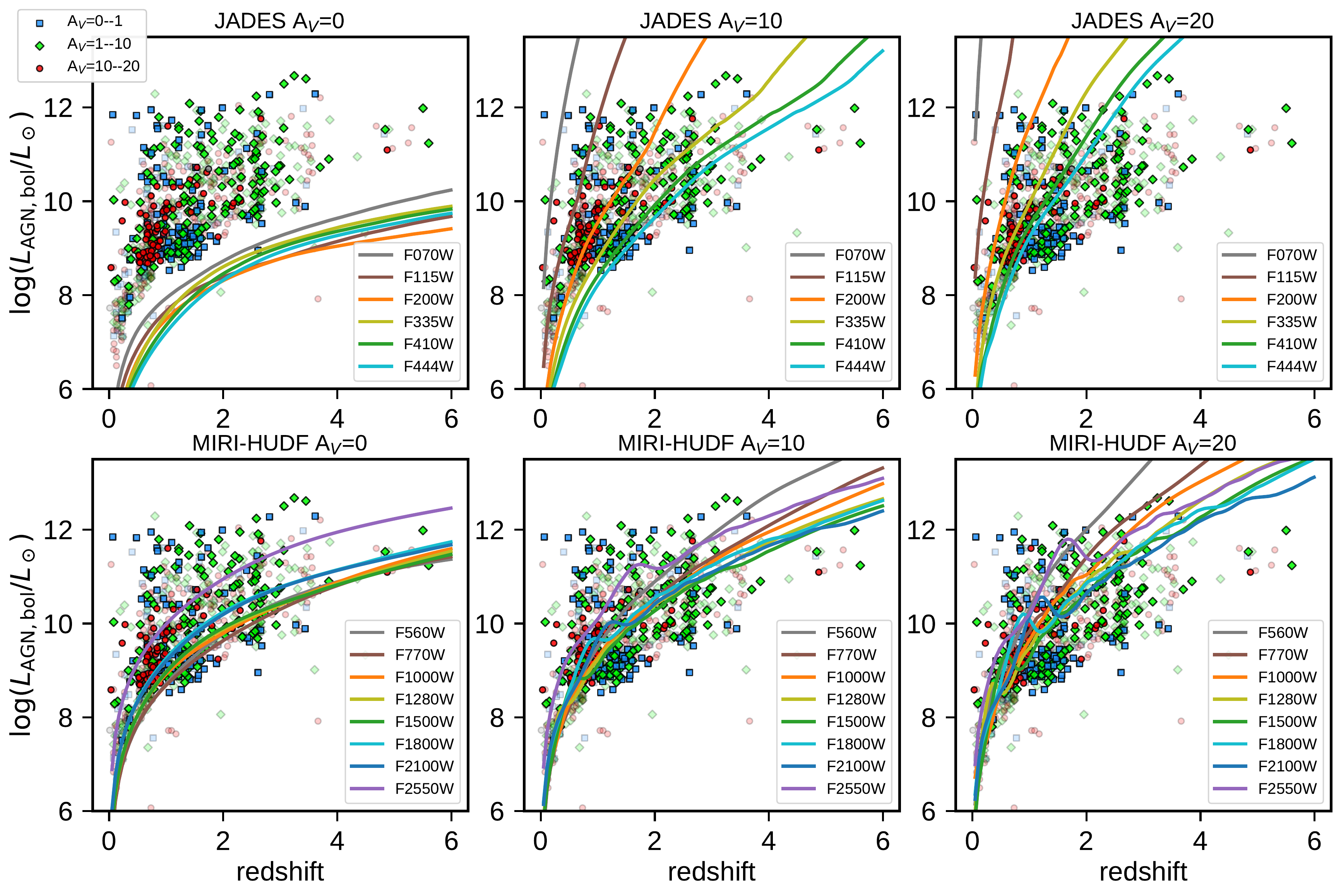}
    \caption{
        { $L_{\rm AGN, bol}$ detection limits for different filters (colorful solid lines) in the JADES
        (top panels) and MIRI HUDF surveys (bottom panels). We show the predictions for 
        unobscured AGN ($A_{\rm V}=0$) on the left, heavily-obscured AGN ($A_{\rm V}=10$, corresponding
        to a $\sim$10~\mum~silicate absorption
        strength $\tau_{10}\sim$1.5) in the middle, and extremely-obscured AGN
        ($A_{\rm V}=20$, corresponding to $\tau_{10}\sim$2.6) on the right. For all the bands, we assume the AGN light contributes 
        20\% of the total flux. In all the panels, we also plot the AGNs
        identified in this work with blue squares for low obscuration ($A_{\rm V}=$0--1), green diamonds for moderate
        obscuration ($A_{\rm V}=$1--10) and red dots for high obscuration ($A_{\rm V}$=10--20). The SED-identified 
        AGNs are denoted in darker colors while the other AGNs are shown in lighter colors. 
        }
    }
  \label{fig:jwst_curve}
    \end{center}
\end{figure*}

\section{Conclusion and Summary}\label{sec:summary}

{ With the updated ultra-deep JVLA images at 3 GHz and 6 GHz and other archival 
measurements in the X-ray, optical and near- to mid-IR of the GOODS-S field, this paper 
has explored the effectiveness of finding obscured AGNs with various
techniques in an effort to reach a complete AGN census from observations.

Compared with previous work, this study has taken full advantage of the deepest multi-wavelength datasets available and offered a relatively unbiased sampling of both unobscured and obscured AGNs.
Given the high sensitivities of our JVLA data, the radio emission of virtually all of
the detected galaxies is dominated by the output of star formation. In cases where AGN dominates the radio 
emission, the selection should still not be affected by obscuration effects
as the radio emission in the GHz range should be optically thin. Thus, we have focused our AGN
selections on the radio-detected sample as it should provide a relatively unbiased sample
in terms of AGN types. In addition, we have also evaluated a secondary sample of AGNs detected through
either X-rays or mid-IR, at which wavelengths the obscuration effects are also believed to be relatively weak,
to compensate any bolometrically bright AGNs that are missed in the radio band. In other words,
our AGN identifications are not limited to sources detected at one band, but all the sources detected in the X-ray, mid-IR or radio.
}

{ Regarding AGN selection techniques, our most important contribution is the introduction
of an improved version of the Bayesian SED fitting program {\it Prospector}.} We have made improvements in a suite of carefully-calibrated
empirical templates of AGN and galaxy dust emission, integrated them with the FSPS
stellar model within {\it Prospector}, and fully tested their operation and fidelity by
fitting various galaxies and AGNs.  A unique addition of this SED model is an
obscuration law for AGNs based on the comparison of unobscured and obscured AGN
SEDs that allows a continuous range of obscuration to be applied to pure Type-1
(unobscured) templates. Our overall approach has advantages in fitting
observations with a limited set of observational constraints and { does not have
serious over-fitting problems compared with other SED fitting tools with too many 
free parameters. The complexities of our templates and SED model can be adjusted as needed
to reproduce the average observed features of each component when the data constraints are limited
and to match the detailed SED variations when the data constrains are relatively rich.}

In this work, we have applied this SED fitting tool to identify both obscured and unobscured AGN candidates 
and characterize their properties in the GOODS-S region where the primary
infrared constraints are IRAC and MIPS photometry. Due to the limited
SED constraints in the mid-IR currently available, we have used a sparse set of
templates selected from past experience to be capable of giving accurate fits
in most cases, and which are designed to avoid giving false positives for AGN
detection. Compared with traditional infrared color-color diagrams, our approach has the 
important advantage to identify strongly obscured AGNs in a significantly more
complete fashion.

%In total, we have identified a total of 907 AGNs in the GOODS-S field,
%including 323 VLA-detected AGNs and 585 VLA-undetected AGNs, with eight
%different selection methods from the X-ray to the radio-band. This is the
%largest AGN sample constructed in the GOODS-S field, taking advantage of { the deepest
%multi-wavelength datasets available}. One key feature of this study is the
%inclusion of all the sources detected in the X-ray, mid-IR and radio, offering
%a relatively unbiased sampling of both unobscured and obscured AGNs with all
%the relevant data. Value-added catalogs that summarize the selection techniques
%and source properties of these AGNs are provided.

Combining the SED fittings and other seven selection methods from the X-ray to the radio-band, we have identified 
a total of 901 AGNs in the GOODS-S field, including 323 JVLA-detected AGNs and 578 JVLA-undetected AGNs. 
Value-added catalogs that summarize the selection techniques and source properties of these AGNs are provided.

These AGNs are identified, using (in order of decreasing percentage yield): (1)
ratio of X-ray to radio luminosity (65\%); (2) X-ray luminosity (36\%); (3)
infrared SED features (32\%); (4) variability (12\%); (5) optical SED features
(11\%); (6) optical spectroscopy (5.3\%); (7) high radio flux (4.8\%); and (8)
flat radio spectrum (2.0\%). However, most methods find AGNs not revealed by any
of the others, so nearly all are essential for the most complete possible
sample and any selection omitting methods will be incomplete and biased.  To illustrate, 
the identified AGN numbers would change if we limited
the sources to be detected at one particular band, with 73\% of the whole
sample detected in the X-ray, 64\% in the mid-IR, and 36\% detected in the
radio. These differences are caused by a mixture of data coverage and
sensitivity, source properties and the effectiveness of the selection
techniques.

Roughly 20\% of the sample appears to be strongly obscured, emphasizing the
importance of searching using SED fitting in the infrared. Compared with the
previous AGN study of X-ray detected sources by \cite{Luo2017}, we have
doubled the AGN number within the same survey footprint and about 26\% of our
AGN sample is not even detected in the deepest {\it Chandra} image. Compared
with \cite{Alberts2020}, we have also doubled the AGN number among
radio-detected galaxies thanks to our improved SED identification as well as
the addition of X-ray/optical variable sources. { Our reported AGN number density
is also notably higher than a previous study of COSMOS field \citep[e.g.,][]{algera2020} despite
the identical depths of JVLA radio observations at similar frequencies, thanks to the much deeper X-ray and mid-IR
coverages of GOODS-S and additional selection techniques adopted in this work.}

The biases in samples using a subset of the search methods or limiting the
selection to sources detected in one band emphasize the desirability of
selecting samples by bolometric luminosity. The most promising bands for this
goal are hard X-ray and infrared near rest-frame 5 $\mu$m. It is possible that
an approach combining these bands would be even more powerful, since the X-ray
energy absorbed in Compton thick cases emerges in the mid-infrared.
Nonetheless, there are AGNs that deviate from the normal cases and that can be
missed either in X-ray or mid-IR (e.g., AGN emission that might be
intrinsically weak in the X-ray or mid-IR), further highlighting the importance
of AGN selections across the full wavelength range.

We have carried out a preliminary study of the properties of the obscured AGN
population and found that (1) there is no strong correlation between the dust
obscuration (traced by optical to mid-IR SED) and gas absorption (traced by the
X-ray band ratios); (2) the obscured AGN fraction in the UV-optical appears to
decrease with increasing AGN bolometric luminosity; however, the obscured
fraction in the X-ray does not show an obvious luminosity dependence; (3) there
is no convincing evidence for redshift evolution of the AGN obscuration fraction in
the X-ray and UV-optical within the redshift range ($z=$0.5--3.5) studied in
this work.

We identify a high density of AGNs on the sky in this field with 5.3
arcmin$^{-2}$ over the whole GOODS-S field and 7.2 arcmin$^{-2}$ near the HUDF,
limited by the depth of all the multi-wavelength data and the effectiveness of
selection techniques. Within the errors, this density and the corresponding AGN
luminosity function agree with the theoretically derived luminosity function
from \citet{Shen2020}. However, in both cases the heavily absorbed cases, e.g.,
Compton thick, are likely to be underrepresented and their numbers are not well
constrained. 

This work builds the foundation for the future AGN studies with upcoming JWST
data.  More obscured AGNs are expected to be discovered by JWST, advancing us
toward a truly complete understanding of the diversity of these objects.

\begin{acknowledgments}
We appreciate the comments and suggestions from the anonymous referee. This work was supported
by NASA grants NNX13AD82G and 1255094. This work is partly based on
observations taken by the 3D-HST Treasury Program (HST-GO-12177 and HST-GO-12328) with the NASA/ESA
Hubble Space Telescope, which is operated by the Association of Universities for Research in Astronomy, Inc., under 
NASA contract NAS5-26555.
\end{acknowledgments}

\software{Dynesty \citep{Speagle2020}, Matplotlib \citep{matplotlib}, Astropy \citep{astropy2013, astropy2018}, Prospector \citep{Johnson2021}, PyBDSF \citep{PyBDSF}}

\appendix

\section{Inputs to the SED Fitting}\label{app:sed_input}

{\it Spitzer} observations launched a variety of new  methods to use infrared
photometry to identify active galactic nuclei (AGNs), most notably IRAC
color-color diagrams proposed by \citet{lacy2004} and \citet{stern2005}.  An
alternative approach used fitting to the IRAC colors to identify objects with
power law SEDs \citep{alonso2006, donley2007}.  These methods have proven quite
powerful and are widely used.  However, because the longest IRAC band is
centered at 8 $\mu$m, they lose sensitivity to obscured AGN at modest redshift,
where they are best suited to find cases with typical Type-1 SEDs or modestly
obscured Type-1 behavior \citep{Donley2012}. However, using the R - [4.5]
color, \citet{hickox2007} found a significant population of obscured AGN in the
Bootes field. At the same time, they show clearly (their Figure 9) how there is
a strong bias against strongly obscured AGN even at moderate redshift. The
additional band at 12 $\mu$m from WISE allowed more complete searches for
obscured nuclei \citep[e.g.,][]{stern2012, hainline2014, hviding2018,
carroll2021}. The addition of the two mid-infrared bands from Akari provided
more possibilities in this direction \citep{lam2019}. However, both of these
datasets are relatively shallow, limiting the results to either very luminous
or relatively nearby examples.

An  issue is that, with deep data in the IRAC wavelength range (3--9
$\mu$m), the color selection becomes degenerate with stellar continua
\citep[e.g.,][]{donley2008, mendez2013}.The limited ability of these methods to
identify moderately to heavily obscured objects is also a weakness, leading to
ambiguous results for such cases. For example, \citet{mendez2013} concluded
that only 10\% of IR-selected AGNs are heavily obscured, but this result
appears to be contradicted by \citet{delmoro2016}, who concluded that 30\% of
the IRAC-selected AGN were too obscured in the X-ray to be detected there.
Indeed, the predictions of a very large population of heavily obscured AGN
\citep[e.g.,][]{ananna2019} suggest that the current methods are only finding a
small fraction. 

A more powerful approach is to fit the photometry with the expected outputs of
the various underlying components - stars, AGN, star-formation powered infrared
excess, extinction. Models of this type have been demonstrated by
\citet{brown2019,azadi2020,poul2020}. However, these  models are characterized
by a large number of free parameters, e.g., 23 are listed by \citet{poul2020},
and/or are strongly based on theoretical models that themselves are subject to
many free parameters and are not constrained before being used in the fits by
matching observations of well-studied AGNs \citep{brown2019,azadi2020}. These
methods are usefully applied to study the nature of well-measured AGNs, but not
optimized for the less-constrained issue of finding missing ones.
{ \citet{yang2021} have developed another modeling approach to find AGNs in JWST
measurements; it is more reliant on theoretical AGN models than our study is. } \citet{assef2011} used
color-color diagrams for the initial AGN identification and then fit templates
to understand the sample in more depth.  \citet{Alberts2020} searched for
hidden AGN by fitting more accurate  templates on much deeper data; our work
builds on those latter results by including the ability to recognize heavily
obscured AGN and using a more sophisticated fitting approach. { The approach of \citet{algera2020} 
is similar to ours; the main differences with it have to do with the deeper observational database 
in the GOODS-S/HUDF, our use of additional AGN identification approachs, and our integration 
of more up-to-date SED templates into a comprehensive fitting approach.}

In general, the ability of such fitting approaches to provide unambiguous
answers depends critically on selecting inputs that provide accurate fits with
a minimum of free parameters. Our approach to this challenge differs from a number of the
other works.  We have used the Bayesian fitting program {\it Prospector} to fit the
stellar continua. The stellar model is from the Flexible Stellar Population
Synthesis code \citep[FSPS;][]{Conroy2009, Conroy2010}. We have assumed a
Kroupa initial mass function and a star formation history (SFH) described by a
delay tau-model $t e^{-t/\tau}$. A \cite{Calzetti2000} attenuation curve is
taken for the stellar extinction. In total, there are five free parameters: (1)
stellar masses formed ($mass$), (2) stellar metallicity ($logzsol$), (3)
stellar extinction level at 5500~\AA ($dust2$), (4) stellar population age
($tage$), (5) e-folding time of the star formation history ($\tau$).

In the GOODS-S region that we are studying, the data over the relevant spectral
range are extensive and {\it Prospector} is thoroughly optimized to provide accurate
fits. We have, in fact, confirmed this capability by fitting sets of data where
we vary the far infrared luminosity by factors of six up and down and challenge
{\it Prospector} to find accurate fits while maintaining energy balance, and in fact
excellent fits were returned. However, {\it Prospector} historically has not included
equally sophisticated capabilities for modeling in the infrared, that is
including emission by dust heated by young stars or by AGNs. 

In the spectral region of interest for our models, the dominant feature in
star-formation-heated dust is aromatic bands, and in fact the bands short of
$\sim$ 15 $\mu$m, since the longer-wavelength SED is shifted beyond reach of
the {\it Spitzer} 24 $\mu$m band (and of JWST/MIRI) at the redshifts of
interest. The general behavior of these bands has little dependence upon
metallicity (although the 17 $\mu$m complex does change significantly), as can
be seen comparing the spectra presented in \citet{hunt2010} with templates such
as from \citet{Rieke2009}. Moreover, at the redshifts ($\sim$ 2)  and
luminosities of interest, the variety of star-formation-heated far infrared
SEDs is much less than locally \citep{rujo2011, rujo2013, shipley2016}.
\citet{DeRossi2018} demonstrate that the \citet{Rieke2009} templates provide a
good fit to the observations of luminous galaxies at the relevant redshifts.

Our templates make two modifications to the \citet{Rieke2009} models: (1) we
extend the spectrum of emission by dust to $\lambda<5\mu$m; and (2) we remove
the stellar emission in the mid-infrared, which is included in the original
templates.  We have re-constructed the near- to mid-IR part of these templates
with more observations based on {\it Spitzer}/IRS and AKARI/IRC spectra and
subtracted (or reduced) the contamination from galaxy stellar emission. We use
this determination of the stellar photospheric emission to determine the
contribution of the emission by dust in the 2--5 $\mu$m range. We then have
an IR luminosity-dependent library of pure galaxy dust emission SEDs. In
principle, with the decompositions of PAH features and dust continuum and the
interpolations of the templates between different IR luminosities, there could
be three parameters to capture the behaviors of SFG dust emission: (1) the
template IR luminosity {\it L\_IR\_temp}, (2) the scaling factor of the SFG
template {\it L\_IR\_obs}, and (3) the relative strength of the PAH emission
features $f_{\rm PAH}$.  However, the region of the spectra of interest for our
fitting is dominated by the 6--12 $\mu$m aromatic bands. These bands are
usually similar in relative strength; although there are extreme cases, typical
variations in relative band strengths are $\sim$ 30\%, 1-$\sigma$
\citep{smith2007}. This approximate invariance persists even to low metallicity
\citep{hunt2010}. In addition, in the 2--15 $\mu$m range, the continua
underlying the aromatic bands are very similar, independent of luminosity.  We
adopt the template for log(L(TIR)) = 11.25, which is closely representative of
those from other relevant luminosities (see Figure~\ref {fig:fitting_example})
and has been shown to give an optimum fit for a large sample of galaxies at
similar redshifts \citep{DeRossi2018}. Thus, we achieve adequate fits to the
star-formation-heated dust with a single normalization free parameter.

For AGNs, the underlying intrinsic SED from \citet{Lyu2017a, Lyu2017b} from
X-ray through the mid-infrared can be represented as one of the ``normal'',
``hot dust deficient (HDD)'' or ``warm dust deficient (WDD)'' templates with a
cutoff near 30$ \mu$m. To minimize the free parameters, we base our fit on the
normal template, which is the most common. The cutoff seems to be a reasonably
universal behavior; even in the rare cases where AGNs have extreme far infrared
outputs, this behavior is attributed to star-formation-heated dust
\citep{kirkpatrick2020}.  In the general, we would account for the emission of
polar dust as in \citet{Lyu2018}.  However, the presence of sufficient emission
from this dust to affect the SED appears to be uncommon at the relatively high
luminosities of most of our sample. For example, of 91 quasars with 0.5 $<$ z
$<$ 3, \citet{xu2015} found that the a warm component resembling those due to
polar dust was needed for only 8. Since this sample was selected at 24 $\mu$m,
over this redshift range the rest wavelengths are where the polar dust emission
is strong, so the sample may even be biased toward a high detection rate. This
possibility is supported by the good fits without polar dust reported by
\citet{Lyu2017b} for 87 PG quasars, selected on the basis of UV excess. We
therefore leave this source component out of our fits. To account for
attenuation of the intrinsic SED in obscured AGN, we have compared the SEDs of
standard unobscured and obscured examples to generate an obscuration curve.
This procedure is similar to the approach to determine interstellar extinction
laws. However, these laws are unlikely to apply directly to obscured AGN
because both of the possibility that the dust grain properties are modified in
the extreme environments, and because the attenuation arises from dust in a
range of extreme environments. {  Derivation of a more appropriate attenuation law for AGNs will be described in more detail
elsewhere. In a useful empirical solution, the AGN component can be based on the AGN template libraries introduced in \cite{Lyu2017a, Lyu2017b,
Lyu2018}.} This approach requires four free parameters to account for (1) the selection and
(2) normalization ({\it L\_AGN}) of an appropriate near-infrared continuum,
described by the strength of the hot dust component ($f_{\rm HD}$); (3) the
strength the polar dust emission ($f_{\rm WD}$); and (4) the obscuration,
characterized by the extinction level at 5500$\AA$ ($\tau_{\rm AGN}$).

\section{Additional Information of AGN demographics in different samples}\label{app:agn-demo-more}

In the main paper, our discussion is focused on the whole AGN sample. Here we
offer additional tables or plots for AGN demographics in different sub-samples
(e.g., detected in radio, X-ray or mid-IR).  See
Figure~\ref{fig:agn_counts_sub} and Table~\ref{tab:method-eff}.

\begin{figure*}[htb]
    \begin{center}
        \includegraphics[width=1.0\hsize]{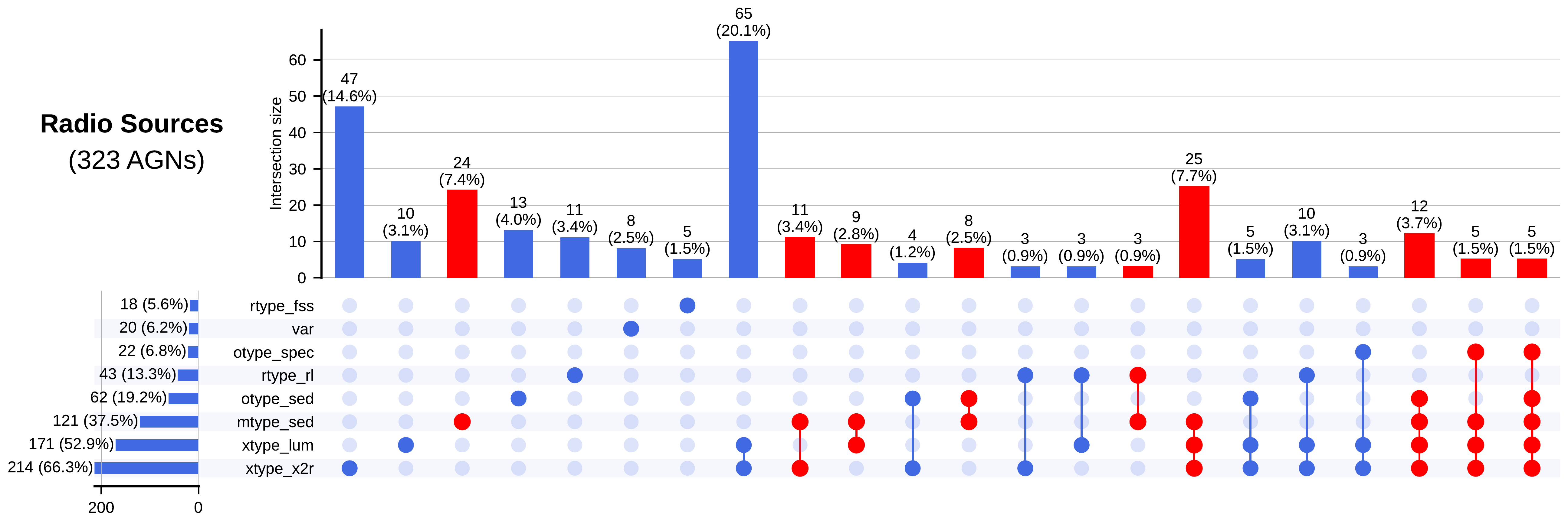}
        \includegraphics[width=1.0\hsize]{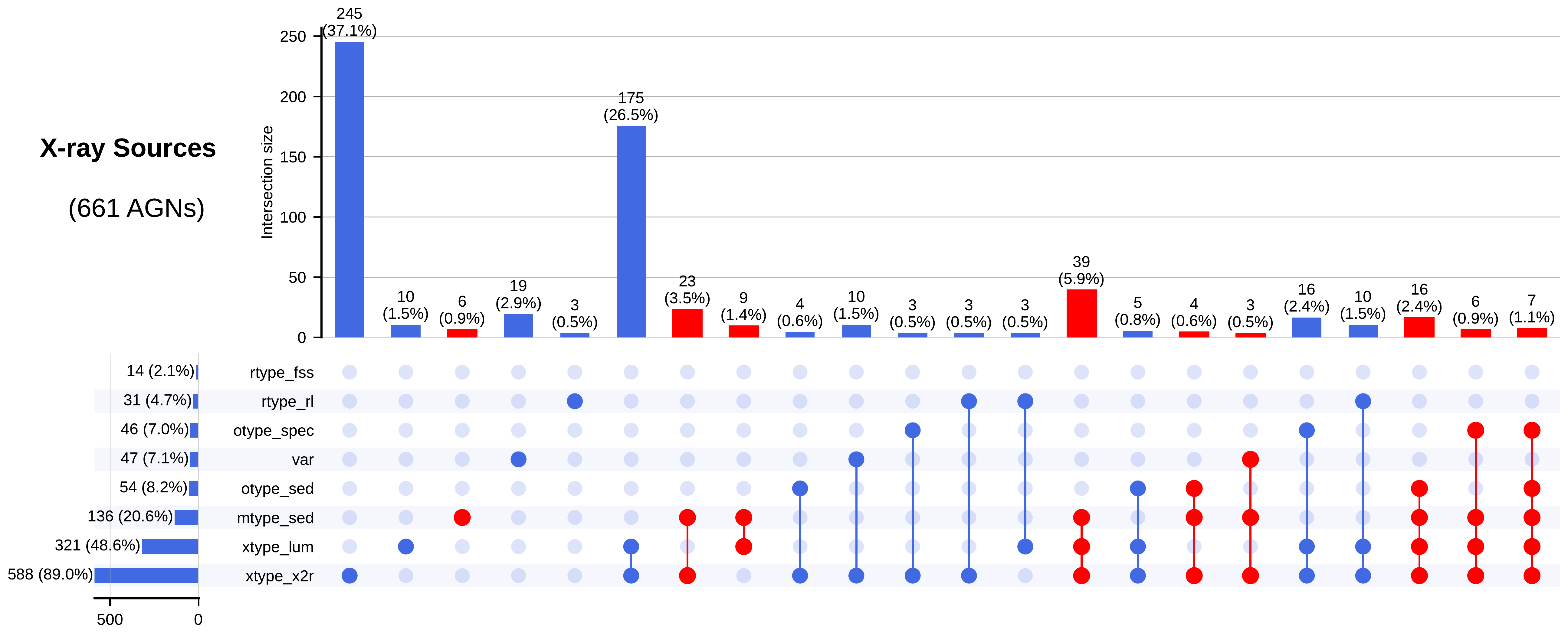}
        \includegraphics[width=1.0\hsize]{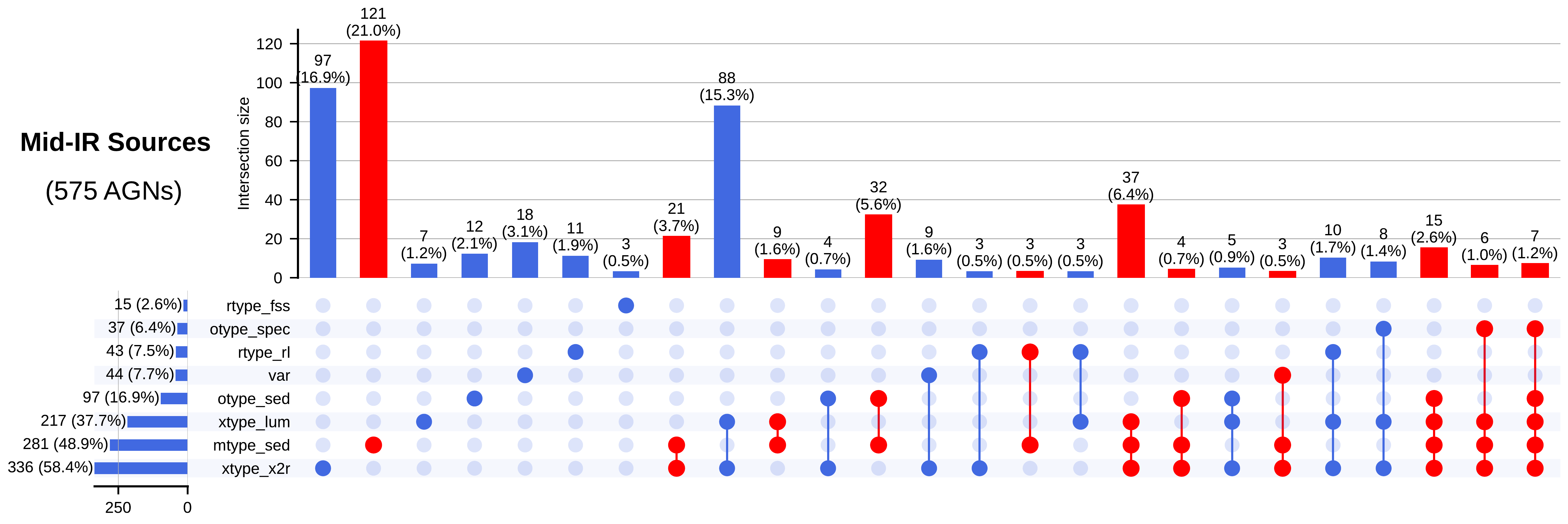}
    \caption{
        Plots similar to Figure~\ref{fig:agn_counts}, but for AGNs detected
        in sub-samples of radio-selected, x-ray-detected or mid-IR parent
        sources.
    }
  \label{fig:agn_counts_sub}
    \end{center}
\end{figure*}

\begin{deluxetable*}{ccccccccc}
    %\tabletypesize{\scriptsize}
    \tabletypesize{\footnotesize}
    \tablewidth{1.0\hsize}
    \tablecolumns{9}
    \tablecaption{Accessibility and effectiveness of the AGN selection methods\label{tab:method-eff}}
    \tablehead{
  \colhead{} &
  \colhead{$[$xtype\_lum$]$} &
  \colhead{$[$xtype\_x2r$]$} &
  \colhead{$[$otype\_sp$]$} &
  \colhead{$[$sed$]$} &
  \colhead{$[$mtype\_irac$]$} &
  \colhead{$[$rtype\_rl$]$} &
  \colhead{$[$rtype\_fss$]$} &
  \colhead{$[$var$]$}
%   \colhead{(1)} &
%   \colhead{(2)} &
%   \colhead{(3)} &
%   \colhead{(4)} &
%   \colhead{(5)} &
%   \colhead{(6)} &
%   \colhead{(7)} &
%   \colhead{(8)} &
%   \colhead{(9)} 
}
\startdata
\multicolumn{9}{c}{AGN in radio sources (323)} \\
\hline
$f_\textrm{yes}$  &  52.9\%  &  66.3\%    &  6.8\%   &   45.2\% &   17.0\% &    13.3\% &   5.6\% &    5.9\% \\
$f_\textrm{acc}$  &  83.3\%  &  83.3\%    & 61.6\%   &   100\% &    91.6\% &    92.0\% &  35.9\% &     100\% \\
 \hline
\multicolumn{9}{c}{AGN in X-ray sources (661)} \\
\hline
$f_\textrm{yes}$  &  48.6\% &   89.0\% &    7.0\% &   22.5\% &      10.7\% &    4.7\% &     2.1\% &     7.0\% \\
$f_\textrm{acc}$  &  100\% &   100\% &  56.6\%   & 100\% &  79.9\% &  37.2\% &  15.9\% &       100\% \\
 \hline
\multicolumn{9}{c}{AGN in mid-IR sample (575)} \\
\hline
$f_\textrm{yes}$  & 37.7\% &  58.4\% &  6.4\%  &  53.0\%   &  17.7\% &   7.5\% &  2.6\% &  7.5\%  \\
$f_\textrm{acc}$  & 68.0\% &  68.0\% &  58.3\% & 100\%    &  90.1\% &  51.7\% & 19.1\% &  100\% \\
 \hline
\multicolumn{9}{c}{AGN in other sources (55)} \\
\hline
$f_\textrm{yes}$  & 0.0\%  &  0.0\%  &   0.0\%   &  0.0\%  &    0.0\%  &    0.0\% &     0.0\% &    100\% \\
$f_\textrm{acc}$  & 0.0\%  &  0.0\%  &   5.5\%   &   100\%  &   9.1\%  &    0.0\% &     0.0\% &     100\% \\
\enddata
%\tablenotetext{1}{All these priors have been sampled linearly.}
%\tablenotetext{2}{The ``uncertainties'' of these median values are 2 $\sigma$ (i.e., 2.5\%, 50\% and 97.5\% quantiles).}
\tablecomments{ $f_\textrm{yes}$ -- the fraction of AGN (in each sample) identified with this method; $f_\textrm{acc}$ -- the fraction of AGN (in each sample) which this method can be applied.
 }
\end{deluxetable*}

\section{Average AGN SED Model}\label{app:agn-sed-model}

To investigate the detection limits of various survey bands, we construct an
SED model for typical AGNs with varying levels of obscuration and compute the
corresponding $L_{\rm AGN, bol}$ that a single survey band can detect as a
function of redshift. Figure~\ref{fig:sed_sens} illustrates this SED model
together with the 5-$\sigma$ sensitivities of the photometric bands used in
this work. The details of the model configurations are given below.

The AGN average intrinsic SED template is developed on the basis of our normal
AGN template, which is representative for the UV to far-IR intrinsic emission
of most AGNs at very broad ranges of luminosity and redshift \citep[see details
and further references in][]{Lyu2017a, Lyu2017b, Lyu2018}. We extended this
template to the hard X-ray and radio bands with the relative AGN emission
strength constrained by the  \cite{Richards2006} quasar templates normalized to
ours at 0.44 $\mu$m. As the \cite{Richards2006} templates only have a few very
limited data points at these wavelengths, we have adopted analytical SED models
to fill the wavelength gaps. In the X-ray (0.1--100 KeV), we assume the
intrinsic spectrum can be described by a cut-off power-law model $f(E)\sim
E^{1-\Gamma}\exp(-E/E_c)$, where $\Gamma$ is the photon index and $E_c$ is the
cut-off energy and their values are assumed to be 1.9 and 300 KeV
\citep[e.g.,][]{Aird2015}. In the radio-band ($\lambda\gtrsim300~\mum$), we
assume a power-law spectrum with $f_\nu\propto\nu^{-0.8}$ for the synchrotron
emission for both radio-loud and { radio-quiet} cases. The relative strengths of
all these model spectra are normalized according to the \cite{Richards2006}
templates.

For the gas obscuration in the X-ray, we compute the attenuated X-ray SEDs at
different gas column densities $N_{\rm H}$ by obscuring our spectrum with the
flux correction factors given by the PIMMS. In
the simulator, we assume the intrinsic X-ray spectrum is a power-law $\propto
E^{-\alpha}$ with the photon index $\alpha$=1.9, and calculate the fraction
between the observed and intrinsic flux (i.e., the flux correction factor) at
different wavelengths for a range of hydrogen column densities. We interpolate
the discrete values of these values and build an analytical model showing how
the X-ray spectrum should vary as a function of $N_{\rm H}$.

The AGN UV-to-IR extinction is treated independently from the X-ray extinction
given the complicated relation between the X-ray gas absorption and UV-to-IR
dust extinction in AGNs (see discussion in Section~\ref{sec:obscured_pop}). We
obscure the AGN intrinsic UV-to-IR SED assuming a simple screen-like dust
geometry and an AGN attenuation curve given by Lyu et al. (in prep.) and
compute the attenuated SEDs from $A_V=0$ to $A_V=20$. The upper limit of $A_V$
yields a silicate absorption strength at $\sim10~\mum$ of 2.6, consistent with
the values for the most heavily obscured AGNs in the mid-IR (e.g.,
\citealt{Hao2007}).

\end{document}